\documentclass[aps,prd,amsmath,floats,floatfix,onecolumn,superscriptaddress,showpacs,nofootinbib]{revtex4}

\usepackage[T1]{fontenc} 
\usepackage{aecompl}
\usepackage{amssymb}
\usepackage{amsmath}
\usepackage{float}
\usepackage{latexsym}
\usepackage{epsfig}
\usepackage{caption}
\usepackage{graphics}
\usepackage{graphicx}
\usepackage{epstopdf}
\usepackage{array}
\usepackage{color}
\usepackage[linktocpage]{hyperref}
\usepackage{subfigure}

\usepackage{multirow}

\begin{document}


\title{Description of the evolution of inhomogeneities on a Dark
  Matter halo with the Vlasov equation}

\author{Paola Dom\'inguez-Fern\'andez}
\email[]{paola.dominguez@uni-bonn.de}
\affiliation{Argelander Institut f\"ur Astronomie, Universit\"at Bonn, Auf dem H\"ugel
71, 53121, Bonn, Germany}

\author{Erik Jim\'enez-V\'azquez}
\email[]{erjive@ciencias.unam.mx}
\affiliation{Instituto de Ciencias Nucleares, Universidad Nacional
  Aut\'onoma de M\'exico, Circuito Exterior C.U., A.P. 70-543,
  M\'exico D.F. 04510, M\'exico}
  
\author{Miguel Alcubierre}
\email[]{malcubi@nucleares.unam.mx}
\affiliation{Instituto de Ciencias Nucleares, Universidad Nacional
  Aut\'onoma de M\'exico, Circuito Exterior C.U., A.P. 70-543,
  M\'exico D.F. 04510, M\'exico}

\author{Edison Montoya}
\email[]{emontoya@uis.edu.co}
\affiliation{Escuela de F\'{\i}sica, Universidad Industrial de Santander,
  A.A. 678, Bucaramanga, Colombia}
  
\author{Dar\'{\i}o N\'u\~nez}
\email[]{nunez@nucleares.unam.mx}
\affiliation{Instituto de Ciencias Nucleares, Universidad Nacional
  Aut\'onoma de M\'exico, Circuito Exterior C.U., A.P. 70-543,
  M\'exico D.F. 04510, M\'exico}
  
\email[]{paola.dominguez@uni-bonn.de}
\email[]{erjive@ciencias.unam.mx}
\email[]{emontoya@uis.edu.co}
\email[]{malcubi@nucleares.unam.mx}
\email[]{nunez@nucleares.unam.mx}


\date{\today}


\bigskip

\begin{abstract}
We use a direct numerical integration of the Vlasov equation in
spherical symmetry with a background gravitational potential to
determine the evolution of a collection of particles in different
models of a galactic halo. Such a collection is assumed to represent a
dark matter inhomogeneity which reaches a stationary state determined
by the virialization of the system.  We describe some features of the
stationary states and, by using several halo models, obtain
distinctive signatures for the evolution of the inhomogeneities in
each of the models.
\end{abstract}




\maketitle


\section{Introduction}
\label{introduction}

The nature of dark matter remains one of the most puzzling questions
in modern cosmology.  Its assumed that elementary particles could be 
the constituents of dark matter for which there have been several experimental
attempts to perform a direct detection of such dark matter particle, but still
a clear evidence for such particles remains absent \cite{Undagoitia:2015gya}. 
Examples of these experiments are the Large Underground Xenon (LUX) experiment
\cite{Akerib:2012ys}, the DAMA/Libra experiment \cite{Bernabei:2008yh}, ZEPLIN
experiment \cite{Akimov:2006qw}, XENON10 experiment \cite{Aprile:679A} or DAMIC
experiment at SNOLAB \cite{Barreto:2011zu}, \cite{Chavarria:2014ika, Aguilar-Arevalo:2016zop, Aguilar-Arevalo:2016ndq}.  Another
approach to determine the nature of dark matter is to find specific
features in the dynamics of visible matter in the presence of the
gravitational potential of a dark matter halo.

Dark matter is usually assumed to be a stable particle with no
electric charge, and with a negligible cross-section with baryonic
particles.  However, it does interact gravitationally with the rest of
the matter (in general relativistic terms, dark matter does curve
space-time).  It is also assumed that dark matter particles only
interact with themselves gravitationally, or that at least any
non-gravitational self-interaction is very weak. This means that a
fluid description of dark matter is not adequate (even though this
description may lead to some intuitive properties
\cite{Barranco:2013wy}). Indeed, the fluid approximation requires that
the mean-free path of particles is much smaller than the
characteristic size of the system, which clearly does not apply to
dark matter particles.  One should mention that there are serious
proposals that consider that dark matter can be described as an
ultra-light scalar field, and such proposals seem to ameliorate some
of the problems related with the usual dark matter paradigm (see
e.g.~\cite{Matos:1999et,Matos:2000xi,Matos:2003pe}).

If one considers dark matter as a collection of non-interactive
particles there are at least two different ways to study their
dynamics.  The standard approach has been to study them using
increasingly sophisticated direct gravitational N-body simulations,
with today can use billions of particles \cite{Navarro:1995iw, Harker:2005um}.
In N-body simulations, each particle represents a very large number of dark matter
particles and the interaction between two particles is computed by assuming that
they have a finite size and density profile, which leads to an effective softening 
force at small scales \cite{Athanassoula:10.1046},\cite{Dehnen:10.1046}. In this way, 
a given density velocity field is represented by
a set of particles \cite{Hockney:1981csup}. For more details on the techniques and
status of these simulations one can refer to \cite{Bagla:2004au}.
Another approach is to use a continuum
approximation based on kinetic theory and the Boltzmann
equation~\cite{BT2}.  In this second approach a collection of
particles is described in a collective way by means of a distribution
function $f$ defined as a probability density in the so-called phase
space, namely the space defined by generalized coordinates and
momenta. When the particles are non-interactive, one obtains the
homogeneous Boltzmann equation, also known as the Vlasov
equation~\cite{BT2}. Such description is nearly equivalent to the
numerical description of dark matter by means of the N-body
simulations, although there are conceptual and quantitative
differences in the two approaches~\cite{Colombi:2015eia}.

Solutions of the Vlasov equation are frequently not easy to
interpret. It is quite easy to solve in general, since by just by
expressing the distribution function as a function of conserved
quantities we automatically obtain an exact, and stationary,
solution~\cite{Andreasson:2005qy}. However, such general solutions
often do not have a clear physical interpretation, and can lead to
misunderstandings~\cite{Shapiro:1983du}. We have found it
productive then, to solve the Vlasov equation dynamically in order to study
the evolution in phase space of an initial distribution function,
which allows for a more clear physical interpretation. Using this
approach, we study the evolution of a spherically symmetric
distribution function in different gravitational potentials, which
describe several models of dark matter halos. The distribution
function is assumed to describe a dark matter inhomogeneity within the
dark matter halo.  The aim of this work is first to develop a robust
numerical integration code for the Vlasov equation, and then use it to test 
the stability of inhomogeneities in dark mater halos, as well as to obtain
non-trivial final stationary states for the initial inhomogeneity.

The paper is organized as follows: In Section~\ref{Sec1} we present a
brief review of kinetic theory, introducing the Vlasov equation. In
Section~\ref{Sec2} we present four different halo gravitational
potentials, and derive a dimensionless Vlasov equation for the
corresponding models. In Section~\ref{Sec3} we describe the code used
to perform the numerical simulations of the evolution of the
distribution function by means of the Vlasov equation. After that, we
present our results for each halo model in Section~\ref{Sec4}.
Finally, we discuss our findings in Section~\ref{Sec5}.


\section{Kinetic theory and Vlasov equation}
\label{Sec1}


\subsection{Distribution function in spherical symmetry}

The particle distribution function $f$ lives in a higher dimensional
phase-space: six dimensional in the case of classical mechanics, and
seven dimensional in relativistic physics (down from eight due to the
constraint on the magnitude of the momenta, the so-called mass-shell
condition $p_\mu p^\mu=-m^2 c^2$, where $m$ is the particle's mass and
$c$ denotes the speed of light). The general explicit invariant volume
element in momentum space in the general relativistic description is
\begin{equation} \label{eq:dw}
d \omega = \frac{d^3 \, p_{*}}{p^0\,\sqrt{-g}} \; ,
\end{equation}
where $p_{*}$ denotes the covariant momenta, $p_i=\partial L /
\partial \dot{x}^i$, with $L$ the Lagrangian, and where $g$ is the
determinant of the spacetime metric. In the Newtonian limit this
volume element takes the form
\begin{equation}
d \omega =  \frac{d^3\,p_*}{\sqrt{h}} \; ,
\end{equation}
where now $h$ is the determinant of the flat metric in the coordinates
under consideration, and the momenta $p_{*}$ corresponds to the
conjugate momenta.  Notice that since in general the volume element in
physical space is given by \mbox{$dV = \sqrt{h} \; d^3 x$}, one finds
that the phase space volume element $dV d \omega$ is in fact
invariant.

Keeping in the Newtonian limit, when using Cartesian coordinates the
invariant volume element in momentum space takes the form $d \omega =
dp_x dp_y dp_z$, and when using spherical coordinates it becomes
\begin{equation}
d \omega=\frac{dp_r \; dp_\theta \; dp_\varphi}{r^2 \sin \theta} \; .
\label{eq:dw0}
\end{equation}

There is in fact another expression that is commonly used for
the invariant volume element in momentum space in spherical
coordinates, namely \mbox{$d \omega = p^2 \sin \theta_p \; dp \;
  d\theta_p \; d\varphi_p$}, where $p$ is the magnitude of the
momentum, and $(\theta_p,\varphi_p$) are angles defined in momentum
space in the same way as ($\theta$,$\varphi$) are defined in the space
sector, so that for example $p_x = p \sin \theta_p \cos \varphi_p, p_y
= p \sin \theta_p \sin \varphi_p, p_z = p \cos \theta_p$. This volume
element can be related to the former expression, Eq.~(\ref{eq:dw0}),
by the Jacobian of the corresponding coordinate
transformation~\cite{Dominguez:2015,Jimenez:2016}.

In the following we will consider a stationary background
gravitational potential with spherical symmetry, and a distribution
function $f(t,x^i,p_i)$ that is also spherically symmetric. Moreover,
instead of the angular conjugated momenta, $p_\theta, p_\varphi$, we
will use the conserved total angular momentum squared, $L^2$, and an
auxiliary angle, $\psi$:
\begin{eqnarray}
L^2 &=& p_\theta^2 + \frac{p_\varphi^2}{\sin^2 \theta}, \label{eq:L2} \\
\psi &=& \arctan\left(\frac{p_\theta \sin\theta}{p_\varphi}\right),
\end{eqnarray}
so that the invariant volume element, Eq.~(\ref{eq:dw}), takes the form
\begin{equation}
d \omega = \frac{1}{2 r^2} \; dp_r \, dL^2\, d\psi
= \frac{L}{r^2} \; dp_r \, dL \, d\psi \; .
\label{eq:dw1}
\end{equation}

Finally, since we will be working with spherical symmetry, the
distribution function will have no dependence on $\psi$. 

We will therefore integrate over $\psi$ so the volume element used for this
work will have the final form
\begin{equation}
d \bar{\omega} = \int \left( d \omega \right) \; d \psi
= \frac{\pi}{r^2} \; dp_r\,dL^2
= \frac{2 \pi L}{r^2} \; dp_r\,dL \; .
\label{eq:dw2}
\end{equation}

In spherical symmetry the distribution function will have the form $f
= f(t,r,p_r,L)$. This implies that in the case of spherical symmetry,
the phase space is effectively tri-dimensional.  At this point it is
important to mention a couple of details about this distribution
function.  First, even though we are in spherical symmetry, the
individual particles can have non-trivial angular momentum. The
spherical symmetry will be maintained as long as at any given point in
space with a certain radial distance $r$ from the origin, the
tangential velocities of the particles are distributed uniformly in
every possible direction.  Second, notice that since $f$ depends on
the angular momentum $L$, and this is defined
through~\eqref{eq:L2}, the distribution $f$ when rewritten in terms of
$(p_r,p_\theta,p_\varphi)$ will in fact depend on the angle $\theta$.
This might seem odd since after all we are in spherical symmetry, but
as we will see below this dependence in fact cancels out in the Vlasov
equation.

We can now define the particle density $\rho_f$ in physical space, the
momentum density $j_f$, and the mean value over momentum space of an
arbitrary function $g$, will take the form
\begin{eqnarray}
\rho_f(t,r) &\equiv& \int f(t,r,p_r,L) \; d \bar{\omega}
= \frac{2 \pi}{r^2} \int f(t,r,p_r,L) \; L \; dp_r \; dL \; , \label{eq:rho1} \\
j_f(t,r) &\equiv& \int p_r f(t,r,p_r,L) \; d \bar{\omega}
= \frac{2 \pi}{r^2} \int p_r f(t,r,p_r,L) \; L \; dp_r \; dL \; ,
\label{eq:curr1} \\ 
\bar{g}(t,r) &\equiv& \int g f(t,r,p_r,L) \; d \bar{\omega}
= \frac{2 \pi}{r^2} \int g f(t,r,p_r,L) \; L\,dp_r\,dL \; . \label{eq:mean}
\end{eqnarray}
The total number of particles can then be defined as
\begin{equation}
N = \int f \; dV d \bar{\omega} = 8 \pi^2 \int f L \; dr dp_r dL
= 4 \pi \int \rho_f r^2 dr \; .
\label{eq:totalN}
\end{equation}
One can also define of the average value of a given physical quantity
over the complete phase space as
\begin{equation}
  \left<g\right> (t) \equiv 8 \pi^2 \int g f(t,r,p_r,L) \; L \; dp_r \; dL \; dr
  = 4 \pi \int \bar{g}(t,r) \; r^2 dr \; .
\label{eq:avg}
\end{equation}
As a further simplifying assumption, we will also assume that all the
particles have the same conserved angular momentum $L_0$, so that,
using a delta function, we can write $f(t,r,p_r,L) =
F(t,r,p_r)\delta(L-L_0)$, obtaining
\begin{eqnarray}
\rho_f(t,r) &=& \frac{2 \pi L_0}{r^2} \int F(t,r,p_r; L_0) \; dp_r \; ,
\label{eq:rho2} \\
j_f(t,r) &=& \frac{2 \pi L_0}{r^2} \int p_r F(t,r,p_r; L_0) \; dp_r \; ,
\label{eq:curr2} \\
\bar{g}(t,r) &=& \frac{2 \pi L_0}{r^2} \int g F(t,r,p_r; L_0) \; dp_r \; , 
\label{eq:mean2} \\
\left<g\right> (t) &=& 8 \pi^2 L_0 \int g F(t,r,p_r; L_0) \; dp_r \; dr \; , 
\label{eq:avg2} \\
N &=& 8 \pi^2 L_0 \int F(t,r,p_r; L_0) \; dr dp_r \; .
\label{eq:totalN2}
\end{eqnarray}

One should also mention the fact that for the particular case when the
motion is purely radial and the angular momentum is zero, $L_0=0$, the
above equations are in fact not correct and the derivation has to be
made again from the top (a naive application of these expressions
would seem to indicate that all the integrals above vanish). That
is, we have to begin with a Dirac delta dependence of the distribution function
in the angular momenta
\begin{equation}
f(t,r,p_r,p_\theta,p_\varphi)= F(t,r,p_r) \; \delta(p_\theta) \; \delta(p_\varphi/\sin 
\theta)
\end{equation}
Transforming the Dirac deltas from coordinates
($p_\theta$,$p_\varphi$) to the new coordinates ($L$,$\psi$) yields
\begin{equation*}
\delta(p_\theta) \; \delta(p_\varphi / \sin \theta)
= \frac{1}{L} \; \delta(L) \; \delta(\psi) \;  .
\end{equation*}
So that the expressions for the density, current, mean value and
average in the case of zero angular momentum become
\begin{eqnarray}
\rho_f(t,r) &=& \frac{1}{r^2} \int F(t,r,p_r) \; dp_r \; , \label{eq:rho3} \\
j_f(t,r) &=& \frac{1}{r^2} \int p_r \; F(t,r,p_r) \; dp_r \; , \label{eq:curr3} \\
\bar{g}(t,r) &=& \frac{1}{r^2} \int g \; F(t,r,p_r) \; dp_r \; , \label{eq:mean3} \\
\left<g\right> (t) &=& 4 \pi \int g \; F(t,r,p_r) \; dp_r \; dr \; , \label{eq:avg3} \\
N &=& 4 \pi \int  F(t,r,p_r) \; dp_r \; dr \; . \label{eq:totalN3}
\end{eqnarray}


\subsection{Vlasov equation}

Let us now consider the Vlasov equation.  In the general case, the
collisionless Boltzmann equation, also known as the Vlasov equation,
has the form
\begin{equation}
\frac{\partial f}{\partial t} + \frac{dx^i}{dt} \; \frac{\partial f}{\partial x^i}
+ \frac{dp^i}{dt} \; \frac{\partial f}{\partial p^i} = 0 \; .
\end{equation}
In spherical symmetry, and with the definitions given above, this
reduces to
\begin{equation}
\frac{\partial f}{\partial t} = - \frac{p_r}{m} \; \frac{\partial f}{\partial r}
- \frac{p_\theta}{m r^2} \; \frac{\partial f}{\partial \theta}
- \dot{p}_r \; \frac{\partial f}{\partial p_r}
- \dot{p}_\theta \; \frac{\partial f}{\partial p_\theta} \; . \label{eq:Vla0}
\end{equation}
Notice that, as mentioned before, even in spherical symmetry the
distribution function has a dependency on $\theta$ when written in
terms of the conjugate momenta.

Now, for a central potential $\Phi(r)$, the equations of motion are
\begin{eqnarray}
\dot{p}_r &=& - \frac{\partial {\cal H}}{\partial r}
= - \frac{\partial \Phi}{\partial r} + \frac{L^2}{m r^3} \; ,  \\
\dot{p}_\theta &=& - \frac{\partial {\cal H}}{\partial  \theta}
= \frac{{p_\varphi}^2 \cos \theta}{m r^2 \sin^3 \theta} \; ,
\end{eqnarray}
where ${\cal H}$ is the Hamiltonian of the system. This implies that
\begin{eqnarray}
\frac{p_\theta}{m r^2} \; \frac{\partial f}{\partial \theta}
+ \dot{p}_\theta \frac{\partial f}{\partial p_\theta}
&=& \frac{p_\theta}{m r^2} \; \frac{\partial f}{\partial \theta}
+ \frac{p_\varphi^2 \cos \theta}{m r^2\sin^3 \theta} \; \frac{\partial f}{\partial p_\theta} 
\nonumber \\
&=& \frac{p_\theta}{m r^2} \left( -\frac{2 p_\varphi^2 \cos\theta}{\sin^3 \theta} \right) 
\frac{\partial f}{\partial L^2}
+ \frac{p_\varphi^2\cos\theta} {m r^2 \sin^3 \theta} \;
\left( 2 p_\theta \right) \frac{\partial f}{\partial L^2} \nonumber \\ 
&=& 0 \; ,
\end{eqnarray}
where we have used the chain rule to calculate the derivatives of $f$
with respect to $\theta$ and $p_\theta$ in terms of its derivative
with respect to $L^2$.  This implies that in spherical symmetry the
Vlasov equation takes the final form:
\begin{equation}
\frac{\partial f}{\partial t}= - \frac{p_r}{m} \; \frac{\partial f}{\partial r} 
+ \left( \frac{\partial \Phi}{\partial r} - \frac{L^2}{m r^3} \right) \frac{\partial f}{\partial p_r} 
\; .
\label{eq:Vla1}
\end{equation}

Since there are no terms with derivatives with respect to $L$ in this
equation, we see that different values of the angular momenta evolve
independently of each other. The Vlasov equation in spherical symmetry
is then effectively two-dimensional. In the corresponding temporal evolution shown below,
for simplicity we will in fact assume that all particles have exactly
the same value of the angular momentum $L_0$, and leave the case of a
distribution of different angular momentum for a follow up paper.


\subsection{Continuity equation}
\label{sec:continuity}

Using the Vlasov equation we can derive the continuity equation for
the particles.  For that we start from the Vlasov equation written as
\begin{equation}\label{eq:apx_b_vlasov}
\frac{\partial f}{\partial t} + \frac{p_r}{m}\frac{\partial f}{\partial r}
- \frac{\partial \Phi_{\rm eff}(r)}{\partial r} \;
\frac{\partial f}{\partial p_r} = 0 \; ,
\end{equation}
where $\Phi_{\rm eff}(r) \equiv
\Phi(r)+ L^2/2 m r^2$ is the effective
potential. Integrating over momentum space we find
\begin{equation}\label{eq:vlasovint1}
\int \frac{\partial f}{\partial t} \; d \bar{\omega}
+ \int \frac{p_r}{m} \; \frac{\partial f}{\partial r} \; d \bar{\omega}
- \int \frac{\partial \Phi_{\rm eff}(r)}{\partial r} \;
\frac{\partial f}{\partial p_r} \; d \bar{\omega} = 0 \; ,
\end{equation}
with $d \bar{\omega}$ the volume element given in equation~\eqref{eq:dw1}.

For now on we will assume that the distribution function is either of
compact support in phase space. The first term of the above equation
is easy to simplify:
\begin{equation}
\int \frac{\partial f}{\partial t} \; d \bar{\omega}
= \frac{\partial} {\partial t} \int f \; d \bar{\omega}
= \frac{\partial \rho_f} {\partial t} \; ,
\end{equation}
where have used the definition of the particle density $\rho_f$,
equation~\eqref{eq:rho1}. In a similar way we find for the second
term:
\begin{equation}
\int \frac{p_r}{m} \; \frac{\partial f}{\partial r} \; d \bar{\omega}
= \frac{2 \pi}{m r^2} \; \int p_r \; \frac{\partial f} {\partial r} \;  L \; dp_r dL
= \frac{2 \pi}{m r^2} \; \frac{\partial} {\partial r} \int p_r f  L \; dp_r dL
= \frac{1}{m r^2} \; \frac{\partial} {\partial r} \left( r^2 j_f \right) \; ,
\end{equation}
where we have now used the definition of the momentum density $j_f$,
equation~\eqref{eq:curr1}.  Finally, the last term can be easily
shown to vanish for a distribution function of compact support:
\begin{equation}
\int \frac{\partial \Phi_{\rm eff}(r)}{\partial r} \;
\frac{\partial f}{\partial p_r} \; d \bar{\omega}
= \frac{2 \pi}{r^2} \; \frac{\partial \Phi_{\rm eff}(r)}{\partial r}
\int \frac{\partial f}{\partial p_r} \; L \; dp_r dL
= \frac{2 \pi}{r^2} \; \frac{\partial \Phi_{\rm eff}(r)}{\partial r}
\int \left( \int \frac{\partial f}{\partial p_r} \; d p_r \right) L \; dL
= 0 \; ,
\end{equation}
where we assumed that $f$ is zero for large values of $|p_r|$.  If we
now define the mass density as $\rho_m := m \rho_f$, the Vlasov
equation integrated over momentum space reduces to
\begin{equation}
\frac{\partial \rho_m}{\partial t} + \frac{1}{r^2} \;
\frac{\partial} {\partial r} \left( r^2 j_f \right) = 0 \; ,
\end{equation}
which is nothing more than the standard continuity equation in
spherical symmetry. By integrating the continuity equation in physical
space it can now be easily shown, by using the divergence theorem, that
the number of particles $N$ defined above in
equation~\eqref{eq:totalN} is conserved in the sense that $\partial N
/ \partial t = 0$.


\subsection{Virial theorem}
\label{sec:virial}

When the distribution function depends only on the conserved energy
and angular momentum, the system is automatically in a steady (or
equilibrium) state. However, for arbitrary initial conditions, the
system can reach equilibrium whenever 
$\rho_{,\,t}=0$.  
The Virial theorem allows us to determine if the system has reached a
steady state. To derive the virial equation, we start from the Vlasov
equation in spherical symmetry, multiply it by $r p_r$, and
integrate over phase space:
\begin{equation}\label{eq:int_vlasov}
  \int r p_r \; \frac{\partial f}{\partial t} \; dV d\bar{\omega}
  + \frac{1}{m} \int r p_r^ 2 \; \frac{\partial f}{\partial r} \; dV d\bar{\omega}
  - \int r p_r \; \frac{\partial \Phi_{\rm eff}(r)}{\partial r} \;
  \frac{\partial f}{\partial p_r} \; dV d\bar{\omega}
  = 0 \;  .
\end{equation}
The first term above can be rewritten as:
\begin{equation}
  \int r p_r \; \frac{\partial f}{\partial t} \; dV d\bar{\omega}
  = \frac{\partial}{\partial t} \int r p_r f \; dV d \bar{\omega}
  = \frac{\partial}{\partial t} \left<  r p_r \right> \; .
\end{equation}
For the second term in equation~\eqref{eq:int_vlasov} we find:
\begin{eqnarray}
 \frac{1}{m} \int r p_r^ 2 \; \frac{\partial f}{\partial r} \; dV d\bar{\omega}
&=& \frac{8 \pi^2}{m} \int p_r^2 \left( r \frac{\partial f}{\partial r} \right) L
\; dr dL dp_r
= - \frac{8 \pi^2}{m} \int p_r^2 f L \; dr dL dp_r \nonumber \\
&=& - \frac{1}{m} \; \left<  p_r^2 \right> = - 2 \left< K_r \right> \; ,
\end{eqnarray}
where in the second term above we integrated by parts over $r$, using
the fact that $f$ has compact support, 
and where $K_r = p_r^2/2m$ is
the radial kinetic energy.

For the third term of~\eqref{eq:int_vlasov} we find:
\begin{eqnarray}
\int r p_r \; \frac{\partial \Phi_{\rm eff}(r)}{\partial r} \;
\frac{\partial f}{\partial p_r} \; dV d\bar{\omega}
&=& 8 \pi^2 \int r \frac{\partial \Phi_{\rm eff}(r)}{\partial r}
\left( p_r \; \frac{\partial f}{\partial p_r} \right) L \; dr dL dp_r \nonumber \\
&=& - 8 \pi^2 \int r \frac{\partial \Phi_{\rm eff}(r)}{\partial r} f L \;
dr dL dp_r \nonumber \\
&=& \left< r F_{\rm eff}(r) \right> \; ,
\end{eqnarray}
where again we have integrated by parts,
but now over $p_r$, and where
we have defined the effective force as \mbox{$F_{\rm eff}(r) = -
  \partial \Phi_{\rm eff}(r) / \partial r$}.  The quantity $\left< r F_{\rm eff}(r)
\right>$ in known as the \textit{virial}.

Collecting our results we find that the integrated Vlasov equation
becomes
\begin{equation}
  \frac{\partial}{\partial t} \left<  r p_r \right> =
  2\left< K_r \right> + \left< r F_{\rm eff}(r) \right> \; .
  \label{eq:virial1}
\end{equation}
The last result is known as the virial theorem. In a steady state the
distribution function is independent of time, so the virial theorem
implies
\begin{equation}
  2 \left< K_r \right> + \left< r F_{\rm eff}(r) \right> = 0 \; .
  \label{eq:virial2}
\end{equation}
Notice that if we rewrite the effective force as
\begin{equation}
  F_{\rm eff}(r) = - \frac{\partial \Phi_{\rm eff}(r)}{\partial r}
  = - \frac{\partial \Phi}{\partial r} + \frac{L^2}{mr^3} \; ,
\end{equation}
then we can rewrite equation~\eqref{eq:virial2} above as
\begin{equation}
  2 \left< K_r \right> + \left< r F_g(r) \right>
+ \left< L^2 / m r^2\right> = 0 \; ,
\end{equation}
with $F_g(r)=-\partial \Phi / \partial r$ the purely gravitational
force. If we now recognize that the last term above is nothing more
than twice the angular kinetic energy, the virial theorem becomes
\begin{equation}
  2 \left< K_{\rm tot} \right> + \left< r F_g(r) \right> = 0 \; ,
  \label{eq:virial3}
\end{equation}
where now $K_{\rm tot}$ is the total kinetic energy.  When a system
has reached an equilibrium state and the above equation is satisfied,
we say that the system has \textit{virialized}.

For the special case when the external potential $\Phi(r)$ behaves as
$r^n$, the virial theorem can be rewritten as
\begin{equation}
2 \left< K_{\rm tot} \right> - n \left< \Phi(r) \right> = 0 \;  .
\label{eq:virial4}
\end{equation}
This is the form frequently found in textbooks (often for the special
case $n=-1$ corresponding to a simple point-particle gravitational
potential).  However, it is not the form we are interested in since
the external potentials we will consider for this work are not as simple
as a power law.~\footnote{When instead of particles moving in an external
  potential we consider the case of self-gravitating particles, then
  the virial theorem can also be derived, and in that case we do find
  that equation~\eqref{eq:virial4} is satisfied with $n=-1$ since the
  mutual forces between the particles are purely gravitational.}


\subsection{Entropy}

The entropy is constant for a system that obeys the Vlasov
equation. Indeed, starting from the definition of entropy
\begin{equation}
S = - \int f \ln f \; dV d \bar{\omega} \; ,
\label{def:S}
\end{equation}
it can be seen that this is a conserved quantity for any system
obeying the Vlasov equation: 
\begin{eqnarray}
\frac{d S}{dt} &=& -\int \left(1 + \ln f \right) \; \frac{\partial f}{\partial t}
\; dV d \bar{\omega} \nonumber \\
&=& \int \left( 1 + \ln f \right) \left[ \frac{p_r}{m} \frac{\partial f}{\partial r} 
- \frac{\partial \Phi(r)}{\partial r} \frac{\partial f}{\partial p_r} \right]
\; dV d \bar{\omega} \nonumber \\
&=& \int \left[ \frac{p_r}{m} \frac{\partial(f\ln f)}{\partial r}
- \frac{\partial \Phi(r)}{\partial r} \frac{\partial(f\ln f)}{\partial p_r} \right]
\; dV d \bar{\omega} \; , \nonumber
\end{eqnarray}
the terms in square brackets can be integrated over phase space and
can easily be seen to vanish if $f$ has compact support, and $f$ and $\Phi$
smooth and finite. Hence
\begin{equation}
\frac{d S}{d t} = 0 \; ,
\end{equation}
so that the entropy is a constant as claimed. This result, though
straightforward, is somewhat counter intuitive since one would expect
the entropy to increase.  However, entropy will only increase for
interacting particles for which the collision term of the Boltzmann
equation is non-zero (this is the content of the well-known $H$
theorem). In fact, one can also easily see that the integral over
phase space of any arbitrary, smooth and finite function of $f$ is also conserved,
with the total particle number $N$ and the total entropy $S$ special cases.

In the case of the Vlasov equation, one does find 
that if we start with a well defined and localized
distribution function $f$, during evolution the system will move to a
new state where the distribution function is dispersed in a large
section of phase space, but the entropy still remains constant.  We
then see that entropy is not a very useful quantity if one wants to
study the way in which the distribution function evolves to fill in a
large region of phase space in this case.  Perhaps it would be more
useful to apply ergodic theory to describe the evolution in space
space of systems that obey the Vlasov equation, but this is a subject
that goes beyond the scope of this work.~\footnote{Notice that
  calculating the entropy is also problematic numerically, as the
  value of $\ln f$ becomes minus infinity in regions where $f$
  vanishes, and takes extremely large negative values in regions where
  $f$ is very small, which can cause serious problems of numerical
  accuracy even if one assumes that $f$ does not quite vanish
  anywhere.}


\section{Halo models}
\label{Sec2}

Having set the definitions and discussed the basic properties of the
Vlasov equation, we will now apply the formalism to describe the
stability of perturbations in galactic halo models. In order to do so, we will use the
Vlasov equation, Eq.~(\ref{eq:Vla1}), in a background gravitational
potential of several halo models, and study the evolution of a given
initial localized distribution function $f$ in that potential. We
imagine such a distribution to represent a small perturbation in the
matter distribution of the halo. We perform several numerical tests
which will allow us to determine the fate of the such initial
perturbation.

\begin{figure}[H]
\centering
\subfigure{\includegraphics[scale=0.60]{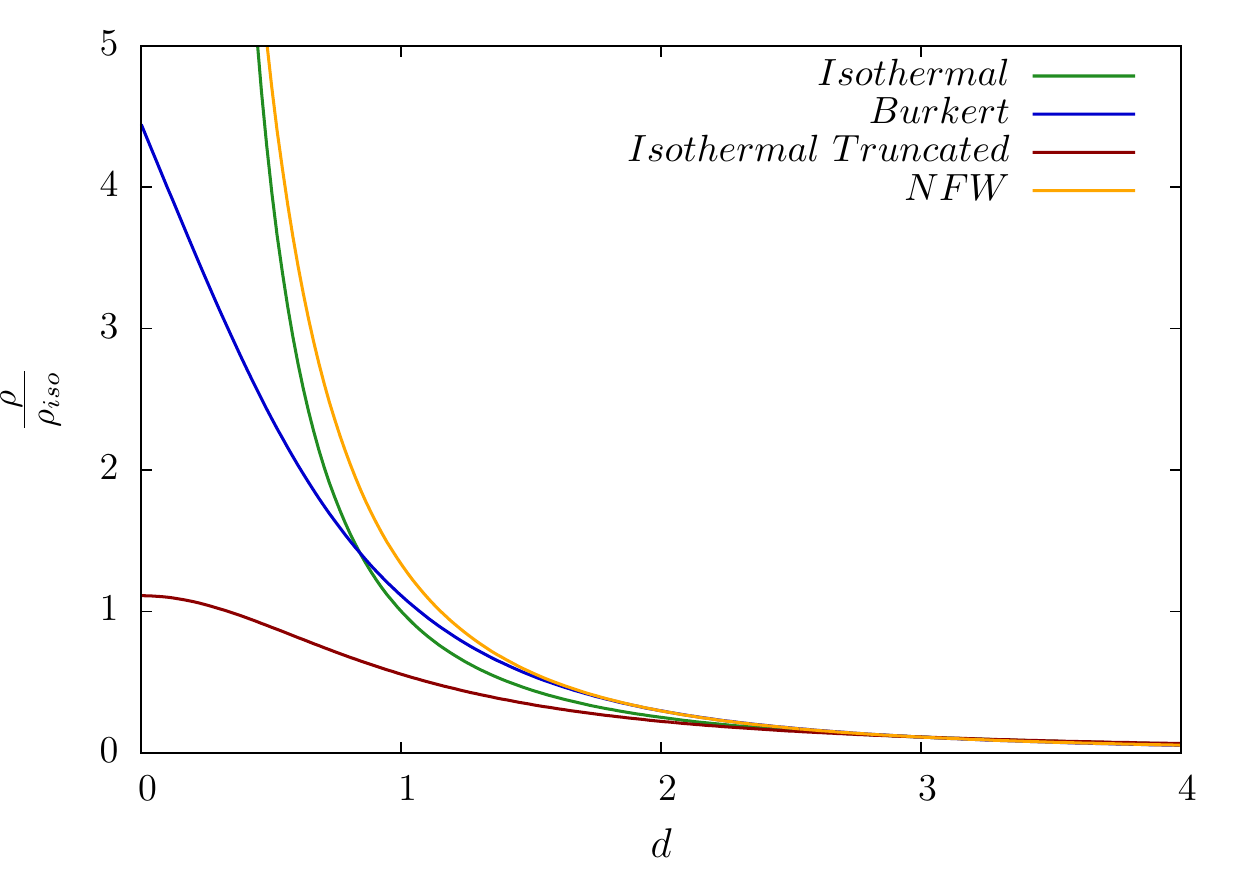}}
\caption{Plots of the density profile $\rho(d)$ of the four halos
  considered. Notice that our chosen calibration is such that the
  densities are equal at $d=3$ in the four cases.}
\label{Fig:rhos}
\end{figure}

We will consider four different dark matter halo models which are
commonly discussed in the literature, namely the isothermal, the
truncated isothermal, the Burkert and the Navarro-Frenk-White
models. All of them will be described by a dark matter density:
\begin{eqnarray}
\rho_{\rm iso} = \frac{{\sigma_0}_{iso}^2}{2 \pi G r^2} \; , &\hspace{1cm}& 
\rho_{\rm iso-tr} = \frac{{\sigma_0}_{iso-tr}^2}{2 \pi G \left(r^2 + {r_s}^2 \right)}
\; , \nonumber \\
\rho_{\rm Burkert} = \frac{{\rho_0}_{\rm Burkert}}{\left(1 + \frac{r}{r_s} \right) \,
\left( 1 + \frac{r^2}{{r_s}^2} \right)} \; , &\hspace{1cm}&  
\rho_{\rm NFW} = \frac{{\rho_0}_{\rm NFW}}{\frac{r}{r_s} \, \left(1 + \frac{r}{r_s}
\right)^2}
\; , \label{rhos}  
\end{eqnarray}
where $\sigma$ is a characteristic velocity, $r_s$ is a characteristic
radius, and $\rho_0$ a characteristic density for each case. In order
to be able to compare the behavior of an inhomogeneity in each halo,
we will start by writing them in similar terms. The characteristic
speed $\sigma_0$ can be related to a characteristic density by means
of a characteristic mass $M_0$ and a pattern radius $R_0$, as
\begin{equation}
{\sigma_0}^2 = \frac{G M_0}{R_0} = \frac{G \left( \frac{4 \pi}{3} \rho_0 {R_0}^3 
\right)}{R_0} = \frac{4 \pi G \rho_0 {R_0}^2}{3} \; ,
\label{eq:sigma_rho}
\end{equation}
where $G$ is Newton's gravitational constant. Writing now the radial
coordinate as multiple of the pattern radius, $r=R_0\,d$, with $d$ a
dimensionless number, it is possible to rewrite the different halo
densities as
\begin{eqnarray}
\rho_{\rm iso} = \frac{2 {\rho_0}_{\rm iso}}{3 d^2} \; , &\hspace{1cm}& 
\rho_{\rm iso-tr} = \frac{2 {\rho_0}_{\rm iso-tr}}{3 d^2 \left(1 + \frac{{d_s}^2}
{d^2}\right)} \; , \nonumber \\
\rho_{\rm Burkert} = \frac{{\rho_0}_{\rm Burkert}}{\left( 1 + \frac{d}{d_s} \right)
\left(1 + \frac{d^2}{{d_s}^2}\right)} \; , &\hspace{1cm}&  
\rho_{\rm NFW} = \frac{{\rho_0}_{\rm NFW}}{\frac{d}{d_s}
\left(1 + \frac{d}{d_s}\right)^2} \; , \label{rhos1}  
\end{eqnarray}
where $r_s$ has been rewritten as $r_s=R_0\,d_s$. Choosing now
$d_s=1$, and demanding that we obtain the same value of the density
$\rho$ for each halo at $d=3$ for the four models (see Fig.~(\ref{Fig:rhos})), 
we find that the four density profiles can be
rewritten in terms of just one characteristic density $\rho_0$:
\begin{eqnarray}
\rho_{\rm iso} = \frac{{\rho_0}_{\rm iso}}{d^2} \; , &\hspace{1cm}& 
\rho_{\rm iso-tr} = \frac{10 {\rho_0}_{\rm iso}}{9 \left(1 + d^2\right)}
\; , \nonumber \\
\rho_{\rm Burkert} = \frac{40 {\rho_0}_{\rm iso}}{9 \left(1 + d\right)¡
\left(1 + d^2\right)} \; , &\hspace{1cm}&  
\rho_{\rm NFW} = \frac{16 {\rho_0}_{\rm iso}}{3 d \left(1 + d\right)^2} \; . 
\label{rhos2}  
\end{eqnarray}

Given the density profile $\rho$, we can find that the halo mass
distribution $m(r)$ as
\begin{equation}
  m(r) = 4 \pi \int \rho r^2 dr \; ,
\end{equation}
so that the gravitational potential is given by $\Phi=-G m(r)/r$.

Where we can also introduce a dimensionless potential $V(d)$ with the
help of the particle mass $m$ and a characteristic speed $\sigma_0$ as
\begin{equation}
  \Phi(r)=m\,{\sigma_0}^2 V(d) \; ,
\end{equation}
In this way, the gradient of the gravitational potential can be
written as
\begin{equation}
  \frac{\partial \Phi(r)}{\partial r}
  = \frac{m {\sigma_0}^2}{R_0} \; \frac{\partial V(d)}{\partial d} \; .
\end{equation}

Making now use of the relation between $\sigma_0$ and $\rho_0$ given
by Eq.~(\ref{eq:sigma_rho}), the dimensionless gradient of the
gravitational potential for each halo becomes:
\begin{eqnarray}
\frac{\partial V_{\rm iso}(d)}{\partial d} = \frac{3}{d} \; , &\hspace{1cm}& 
\frac{\partial V_{\rm iso-tr}(d)}{\partial d} = \frac{10 \left(
  1 - \frac{\arctan\left(d\right)}{d}\right)}{3 d} \; , \nonumber \\
&& \label{eqs:phisd} \\
\frac{\partial V_{\rm Burkert}(d)}{\partial d}
= \frac{10 \left(\ln\left(\left(1+d^2\right) \left(1+d\right)^2\right) 
- 2 \arctan\left(d\right)\right)}{3 d^2} \; , &\hspace{1cm}&  
\frac{\partial V_{\rm NFW}(d)}{\partial d}
= \frac{16 \left(\ln\left(1+d\right) - \frac{d}{1+d}\right)}{d^2}
\; . \nonumber  
\end{eqnarray}

At this point it is also convenient to rewrite the Vlasov equation,
Eq.~(\ref{eq:Vla1}), in a dimensionless form.  As already mentioned,
defining a pattern mass $M_0$ and a pattern distance $R_0$ allows us
to define a pattern density $\rho_0$ and from it a pattern velocity
$\sigma_0$.  Considering that we are dealing with a single type of
particles characterized by a mass $m$, we can continue this idea and
construct all the needed pattern quantities, {\it i.e.} a pattern time
$T_0=R_0/\sigma_0$, a pattern radial momentum $p_0=m \sigma_0$, and a
pattern angular momentum $l_0=m\,\sigma_0\,R_0$.  Rewriting then the
time as $t=T_0\,{\cal T}$, the radial momentum as $p_r=p_0\,{\cal
  P}_r$, and the angular momentum as $L=l_0\,{\cal L}$, we obtain the
following dimensionless Vlasov equation:
\begin{equation}
  \frac{\partial \cal{F}}{\partial {\cal T}} = - {\cal P}_r \,
  \frac{\partial \cal{F}}{\partial d} + \left(\frac{\partial
    V(d)}{\partial d} - \frac{{\cal L}^2}{d^3}\right) \frac{\partial
    \cal{F}}{\partial {\cal P}_r} \; ,
  \label{eq:Vla_dimless}
\end{equation}
where we have defined $f=f_0\,\cal{F}$, with $\cal{F}$ a dimensionless
distribution function, and $f_0$ a normalization constant with units of probability density in phase 
space. 

\begin{figure}[H]
\centering
\subfigure{\includegraphics[scale=0.60]{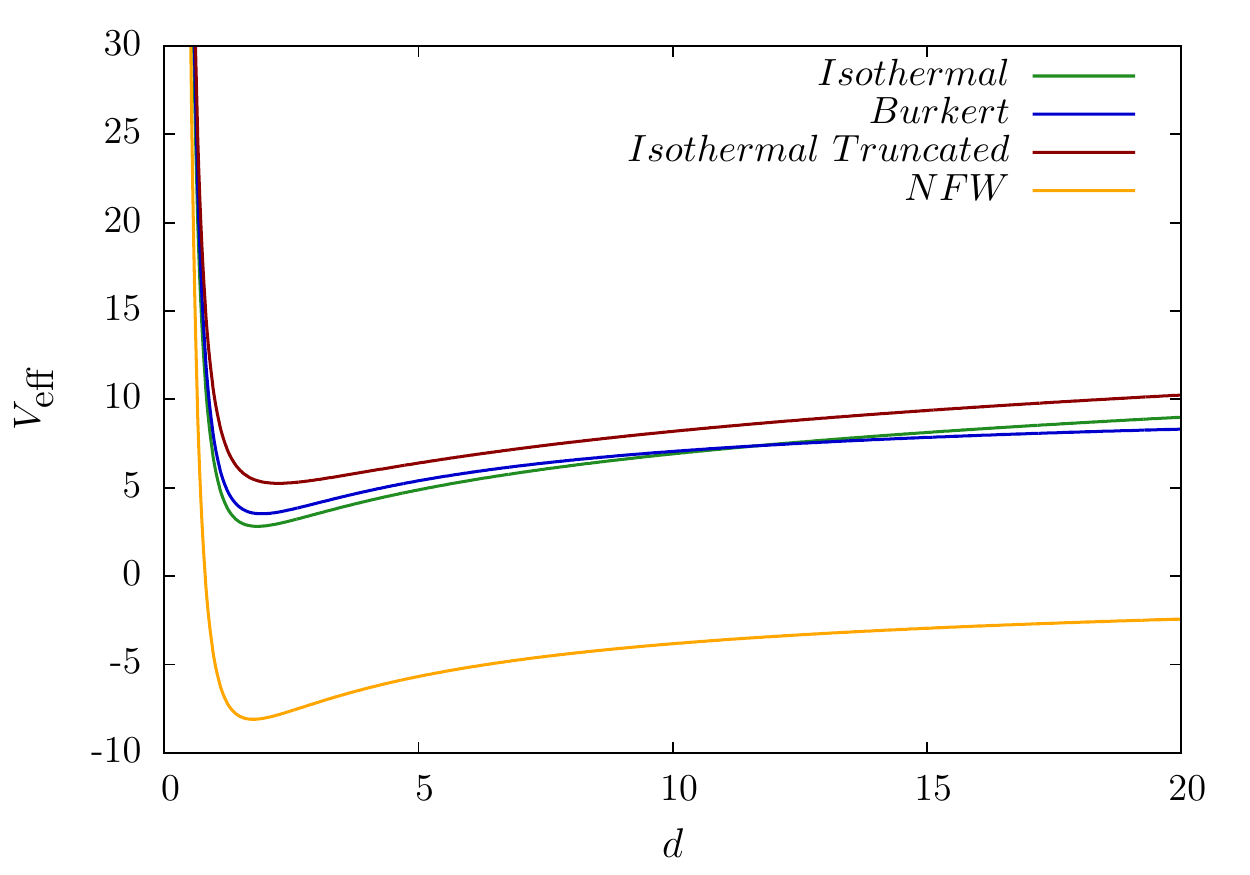}}
\caption{Plots of the effective dimensionless potential of the four
  halos considered with ${\cal L}=3.5$.}
\label{Fig:Phis_ef}
\end{figure}

For each halo model, we use the dimensionless gradient of the
gravitational potential given by Eqs.~(\ref{eqs:phisd}), in order to
evolve the distribution function. The corresponding dimensionless
potentials are given by:
\begin{eqnarray}
&& V_{\rm iso}(d) = 3 \ln(d) \; , \hspace{7cm} 
  V_{\rm iso-tr}(d) = \frac53 \left( \ln(1+d^2)
  + 2 \frac{\arctan(d)}{d} \right)\,, \nonumber \\
&& \label{eqs:phis} \\
&& V_{\rm Burkert}(d) = \frac{10}{3 d} \left(2 \left(1+d\right)
\left( \arctan\left(d\right) - \ln \left(1+d\right) \right)
- \left(1-d\right) \ln\left(1+d^2\right) \right) \; , 
\hspace{1cm}  
V_{\rm NFW}(d) = -16 \frac{\ln(1+d)}{d} \; . \nonumber  
\end{eqnarray}
The dimensionless effective potential is obtained by adding the
centrifugal term. Fig (\ref{Fig:Phis_ef}) shows the plots for the
effective potential in each case.

\begin{figure}[H]
\centering
\subfigure{\includegraphics[scale=0.60]{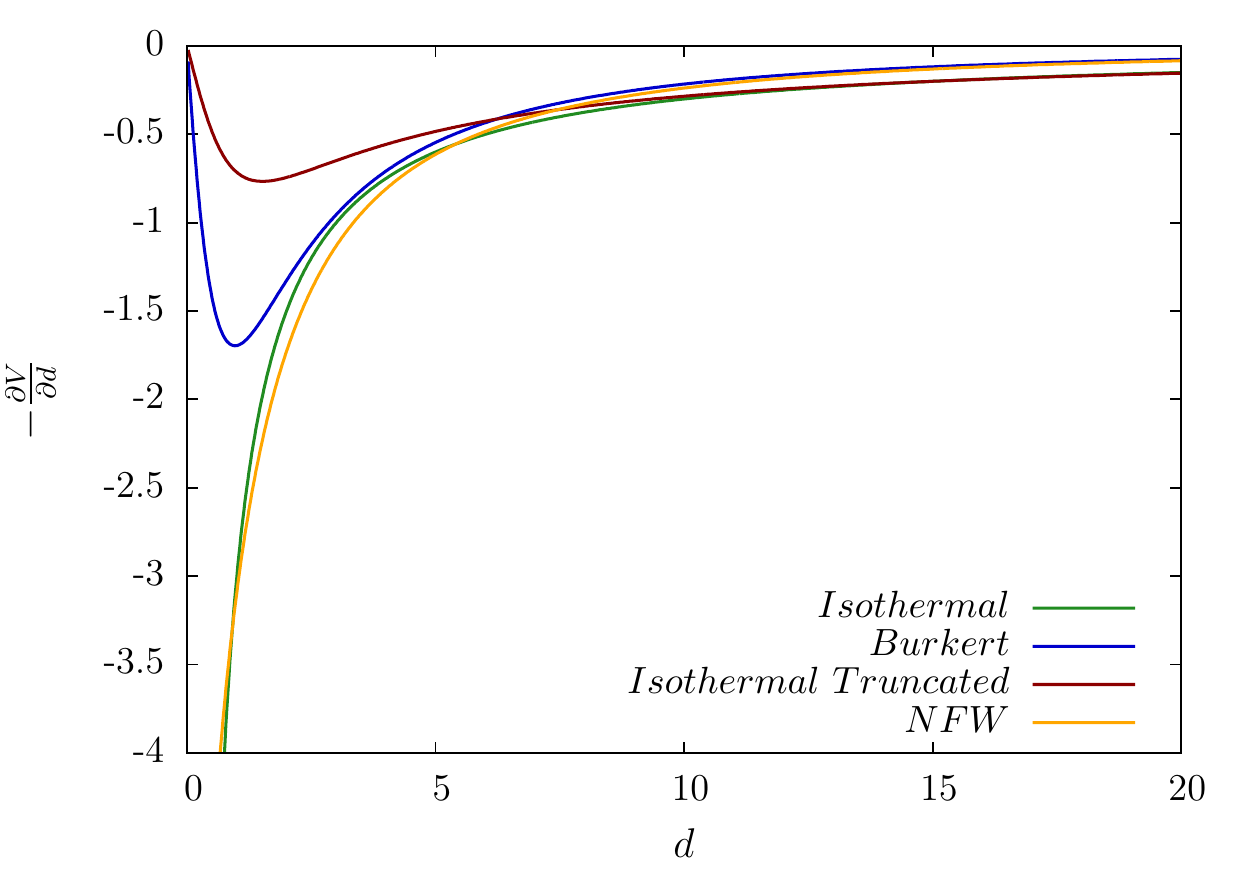}}
\caption{Plots for minus the derivative of the dimensionless
  gravitational potential $V(d)$ for each of the four halos
  considered.}
\label{Fig:dPhis}
\end{figure}

\begin{figure}[H]
\centering
\subfigure{\includegraphics[scale=0.60]{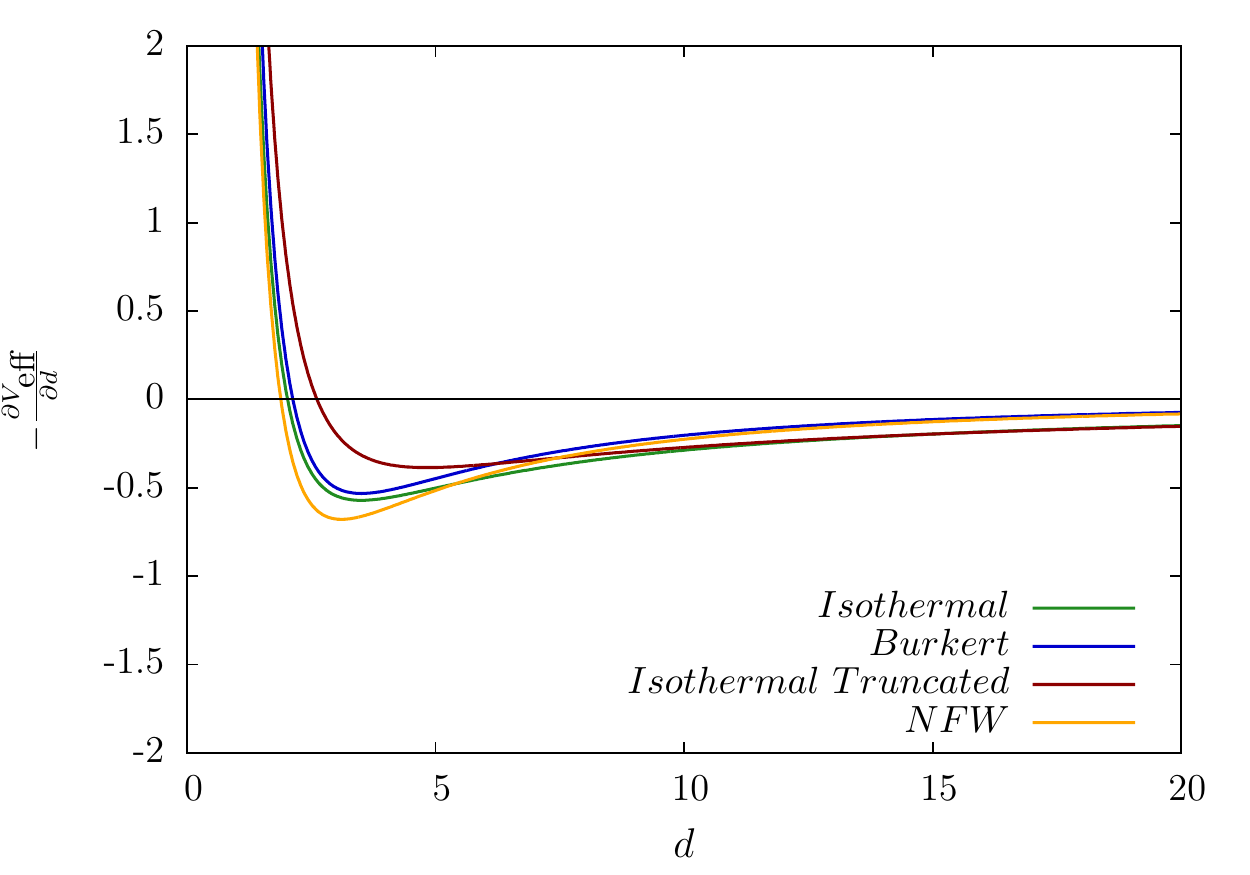}}
\caption{Plots for the derivative of the effective dimensionless
  gravitational potential for the four halos considered with ${\cal
    L}=3.5$.}
\label{Fig:dPhis_ef}
\end{figure}

The profiles for the potential gradients are shown in
Figure~(\ref{Fig:dPhis}), and the corresponding graphs for the
gradient of the effective potential (including the centrifugal term)
are shown in Figure~(\ref{Fig:dPhis_ef}). 

\section{Numerical code}
\label{Sec3}

In this section, we describe our \textit{Vlaso-ollin} code, which evolves the 
distribution function by means of the Vlasov equation, Eq.(\ref{eq:Vla_dimless})
for several background gravitational potentials.


\subsection{Flux conservative methods}

The Vlasov equation~\eqref{eq:Vla_dimless} can be trivially rewritten
in conservation form as:
\begin{equation}
  \label{eq:Vla_cons}
\frac{\partial \cal{F}}{\partial {\cal T}} + \frac{\partial}{\partial d}
\left[ {\cal P}_r {\cal F} \right] + \frac{\partial}{\partial {\cal P}_r}
\left[ \left(\frac{\partial V(d)}{\partial d}
- \frac{{\cal L}^2}{d^3} \right) {\cal F} \right] = 0 \; .
\end{equation}
We will now define the fluxes in the $d$ and ${\cal P}_r$ directions,
$F_d$ and $F_{{\cal P}_r}$, as
\begin{equation}
F_d \equiv {\cal P}_r {\cal F} \; , \quad
F_{{\cal P}_r} \equiv \frac{\partial V_{\rm eff}(d)}{\partial d} {\cal F} \; ,
\end{equation}
where as before
\begin{equation}
V_{\rm eff} \equiv V(d) + \frac{1}{2}\frac{{\cal L}^2}{d^2} \; .
\end{equation}
The Vlasov equation then becomes
\begin{equation}
  \frac{\partial \cal{F}}{\partial \cal{T}}
  + \frac{\partial F_d}{\partial d}
  + \frac{\partial F_{{\cal P}_r}}{\partial {\cal P}_r} = 0 \; .
\end{equation}

We have written a numerical code to solve the last equation using a
flux conservative finite difference method~\cite{Leveque92}.  A
crucial consequence of a using conservative method for the Vlasov
equation is that the numerical integration will conserve the total
number of particles $N$ exactly (up to machine round-off error), so
that any change in the number of particles will be the result of
particles leaving (or entering) the computational domain through the
boundaries.

In our code phase-space is discretized in a rectangular grid of size
$N_r \times N_p$, with $0 \leq d \leq d_{max}$ and \mbox{$-{\cal
    P}_{max} \leq {\cal P} \leq {\cal P}_{max}$}, and with grid sizes
$\Delta d$ and $\Delta {\cal P}$.  The code uses a method of lines,
with a three-step iterative Crank-Nicholson time integrator. We use a
flux-limiter method for the calculation of the fluxes $F_d$ and
$F_{{\cal P}_r}$ at the cell interfaces.  The limiter methods minmod,
van Leer, superbee and monotonized-centered (MC)~\cite{Leveque92} were
tested for the special case of an external potential corresponding to
a constant density star, and particles with zero angular momentum, and
the MC limiter was found to show the best results in both preserving
the number of particles and preserving the form of the density profile
for stationary solutions: minmod is very diffusive in the sense that
stationary profiles tend to diminish in amplitude and become wider
(while conserving the integral), while the van Leer and superbee
limiters, though less diffusive, in general deform the stationary
profiles. The reason why we found it necessary to use sophisticated
high resolution flux-limiter methods is because such methods can be
shown to be total-variation-diminishing (TVD), and this guarantees
that the solution has no spurious oscillations. This ensures that in
regions where the distribution function $f$ is close to zero it
remains well behaved and does not become negative (which would be
unphysical, but can easily happen with numerical methods that are not
TVD). The limiter methods we use are second order accurate everywhere
except at extrema 
of the distribution function, where they become only
first order.


\subsection{Boundary conditions}

There are two different types of boundaries to consider in our
simulations: the external boundaries of the computational domain at
$|{\cal P}|={\cal P}_{max}$ and $d=d_{max}$, and the internal boundary
at $d=r=0$.

For the external boundaries the boundary condition we use depends on
the sign of the coefficient of the corresponding derivative.  That is,
in the radial direction we consider the sign of the momentum ${\cal
  P}_r$: For negative values of ${\cal P}_r$, corresponding to
particles that would enter the computational domain, we just set the
distribution function $f$ to zero, while for positive values of ${\cal
  P}_r$, corresponding to particles leaving the domain, we calculate
the fluxes all the way to the boundary using one-sided differences.
In the momentum directions we do the same thing, but in this case we
consider the sign of the force term $\partial V_{\rm eff}(d)/\partial
d$.

The boundary at $d=r=0$ is of a different type and corresponds to the
origin of the radial coordinate.  Notice that at $r=0$ the centrifugal
term becomes singular, and the gravitational force can also become
singular for some of the halo models we are considering (the
isothermal and NFW models). To avoid dealing with singular quantities
we use a finite differencing grid that staggers the origin, so that
the first grid point is located at $d = \Delta d/2$.  We add a
fictitious point at $d = -\Delta d/2 $ in order to be able to
calculate derivatives at $d=0$. This fictitious point also allows us
to impose adequate parity conditions on the different quantities.

Now, in the case of non-zero angular momentum the particles are never
allowed to reach the origin so that the distribution function $f$
should be zero there.  But in the case of zero angular momentum there
is nothing to stop them from reaching $r=0$.  If moreover, the
gravitational potential is regular there (as in the case of the
truncated-isothermal and Burkert models), the particles that reach the
origin can in principle just come back on the opposite side and
oscillate around the origin.  This is in fact an interesting case, as
particles that reach the origin with a given negative radial momentum
will move back out of the origin with a positive radial momentum. In
order to capture this behavior we use the fictitious points at $d =
-\Delta d/2 $ to impose the boundary condition $f(-d,-{\cal
  P}_r)=f(d,{\cal P}_r)$.  We find that when doing so, particles
with zero angular momentum in a regular potential just describe orbits
around the origin of phase space, in a very similar way as one would
expect for the case of a simple harmonic oscillator.

When the angular momentum is non-zero, particles coming from infinity
should reach a finite minimum radius, so in principle one would not
need a special boundary condition at $d=0$.  In practice, however, we
have found that in this case the distribution function $f$ might still
attain small non-zero values at $d=r=0$ due to numerical errors, and
using the boundary condition described above results in robust
evolutions.

There is a final important point that should be made about the
behavior of the distribution function $f$ at the origin for the case
of zero angular momentum. From the definition of the particle density
$\rho_f$ and the momentum density $j_f$, equations~\eqref{eq:rho1}
and~\eqref{eq:curr1}, one can easily see that if the distribution
function $f$ does not vanish at the origin then $\rho_f$ and $j_f$ will
be singular there (in fact, $f$ must vanish at least as $r^2$).  This
actually makes perfect physical sense: if an infalling spherical shell
of matter reaches the origin at the same time, then the matter density
will clearly become infinite there as all the particles are now at a
single point.  For an external (regular) gravitational potential this
represents no problem since we can just interpret the distribution
function statistically: every individual particle behaves
independently of the others and will just oscillate around the origin.
But if we were to consider the case of self-gravitating particles,
then as they reach the origin the mass density will become infinite
there, so that the gravitational potential will become singular, and
our description in terms of the Vlasov equation will break down. All
the simulations shown below will be for the case of an external
gravitational potential (the particles have no self-gravity) and
non-zero angular momentum, so that this problem will never arise, but
it is something that should be kept in mind for the study of systems
of self-gravitating particles with the Vlasov equation.


\subsection{CFL condition}

Since we are using an explicit scheme, in order to get a stable
evolution we need to satisfy the Courant-Friedrichs-Lewy (CFL)
condition.  During the evolution we then choose the
time step as $\Delta t=\min(\Delta t_d,\Delta t_p)$, with
\begin{equation}
  \Delta t_d = \frac{C\Delta d}{{\cal P}_{max}} \; , \qquad
  \Delta t_p = \frac{C\Delta {\cal P}}{|F|_{max}} \; .
\end{equation}
where $F \equiv \partial V_{\rm eff}(d)/\partial d$ and $C$ a number
of order 1 which in two dimensions must be such that $C\leq
1/\sqrt{2}$. In all the simulations shown below we choose $C=0.7$ and
$P_{max}=6$.  The maximum value of the force $F_{max}$ is calculated a
$t=0$ and does not change in time since the background potential is
fixed.

The fact that the centrifugal term ${\cal L}/d^3$ in the effective
force diverges at $d=0$ is a serious problem and forces us to use very
small values for $\Delta t$ in order to satisfy the CFL condition.  In
order to avoid this problem we introduce a small parameter $\epsilon$
in the denominator that modifies the centrifugal potential for small
values of $d$, and the modified centrifugal force is obtained by deriving the new potential.
\begin{eqnarray}
\frac{{\cal L}^2}{d^2} &\longrightarrow& \frac{{\cal
    L}^2}{d^2+\epsilon^2} \; , \nonumber \\     
\frac{{\cal L}^2}{d^3} &\longrightarrow& \frac{{\cal
    L}^2\,d}{(d^2+\epsilon^2)^2} \; .        
\end{eqnarray}

For a given value of the angular momentum ${\cal L}$ we choose $\epsilon$ in
such a way as to guarantee that the centrifugal force is minimally
modified at the smallest radius that can be reached by a particle with
that angular momentum.  Assuming that the particle starts far away
with initial momentum ${\cal P}_i$, and that for small values of $d$ the
centrifugal force dominates over the gravitational force, then
conservation of energy implies that the minimum radius the particle
can reach is such that ${\cal L}^2/d_{\rm min}^2 \sim {{\cal P}_i}^2 / 2 m$, which
implies $d_{\rm min} \sim \sqrt{2 m} \; {\cal L} / | {\cal P}_i |$.  In our
simulations we typically take $\epsilon=d_{min}/10$.
  
This is clearly not an ideal solution as in involves modifying the
effective force so that a small error is introduced for small radii,
and in practice still results in values of $\Delta t$ that are much
smaller than $\Delta d$, so that our code is rather slow.  We are
exploring ways to improve our code, perhaps by using an operator
splitting method with an implicit scheme for the centrifugal term.  We
will report on this elsewhere.


\subsection{Initial data}

For our simulations we consider an initial localized distribution of
particles in phase space, in the background gravitational potential of
the different halo models.  We interpret this as a inhomogeneity in
the dark matter halo, and use the Vlasov equation to determine the
final state of such perturbation.

We consider an initial phase space distribution $f$ given as
\begin{equation}
  f(t=0) = \frac{N_0}{8 \pi^3 L_0 \sigma_d \sigma_{\cal P}} \left(
  e^{-\frac{(d-d_0)^2}{{\sigma_d}^2}}
  e^{-\frac{({\cal P}_d-{{\cal P}_r}_0)^2}{{\sigma_{\cal P}}^2}} +
  e^{-\frac{(d+d_0)^2}{{\sigma_d}^2}}
  e^{-\frac{({\cal P}_d+{{\cal P}_d}_0)^2}{{\sigma_{\cal P}}^2}} \right) \; .
\label{eq:f0}
\end{equation}
This is normalized so that the total number of particles is $N_0$, and
in all our simulations we take $N_0=1$. Notice that, when the angular
momentum is zero, the normalization factor in the initial distribution
is instead $1/(4 \pi^2 \sigma_r \sigma_{\cal P})$.

\begin{figure}[H]
\centering
\includegraphics[scale=0.45,angle=0]{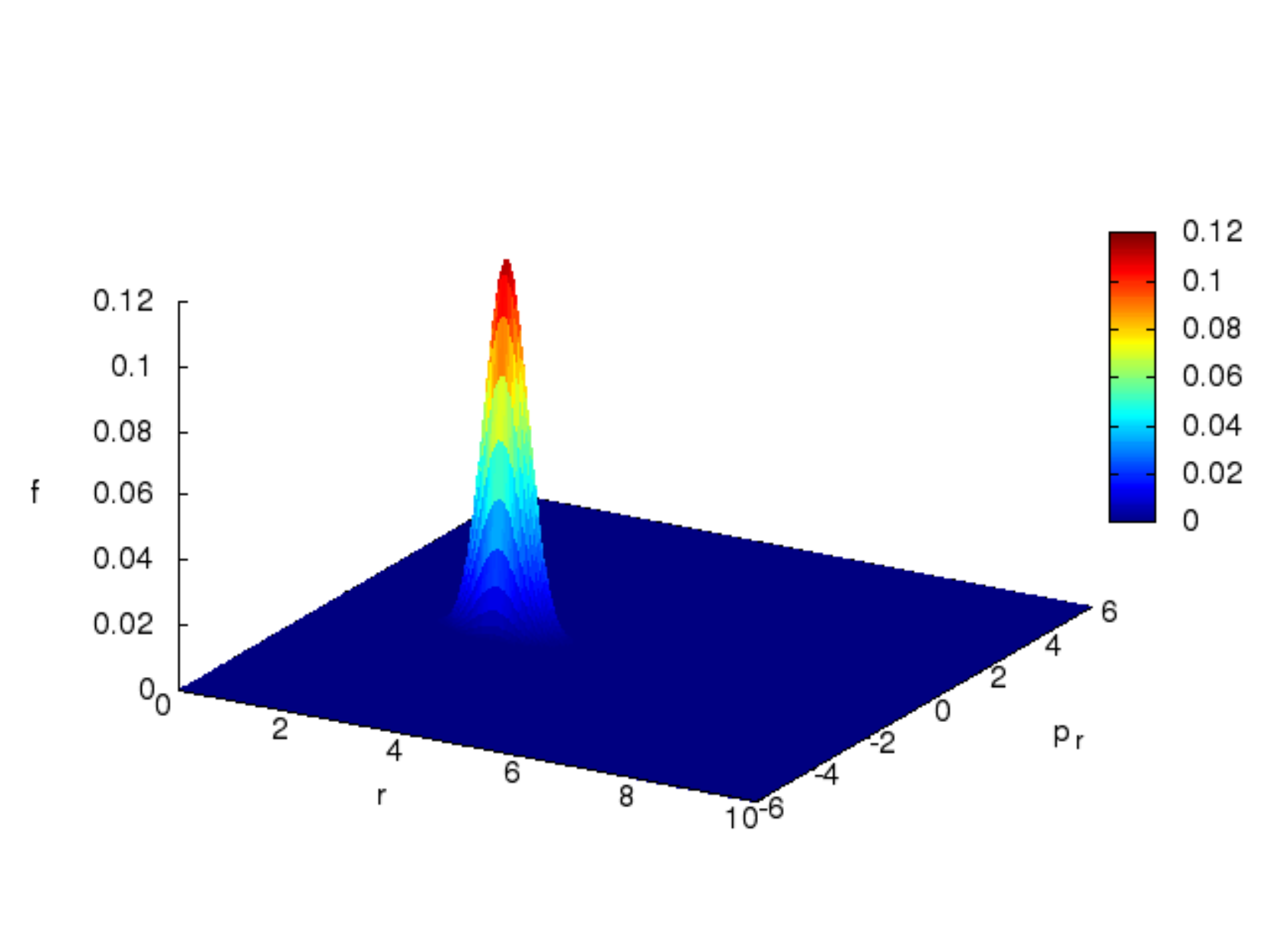}
\caption{Initial distribution function in phase space for $N_0=1$,
  $d_0=3.0$, ${{\cal P}_d}_0=-0.5$, $\sigma_d=0.5$, and $\sigma_{\cal P}=0.25$.}
\label{Fig:f0}
\end{figure}

The above initial data represents a spherical shell of particles, all
of which have the same angular momentum $L_0$, with a Gaussian
distribution centered at a radius $d_0$ and radial momentum ${{\cal P}_d}_0$,
with widths $\sigma_d$ and $\sigma_{\cal P}$ respectively.  The initial data
is constructed so that the symmetry $(d,{\cal P}) \to (-d,-{\cal P})$ discussed
above when we described the boundary conditions is preserved
(incidentally, this form of the initial data also guarantees that the
normalization is exact when integrating over the radial coordinate
from 0 to infinity).  We remind the reader at this point that even if
all the particles have the same angular momentum $L_0$, by
construction their individual motions are uniformly distributed in all
angular directions, so that the overall spherical symmetry is
preserved.  For the simulations shown below we take the following
parameters for the initial data: $d_0=3.0$, ${{\cal P}_d}_0=-0.5$,
$\sigma_d=0.5$, and $\sigma_{\cal P}=0.25$. The corresponding initial data is
shown in Figure~(\ref{Fig:f0}).

\section{Results}
\label{Sec4}

The time evolution of the initial distribution function $\mathcal{F}$
is given by solving equation~\eqref{eq:Vla_cons} in the numerical
mesh. We perform a series of simulations for different values of the
angular momentum, $L=2.0,2.5,3.0,3.5$.  We have not considered smaller
values of the angular momentum because in those cases we have found
that the particles fall very rapidly to the center, while larger vales
of the angular momentum result in the particles escaping the
computational domain. In all our evolutions we also fix the mass of
the individual particles to $m=1$.

For the simulations shown below we considered two different
resolutions $\Delta d = \Delta {\cal P} = 0.05$ and $\Delta d = \Delta
{\cal P} = 0.1$. The boundaries of the computational domain
are located at $d_{max}=20$ and ${\cal P}_{max}=6$. This range is
chosen so that at the higher resolution we use only a very small
fraction of the particles has escaped through the boundaries at the
end of the simulation.  As we will see below, that particles that
escape do so mostly because of numerical dissipation.


\subsection{Time evolution for $\mathcal{L}=3.5$}

As an example of the dynamics of the Vlasov equation, in this section
we will concentrate on presenting the results of the time evolution of
the distribution function for the case of our largest value of the
angular momentum, namely $\mathcal{L}=3.5$.

\begin{figure}[H]
\centering
\subfigure{\includegraphics[scale=0.34]{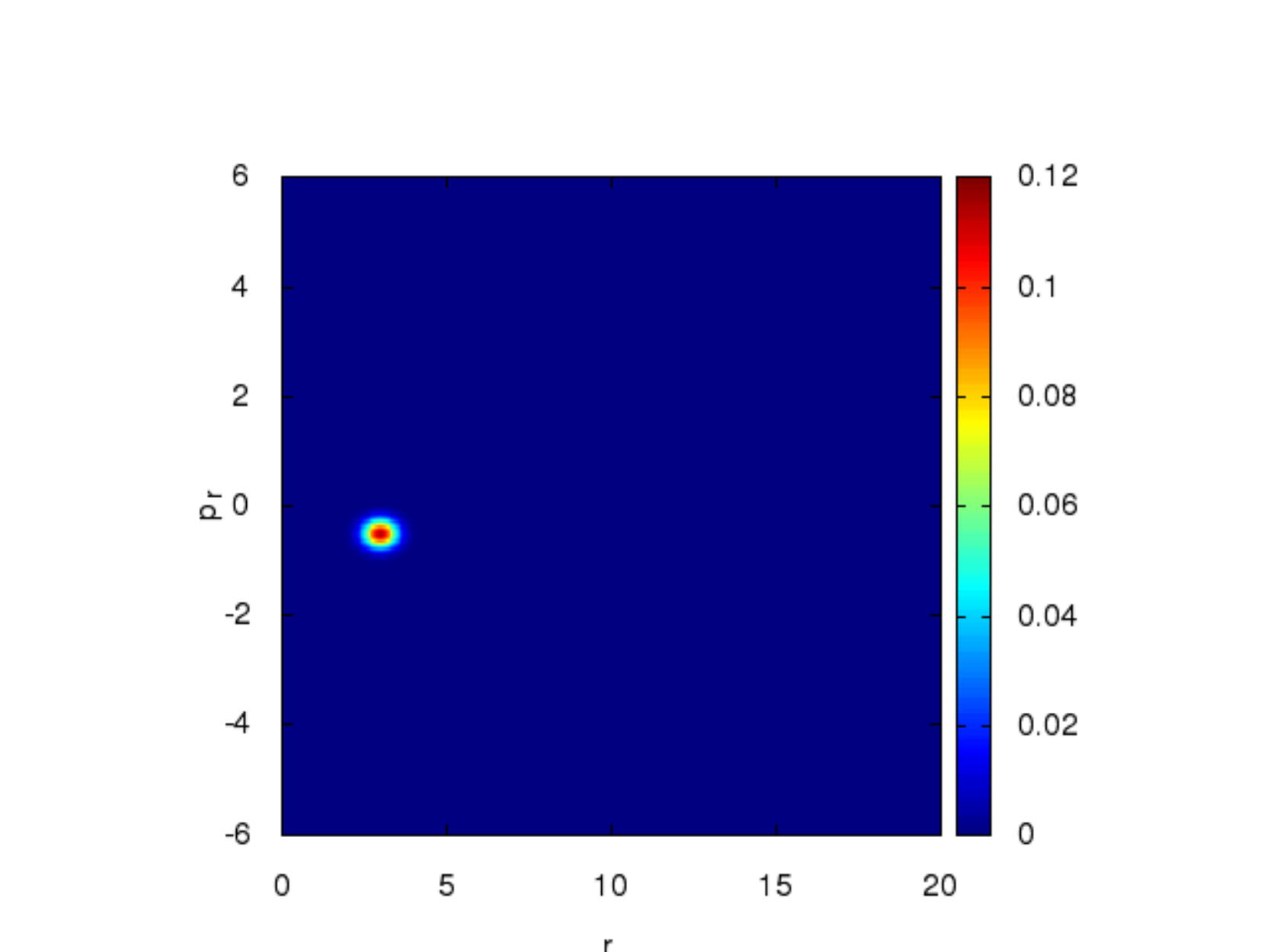}} 
\subfigure{\includegraphics[scale=0.34]{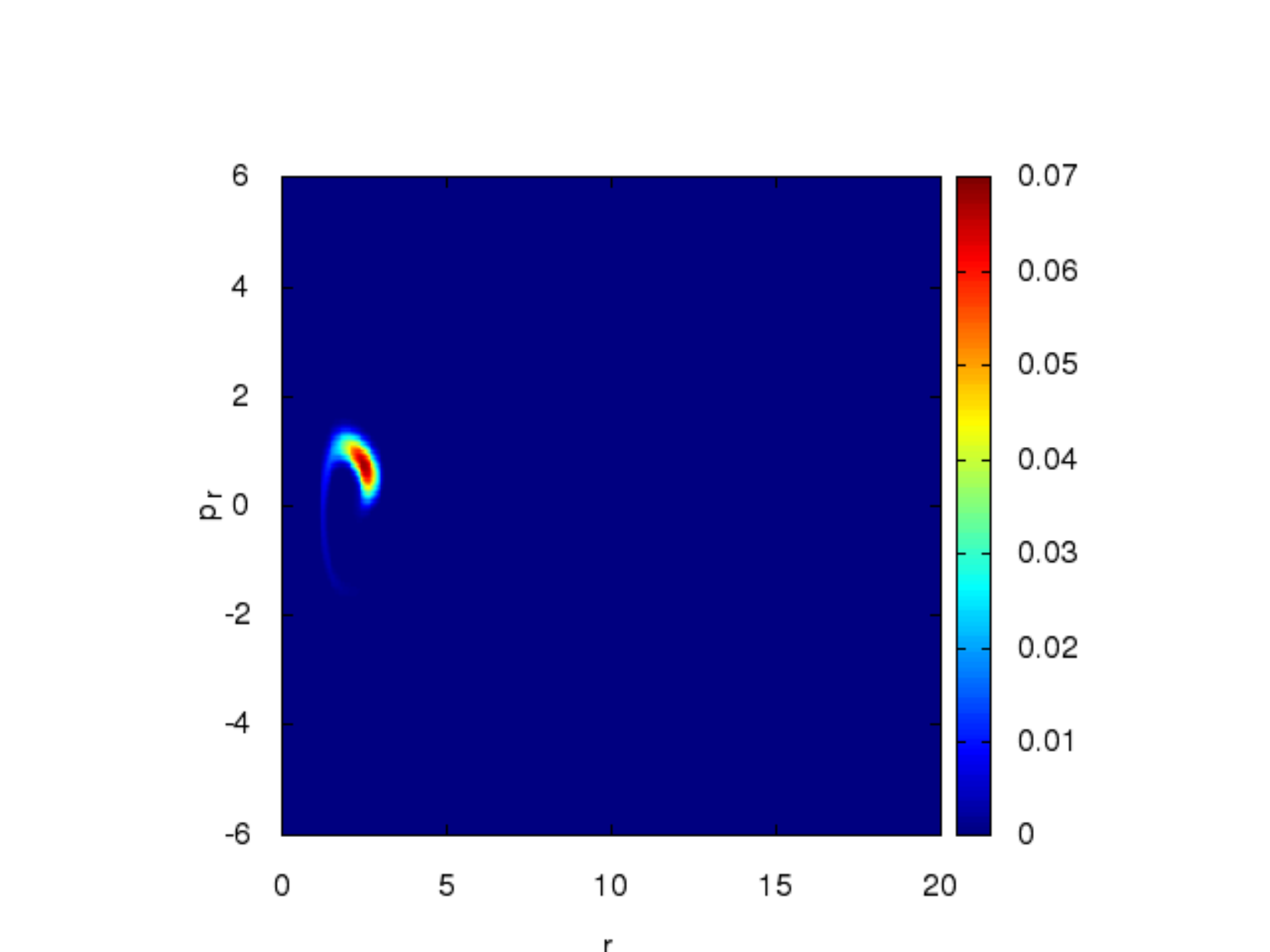}} 
\subfigure{\includegraphics[scale=0.34]{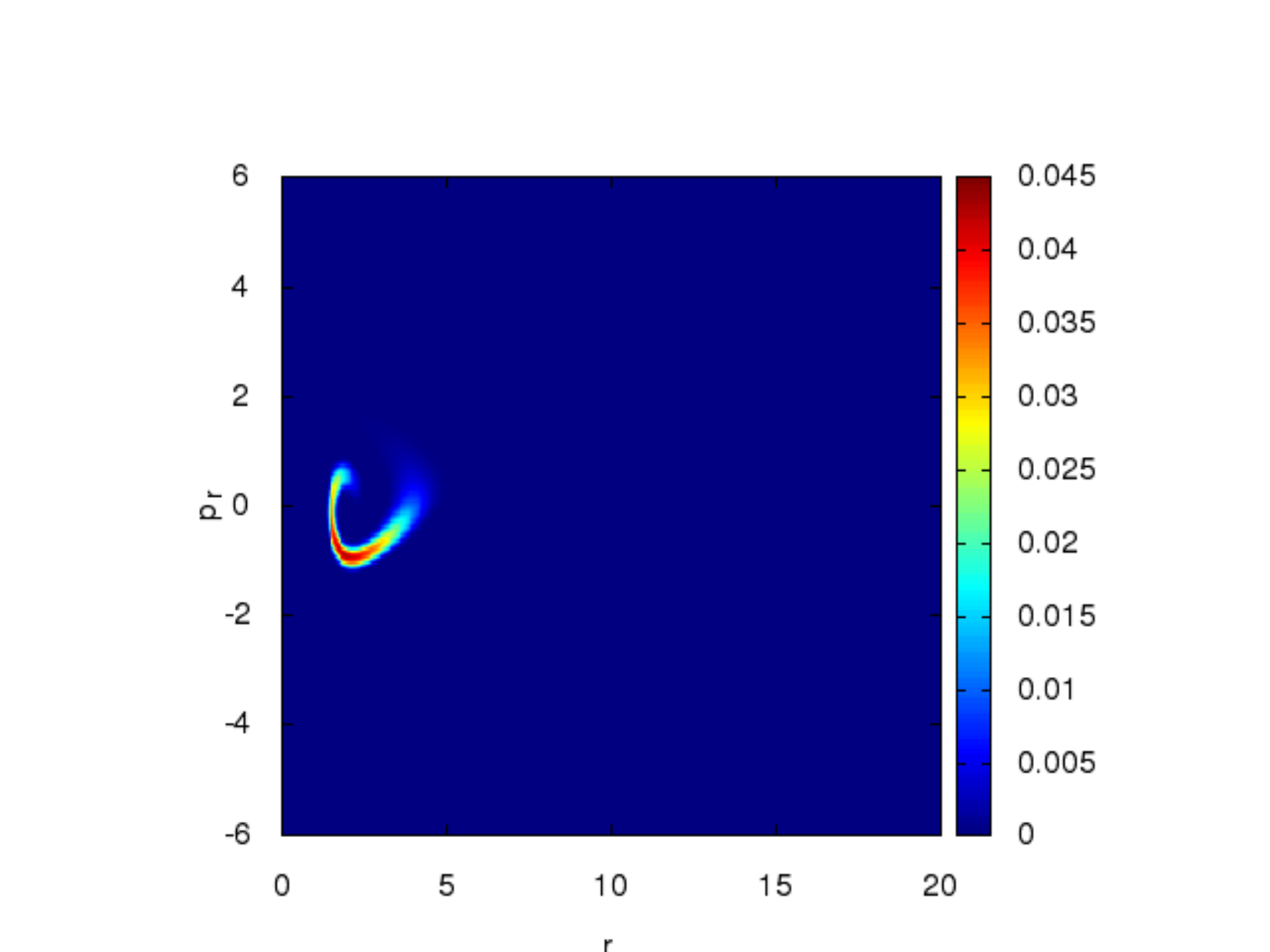}} 
\subfigure{\includegraphics[scale=0.34]{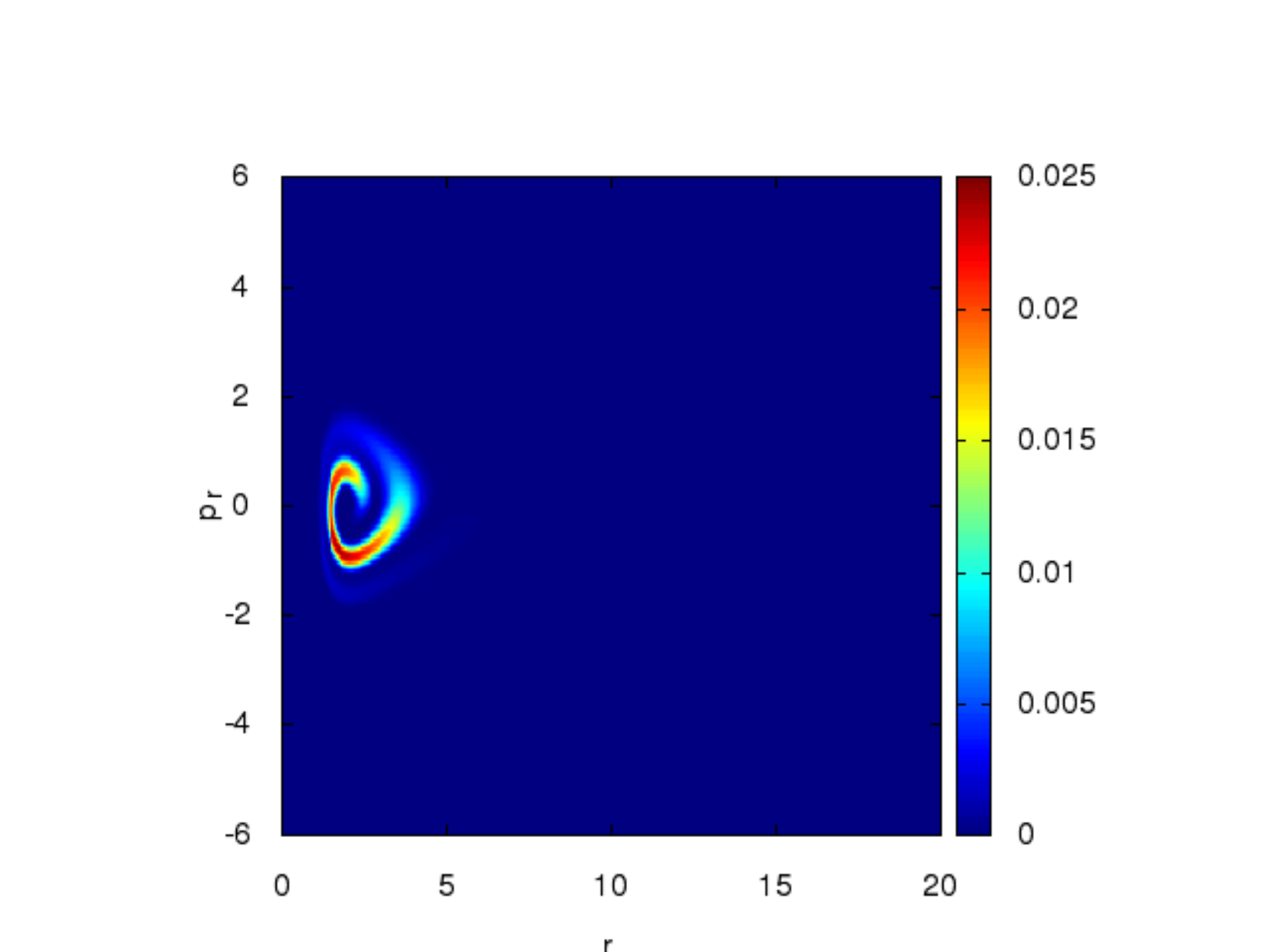}}
\subfigure{\includegraphics[scale=0.34]{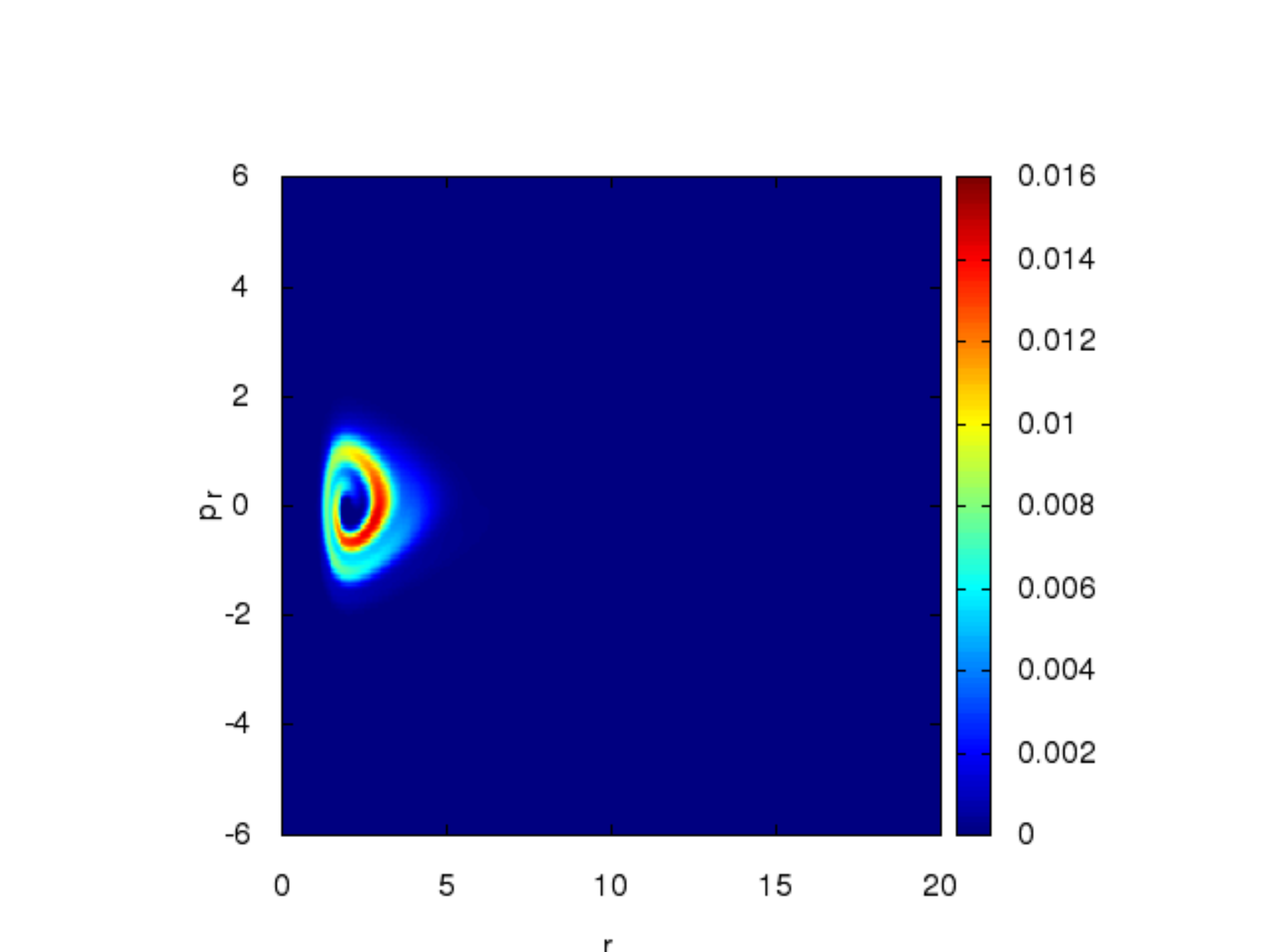}}
\subfigure{\includegraphics[scale=0.34]{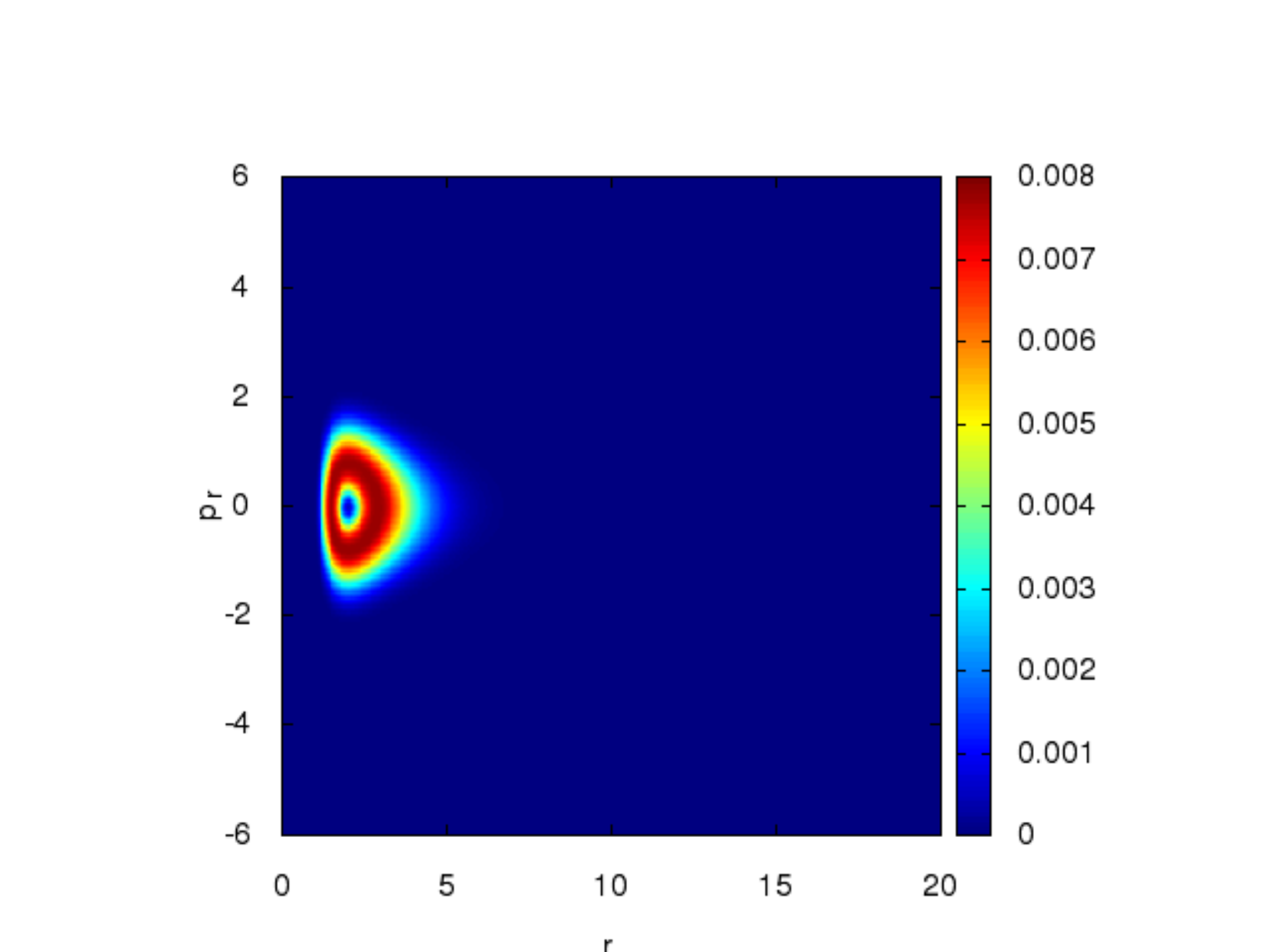}}
\caption{Time evolution of the distribution function $\mathcal{F}$ for
  the isothermal model with angular momentum $\mathcal{L}=3.5$. The
  different panels correspond to times $\mathcal{T}=0, 9.33, 18.84,
  30.79, 45.71, 200.39$.}
\label{pics:PS_iso}
\end{figure}

\begin{figure}[H]
\centering
\subfigure{\includegraphics[scale=0.34]{fig06.pdf}}
\subfigure{\includegraphics[scale=0.34]{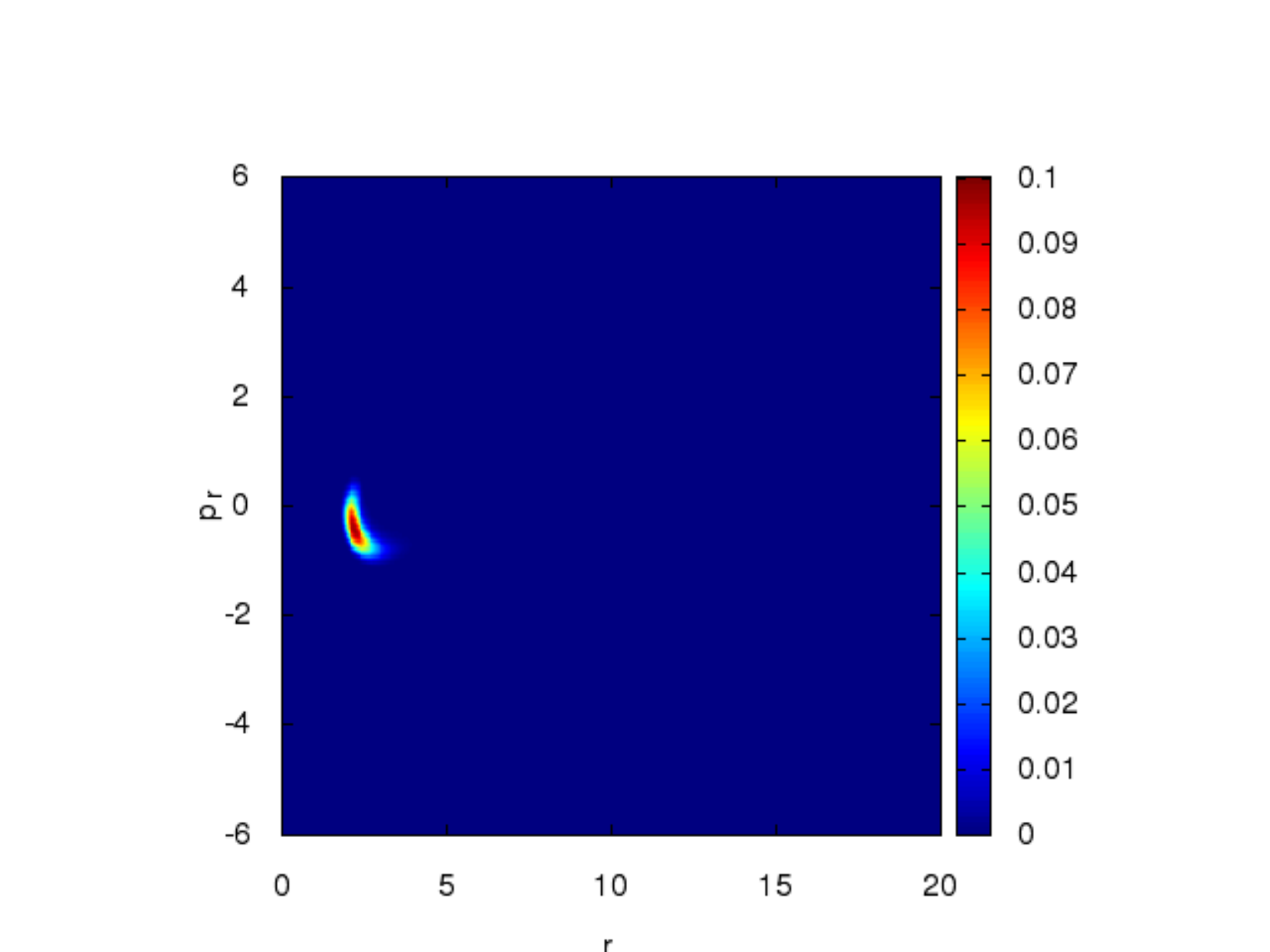}}
\subfigure{\includegraphics[scale=0.34]{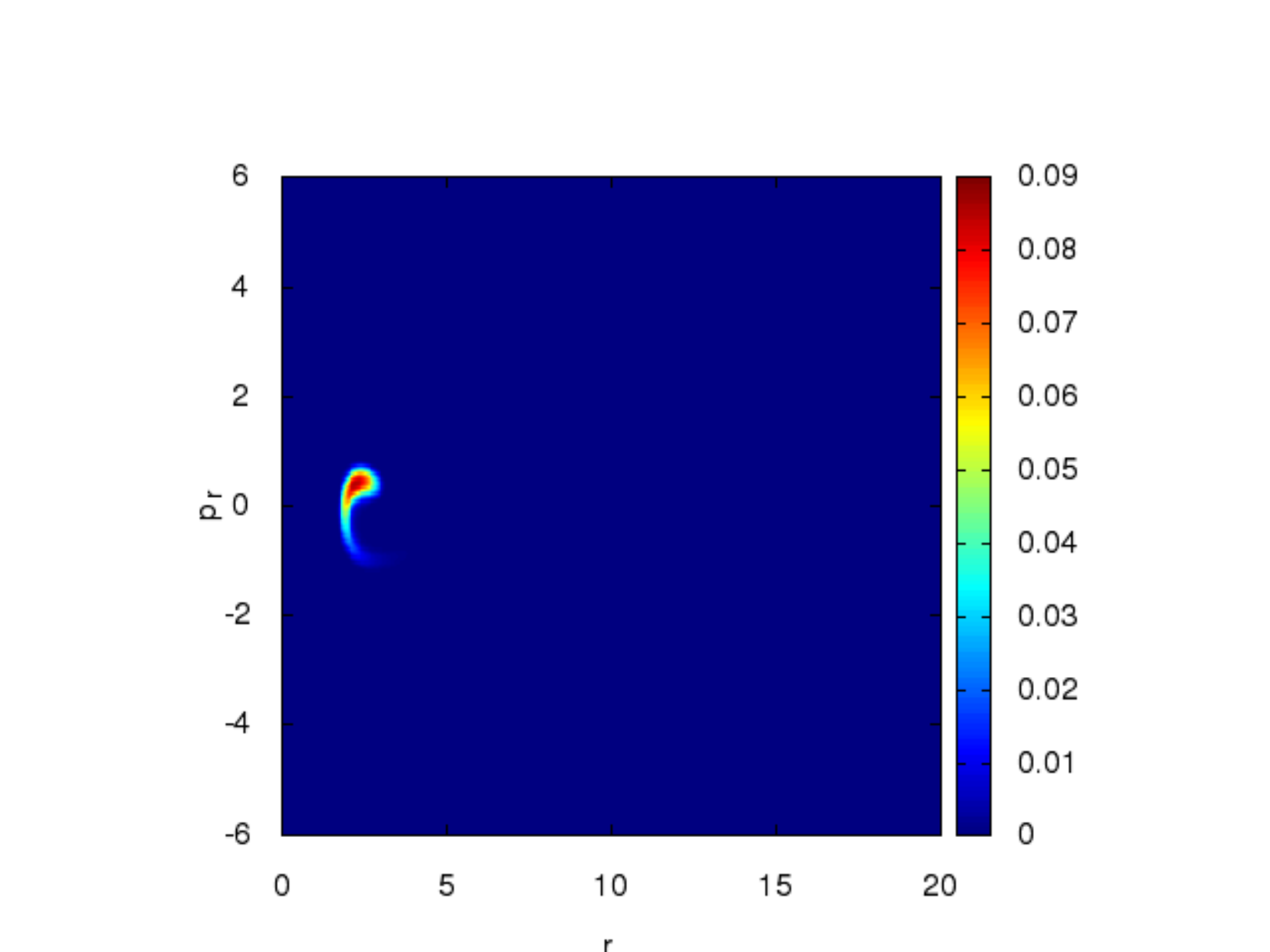}}
\subfigure{\includegraphics[scale=0.34]{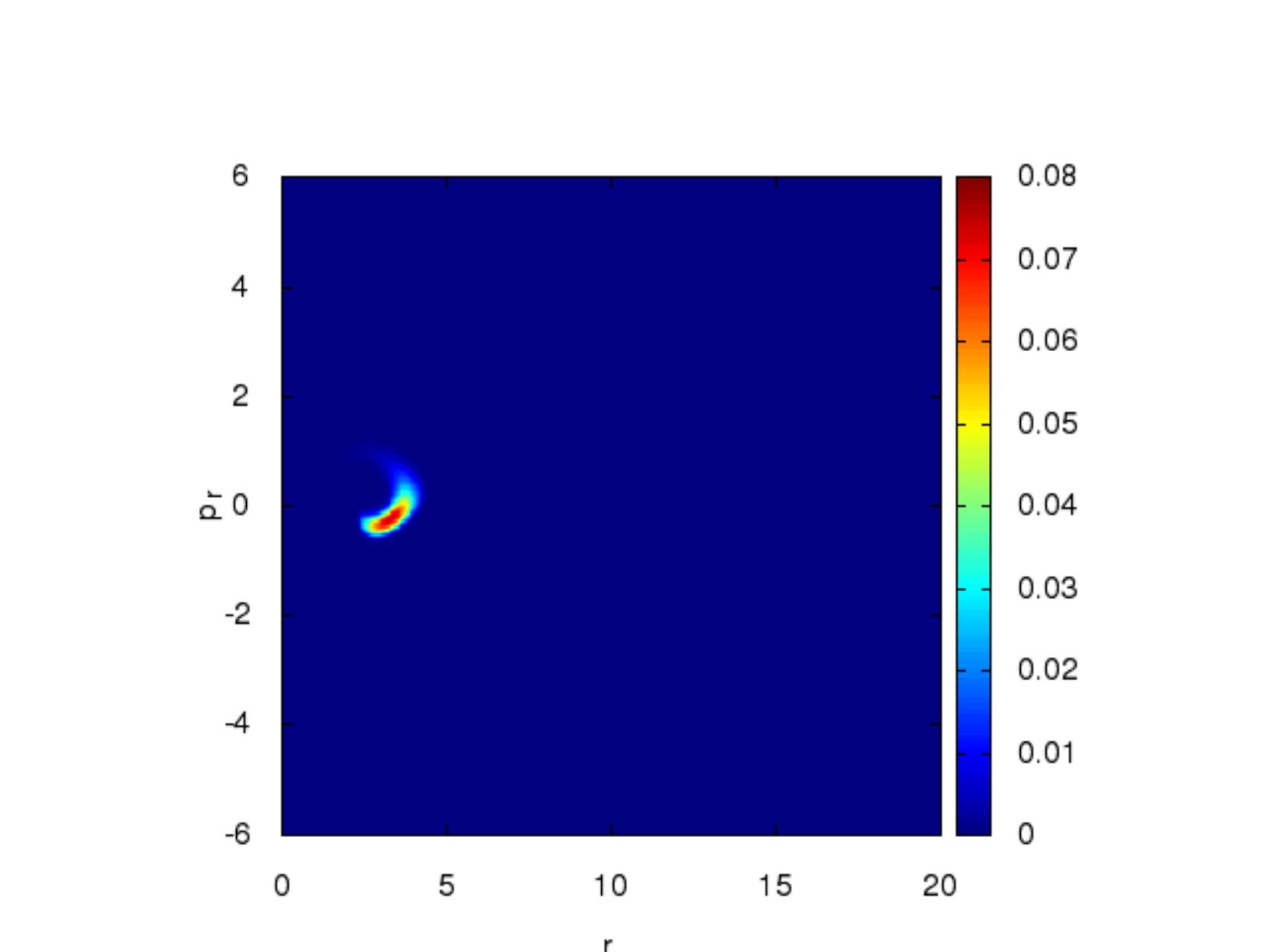}}
\subfigure{\includegraphics[scale=0.34]{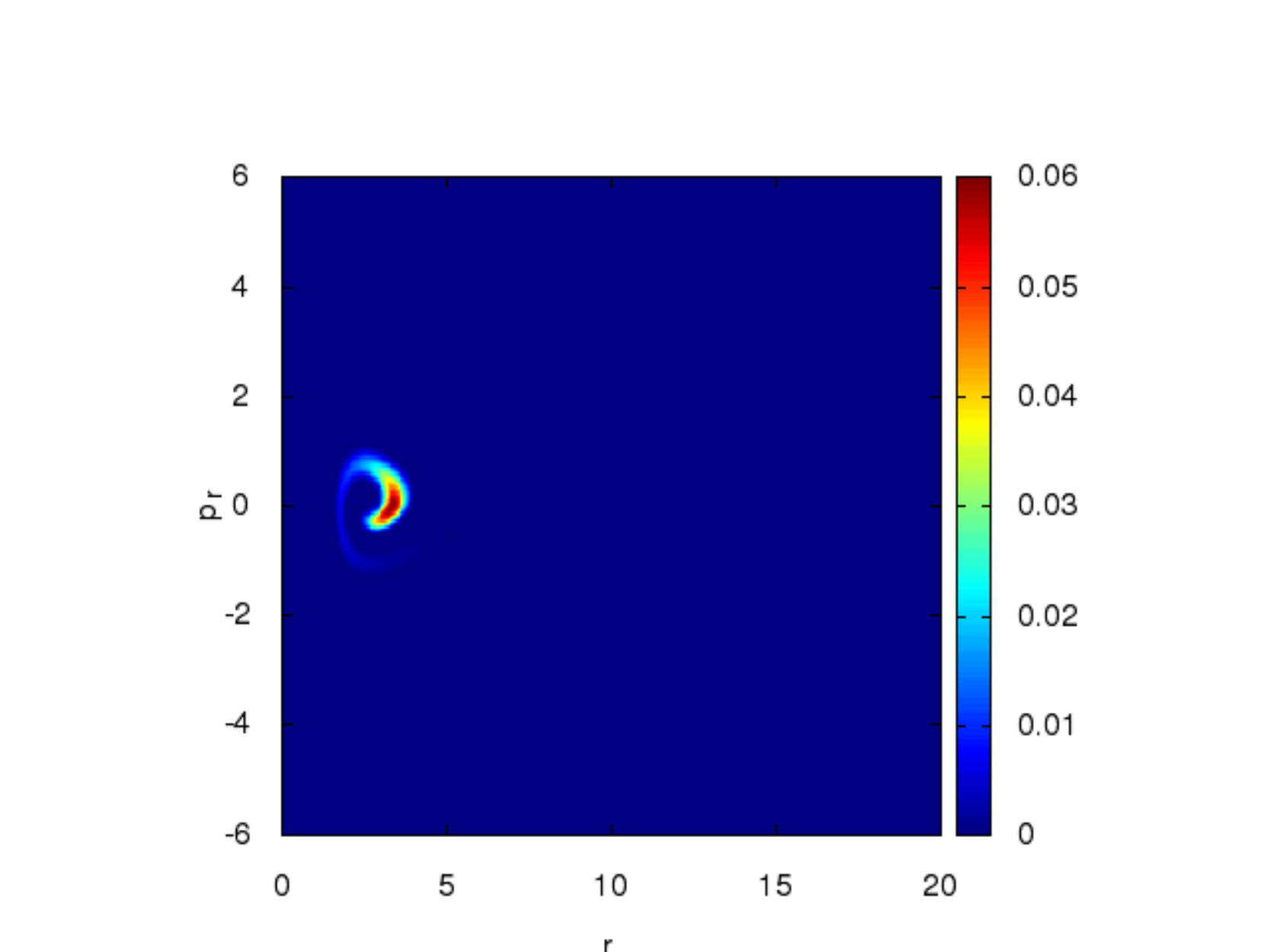}}
\subfigure{\includegraphics[scale=0.34]{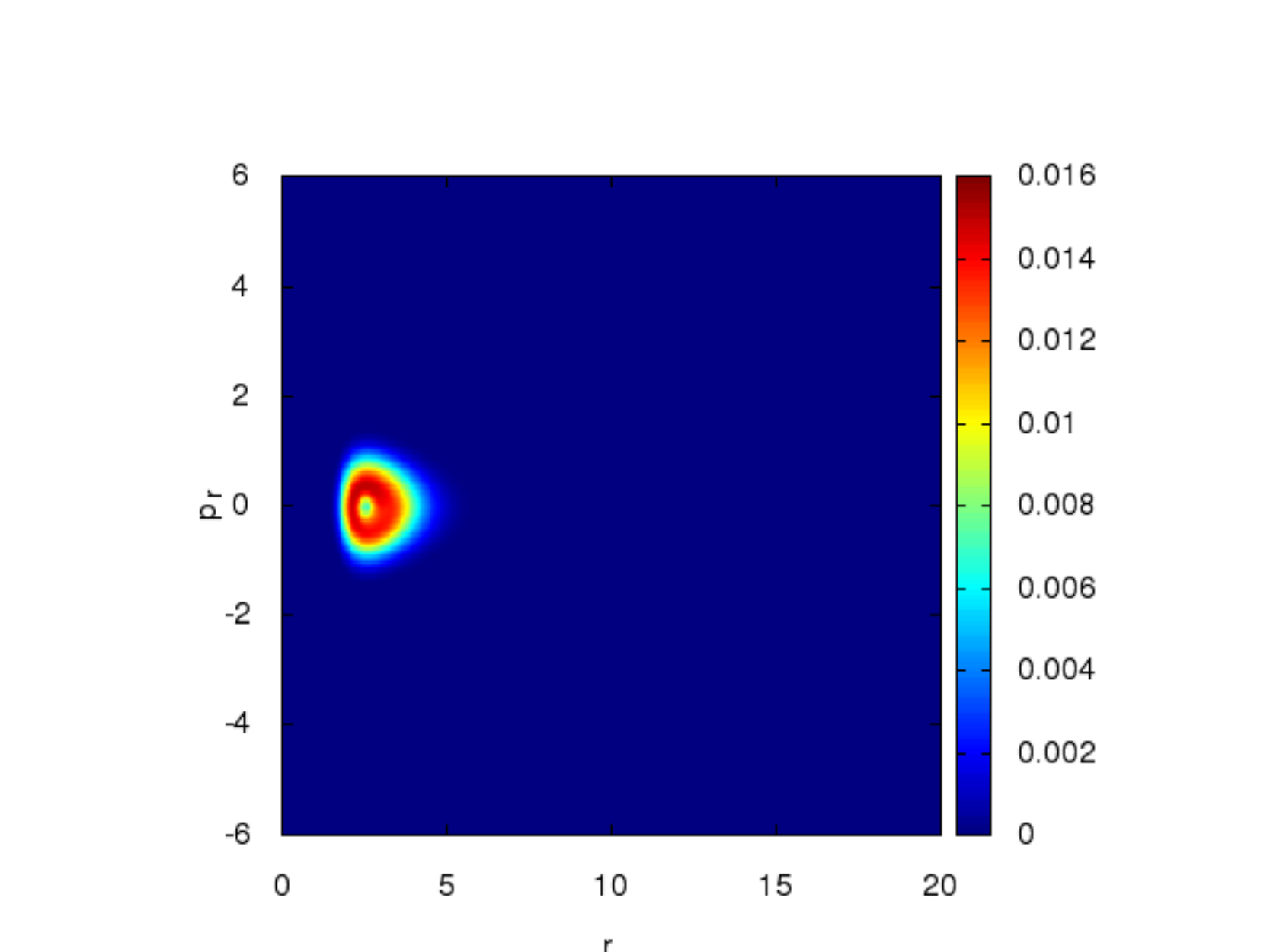}}
\caption{Time evolution of the distribution function $\mathcal{F}$ for
  the truncated isothermal model with angular momentum
  $\mathcal{L}=3.5$. The different panels correspond to times
  $\mathcal{T}=0, 9.33, 18.84, 30.79, 45.71, 259.38$.}
\label{pics:PS_trun}
\end{figure}

\begin{figure}[H]
\centering
\subfigure{\includegraphics[scale=0.34]{fig06.pdf}}
\subfigure{\includegraphics[scale=0.34]{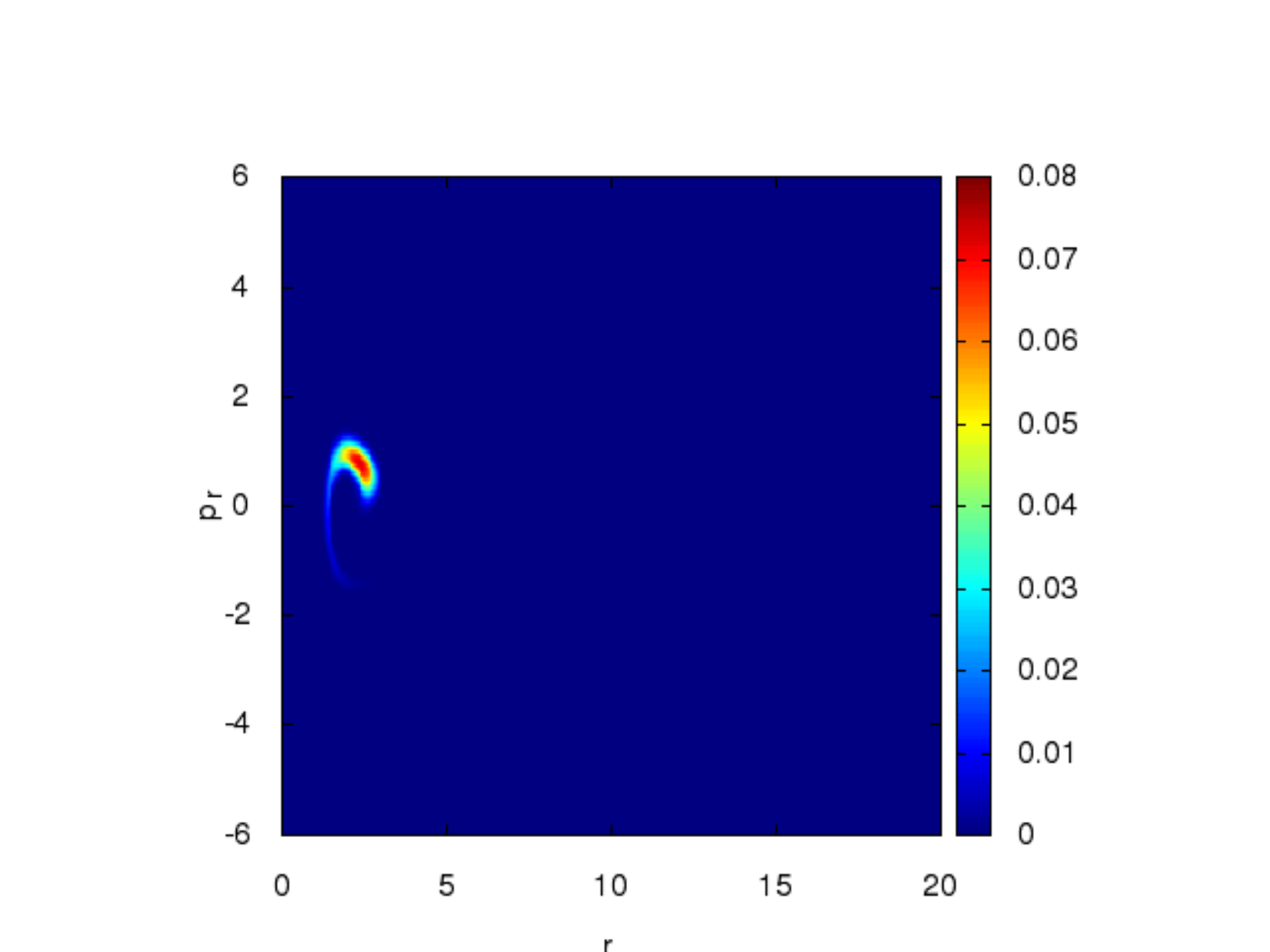}}
\subfigure{\includegraphics[scale=0.34]{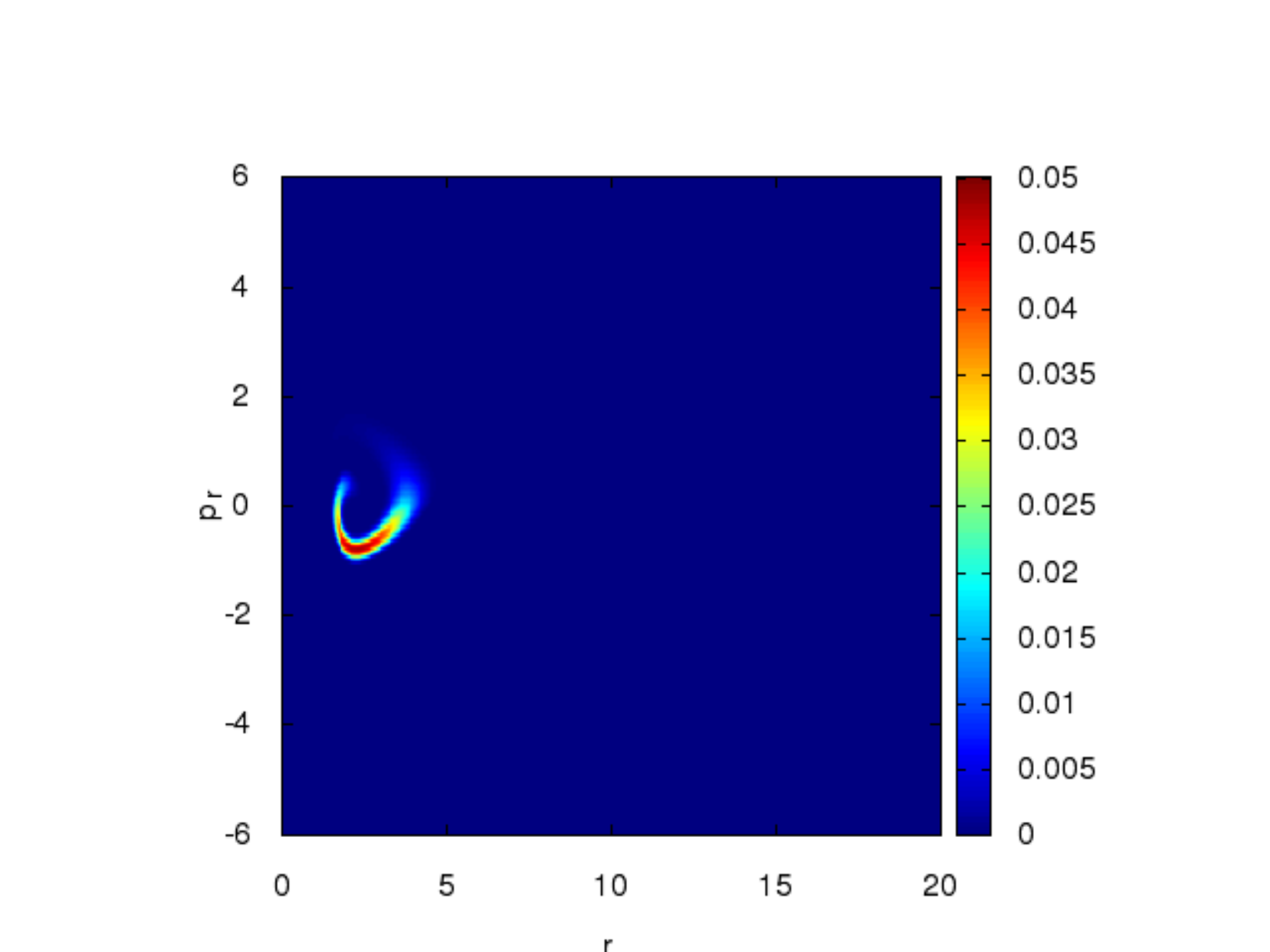}}
\subfigure{\includegraphics[scale=0.34]{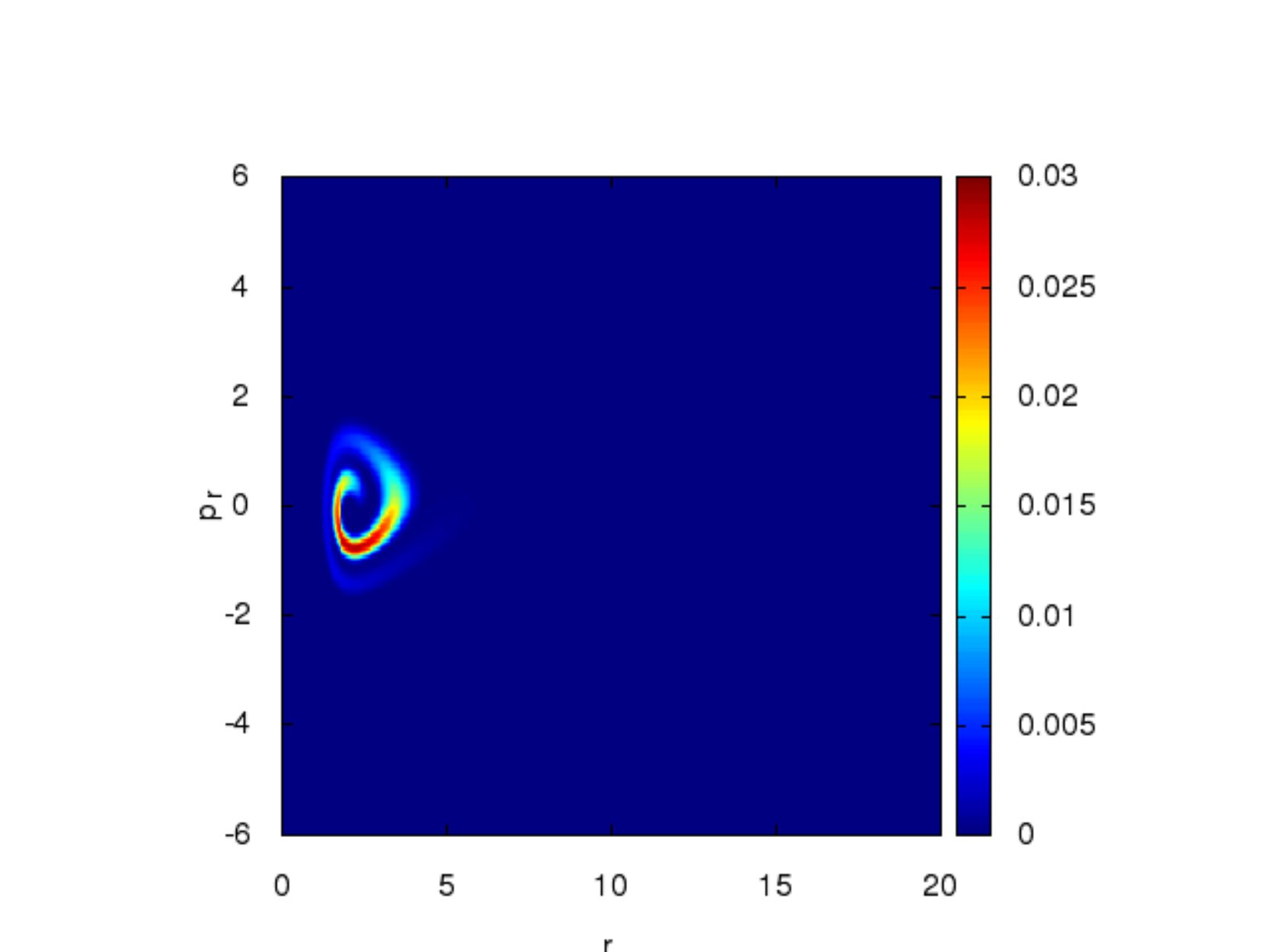}}
\subfigure{\includegraphics[scale=0.34]{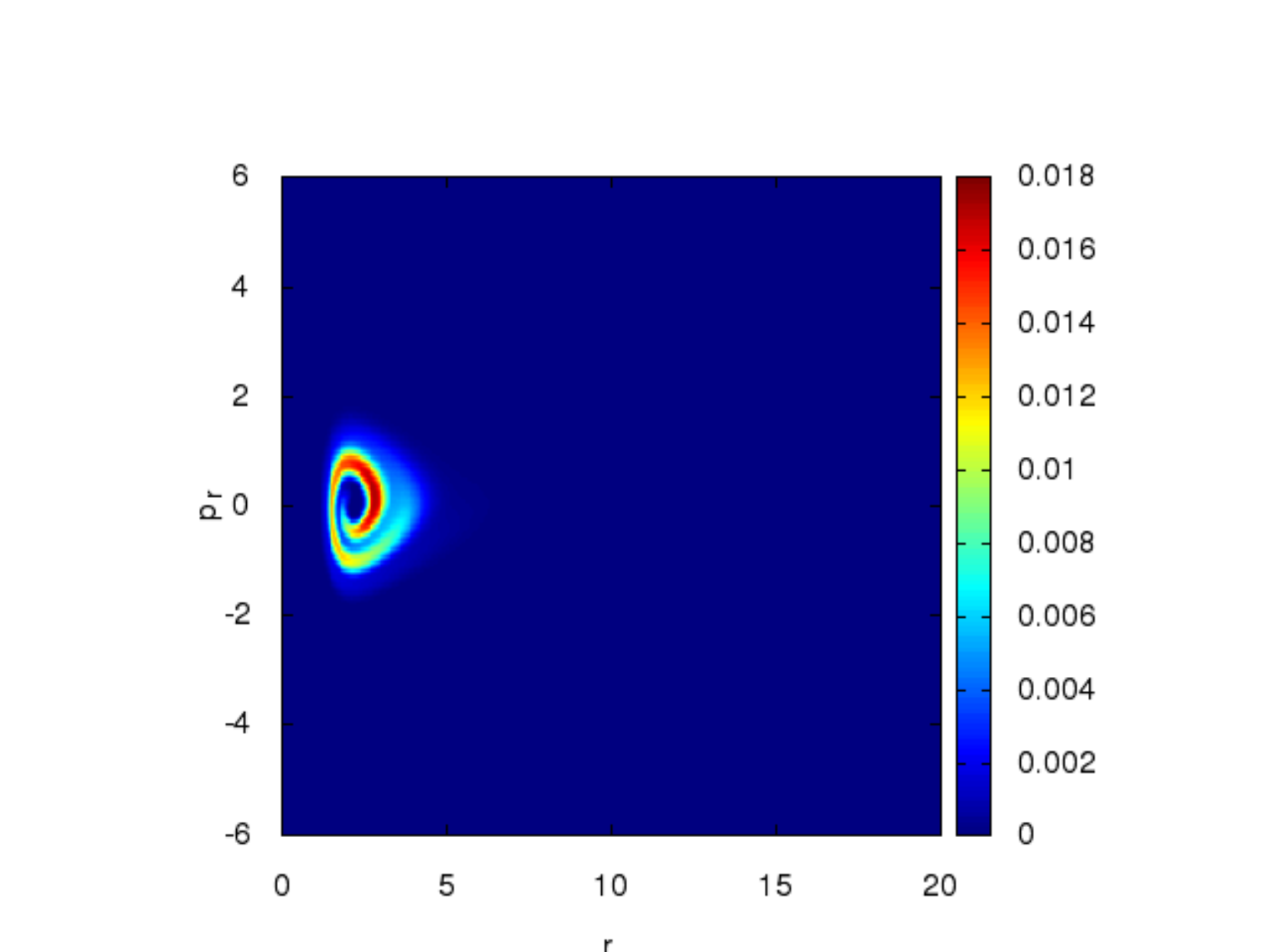}}
\subfigure{\includegraphics[scale=0.34]{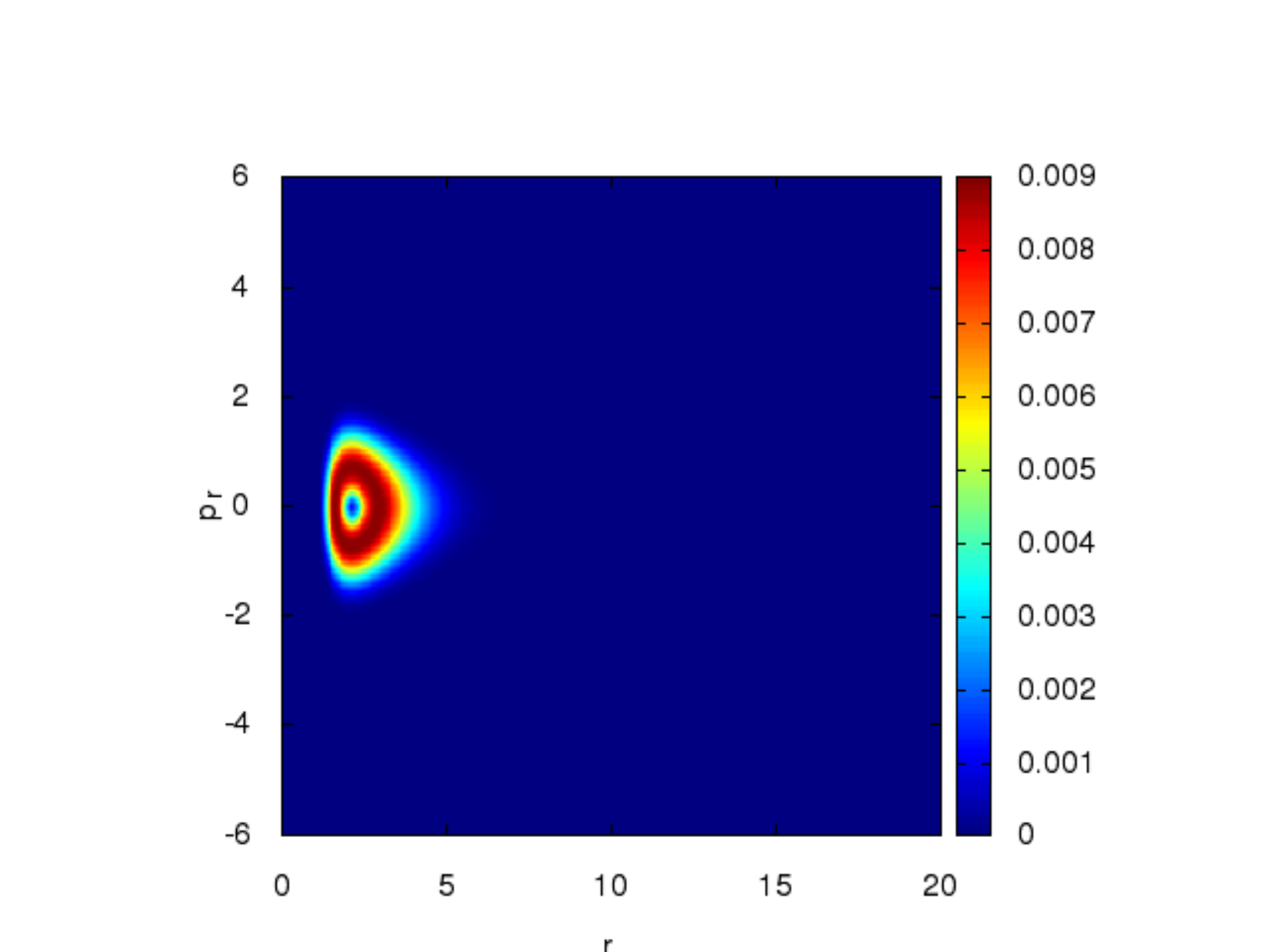}}
\caption{Time evolution of the distribution function $\mathcal{F}$ for
  the Burkert model with angular momentum $\mathcal{L}=3.5$. The
  different panels correspond to times $\mathcal{T}=0, 9.33, 18.84,
  30.79, 45.71, 196.85$.}
\label{pics:PS_bur}
\end{figure}

\begin{figure}[H]
\centering
\subfigure{\includegraphics[scale=0.34]{fig06.pdf}}
\subfigure{\includegraphics[scale=0.34]{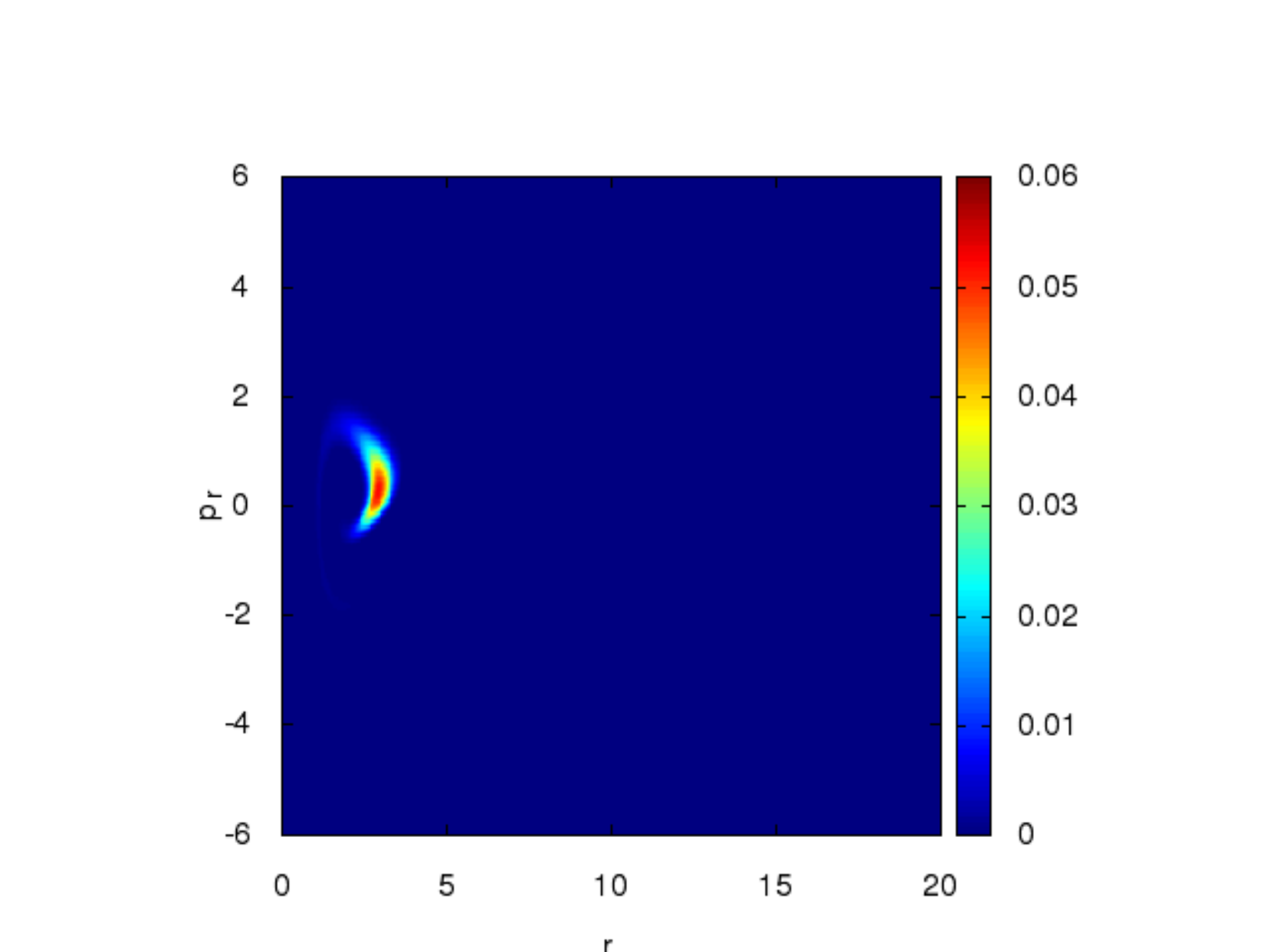}}
\subfigure{\includegraphics[scale=0.34]{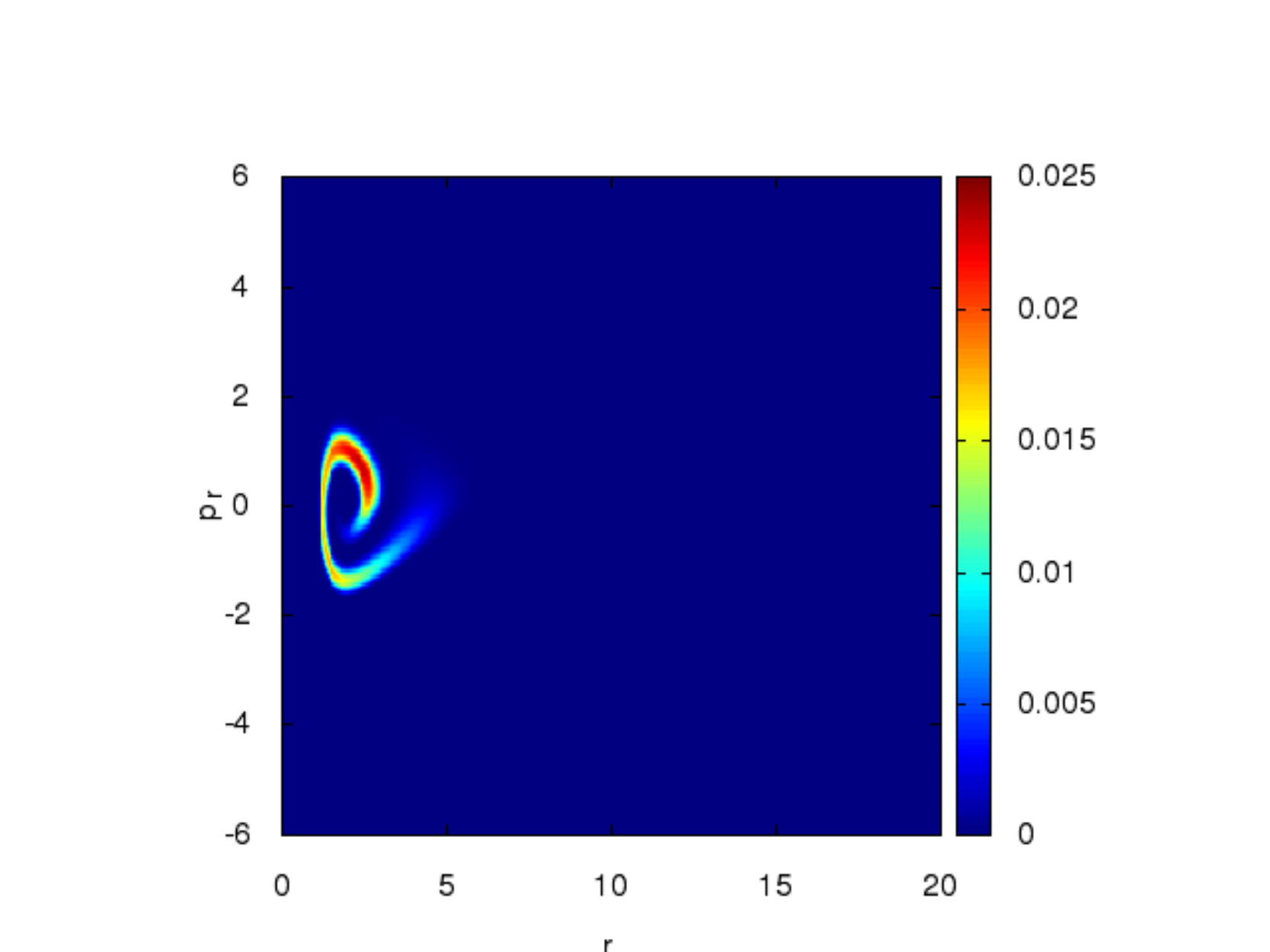}}
\subfigure{\includegraphics[scale=0.34]{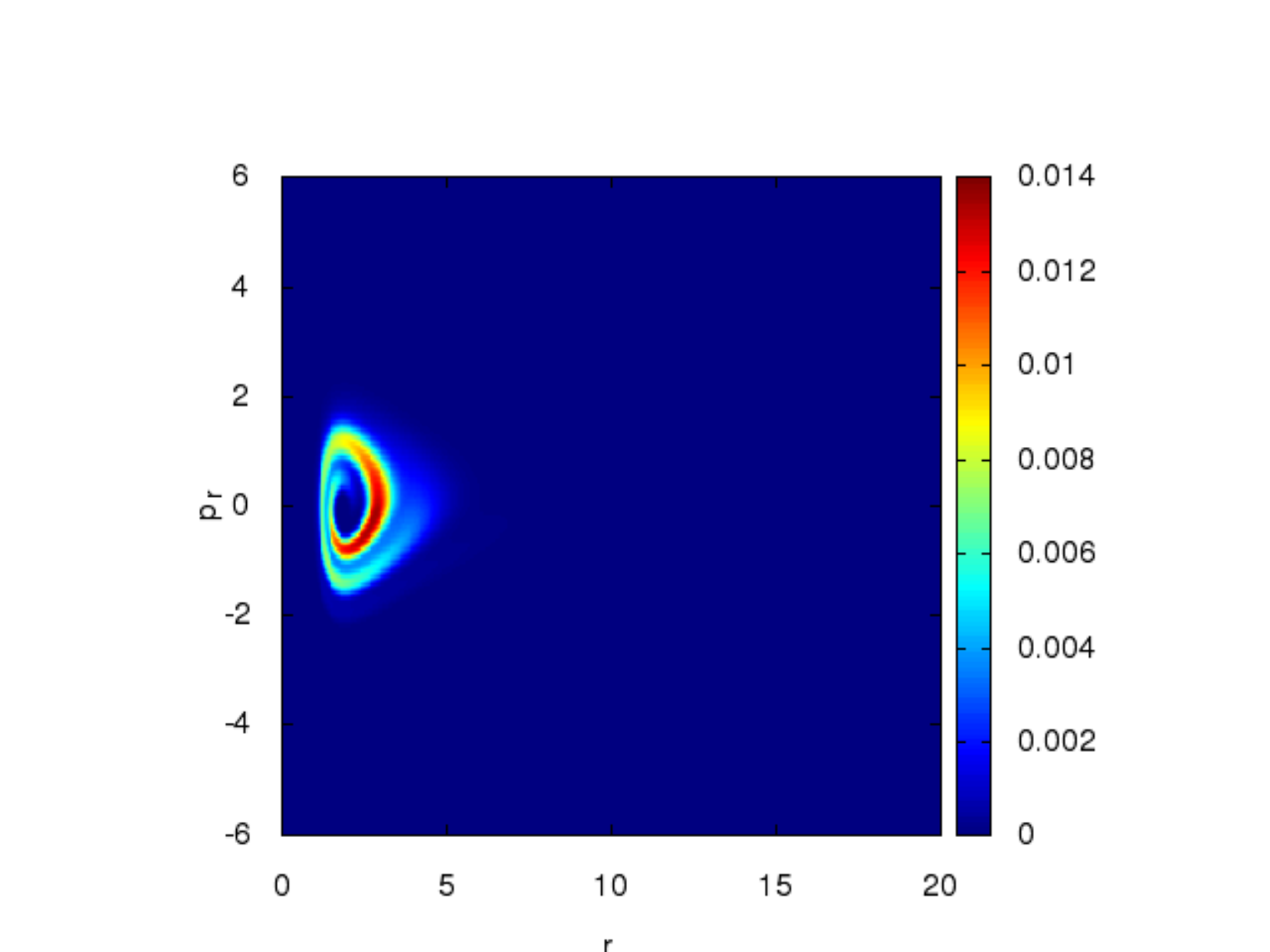}}
\subfigure{\includegraphics[scale=0.34]{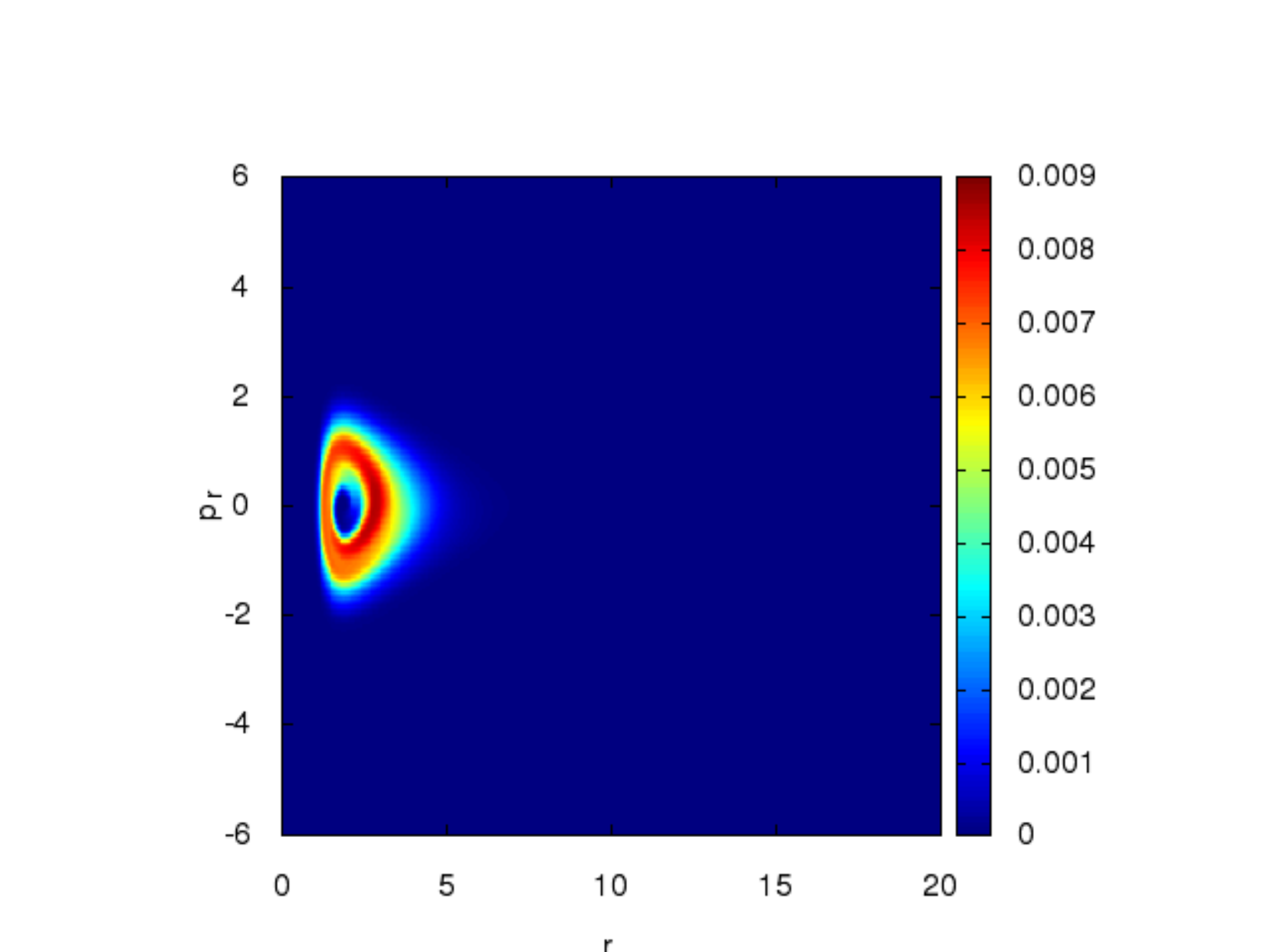}}
\subfigure{\includegraphics[scale=0.34]{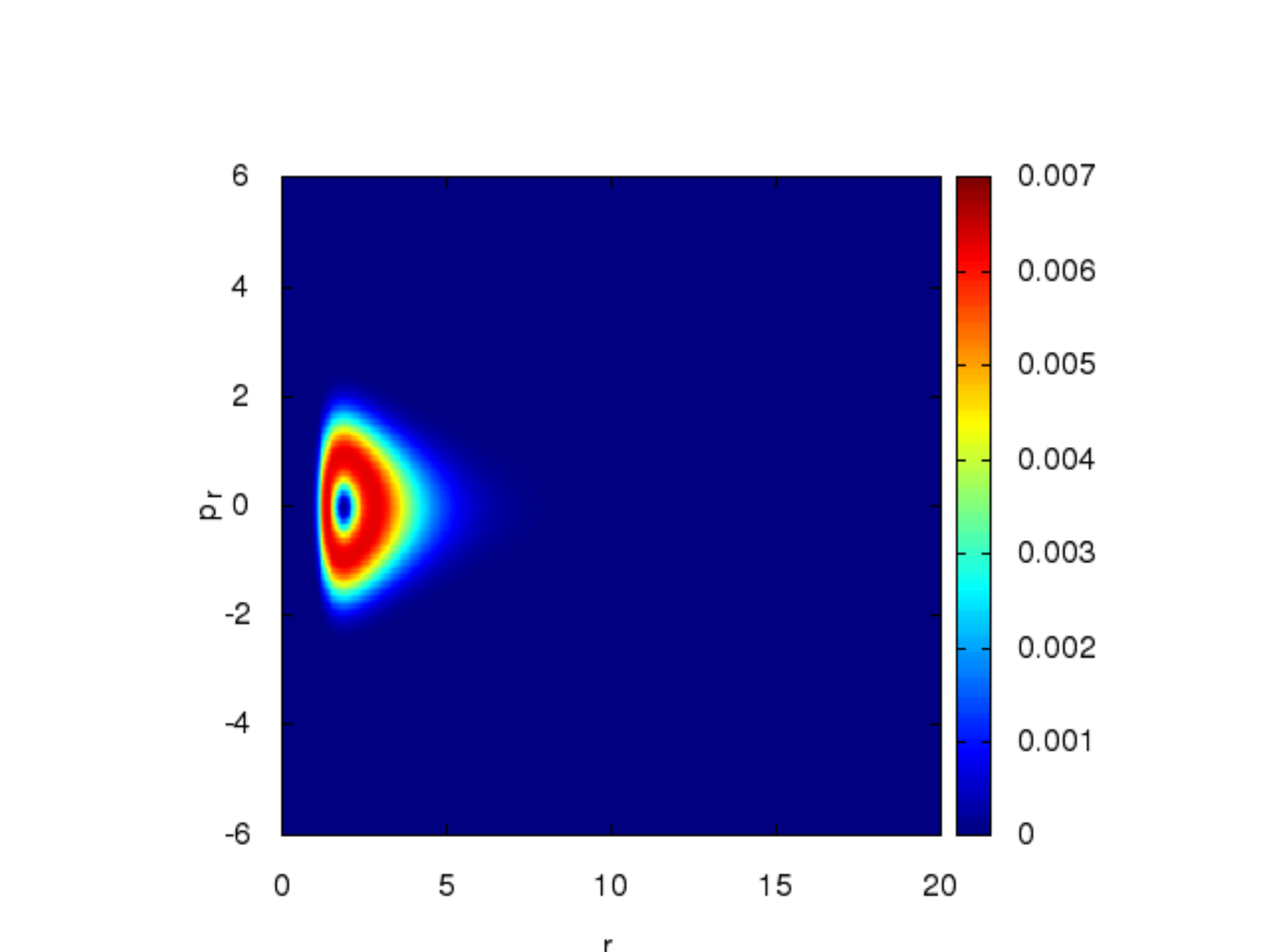}}
\caption{Time evolution of the distribution function $\mathcal{F}$ for
  the NFW model with angular momentum $\mathcal{L}=3.5$. The different
  panels correspond to times $\mathcal{T}=0, 9.33, 18.84, 30.79, 45.71, 189.20$.}
\label{pics:PS_NFW}
\end{figure}

\begin{figure}[H]
\centering
\subfigure{\includegraphics[scale=0.35]{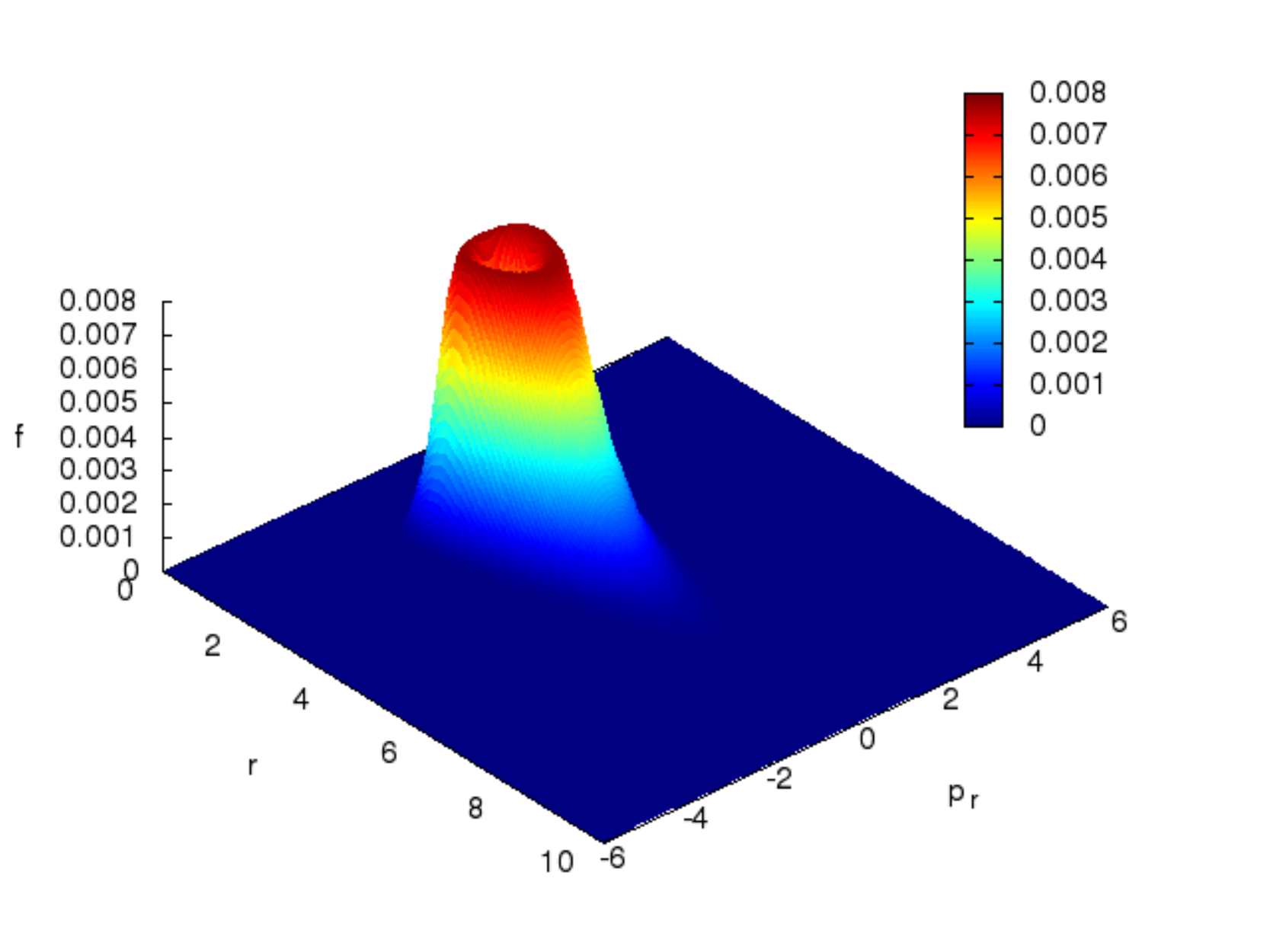}}
\subfigure{\includegraphics[scale=0.35]{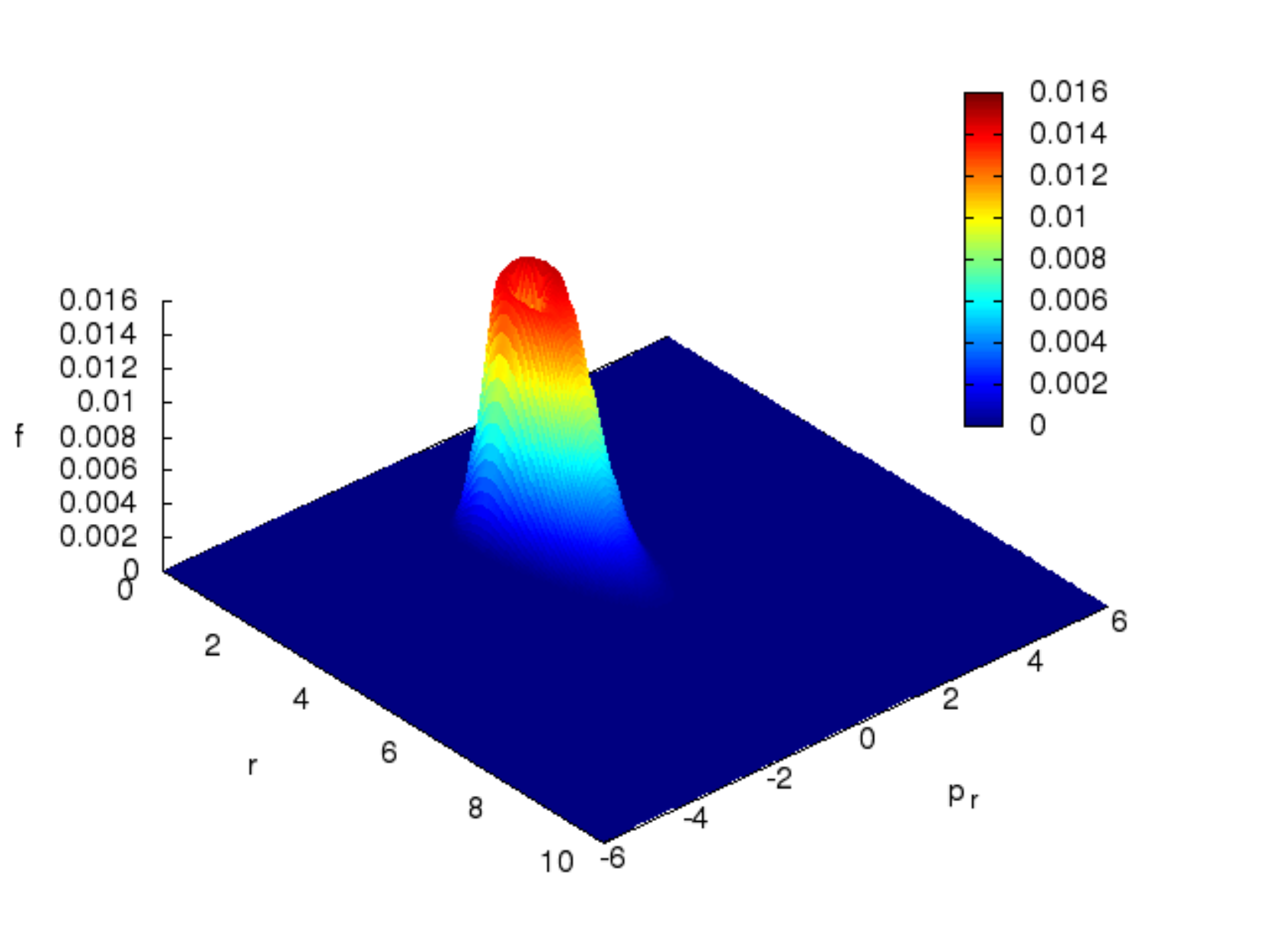}}
\subfigure{\includegraphics[scale=0.35]{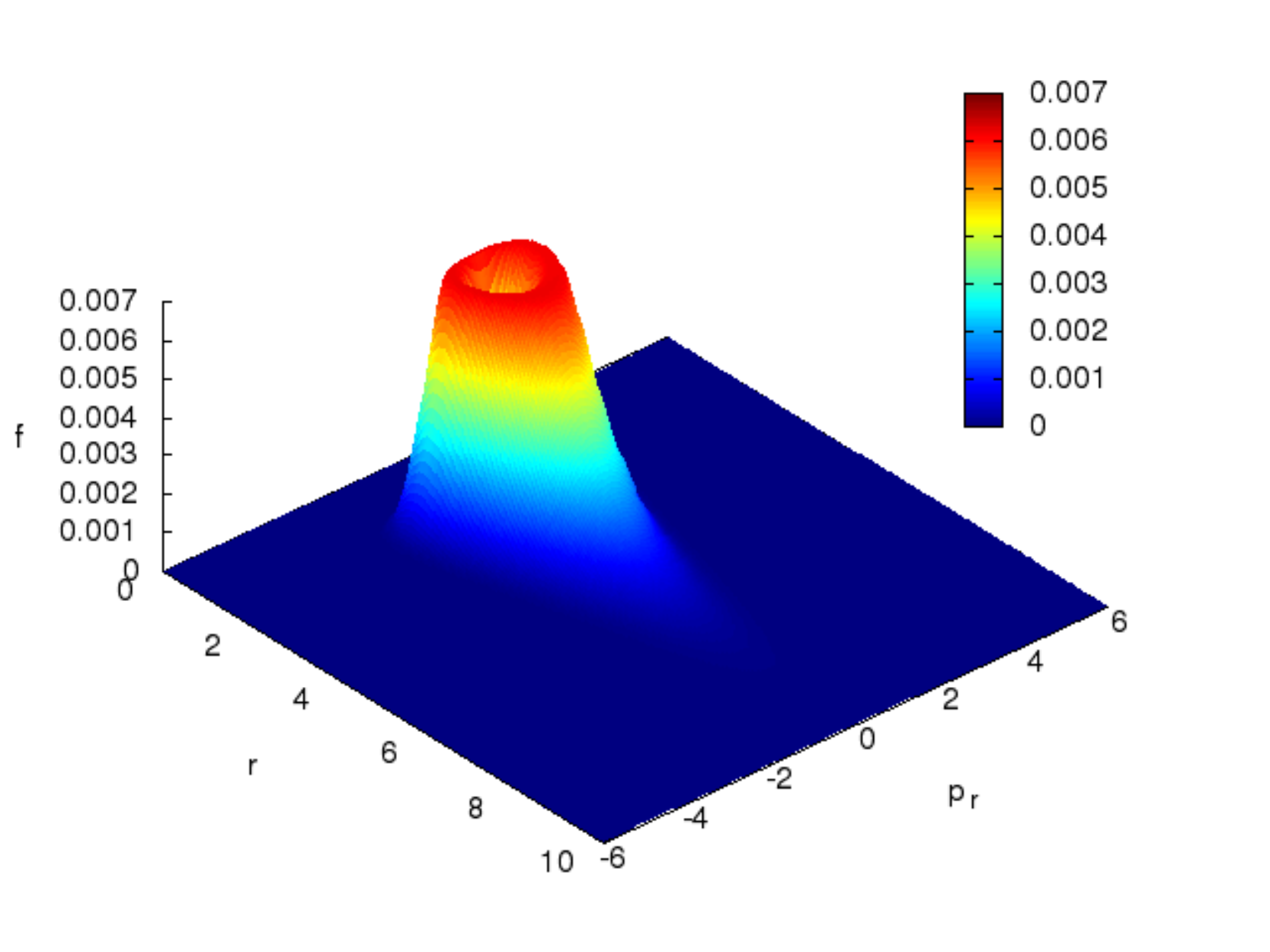}}
\subfigure{\includegraphics[scale=0.35]{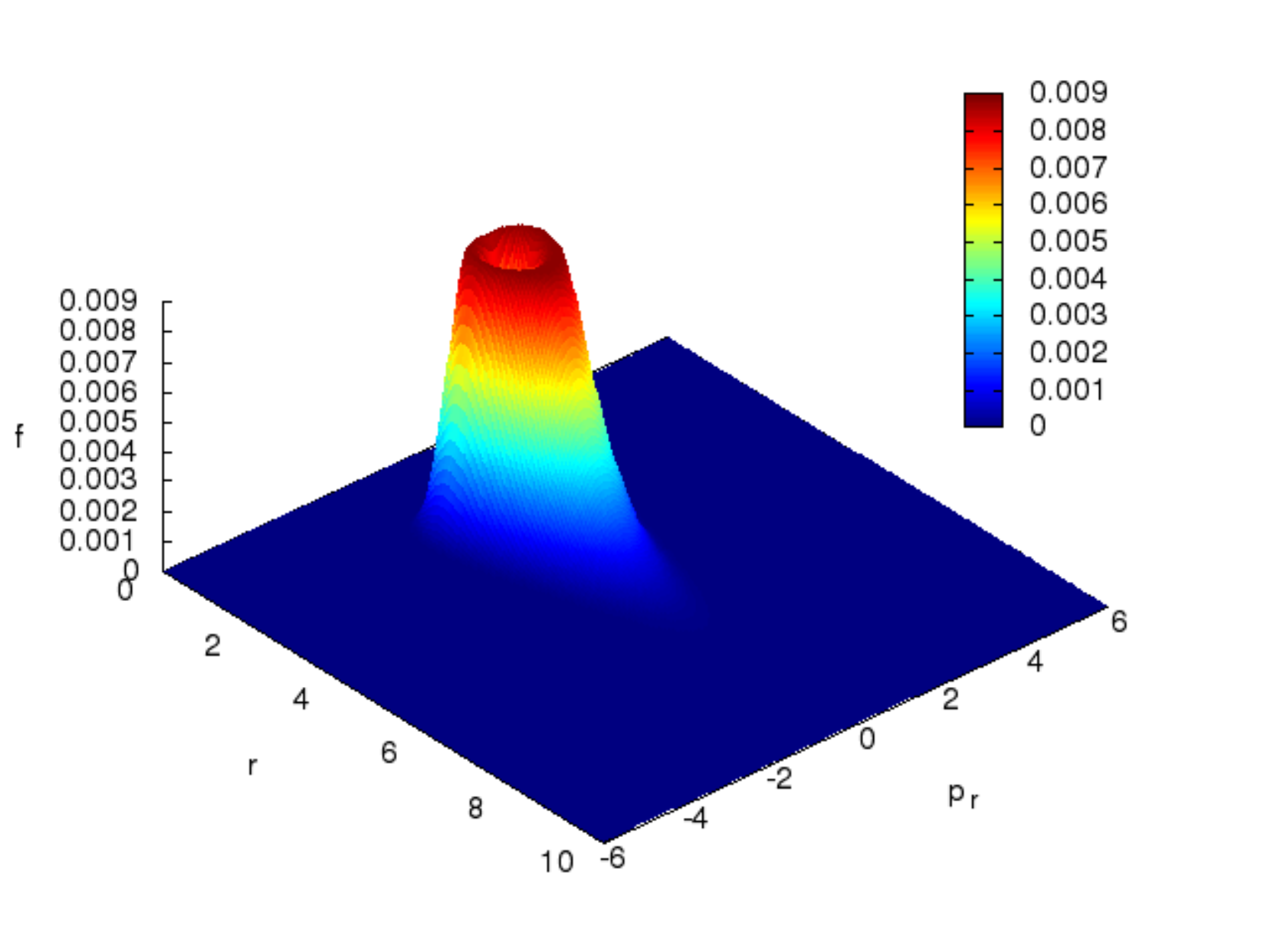}}
\caption{Final virialized distribution function for the four different
  halo models with angular momentum $\mathcal{L}=3.5$. Starting in the
  top left panel: isothermal, truncated isothermal, Burkert and
  NFW.}
\label{Fig:FD_all_ff}
\end{figure}

\begin{figure}[H]
\centering
\subfigure{\includegraphics[scale=0.2, angle=0]{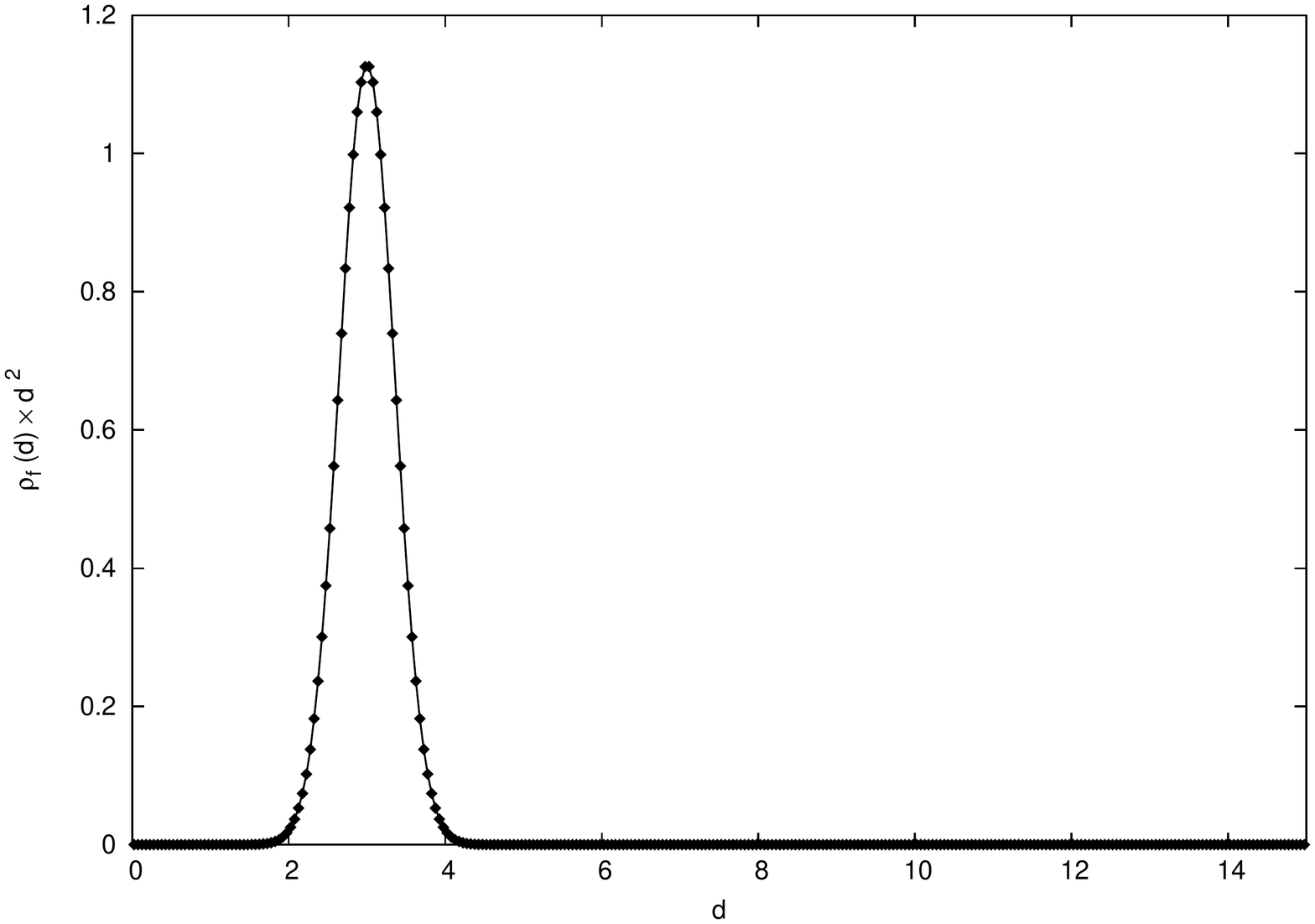}}
\subfigure{\includegraphics[scale=0.2, angle=0]{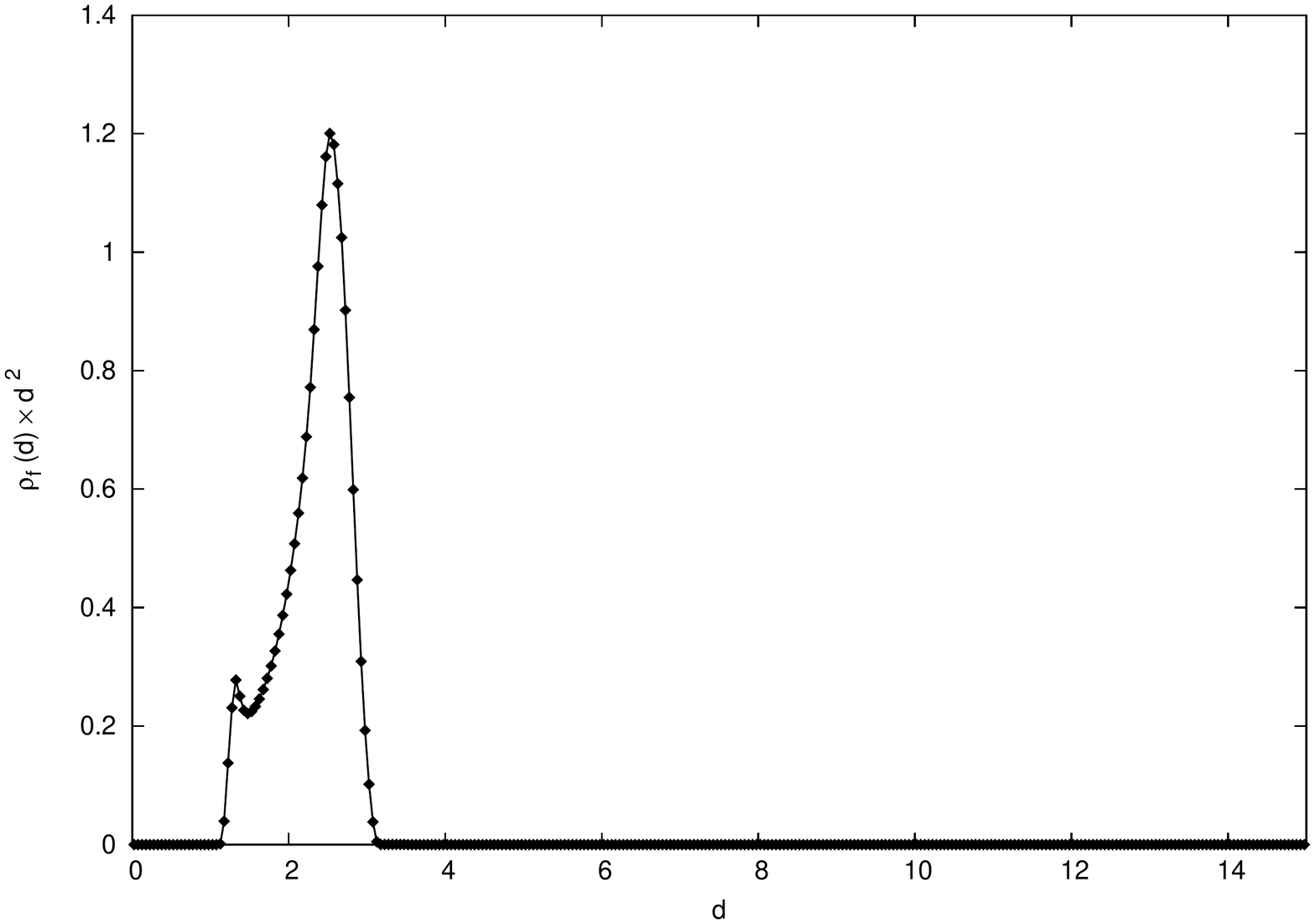}}
\subfigure{\includegraphics[scale=0.2, angle=0]{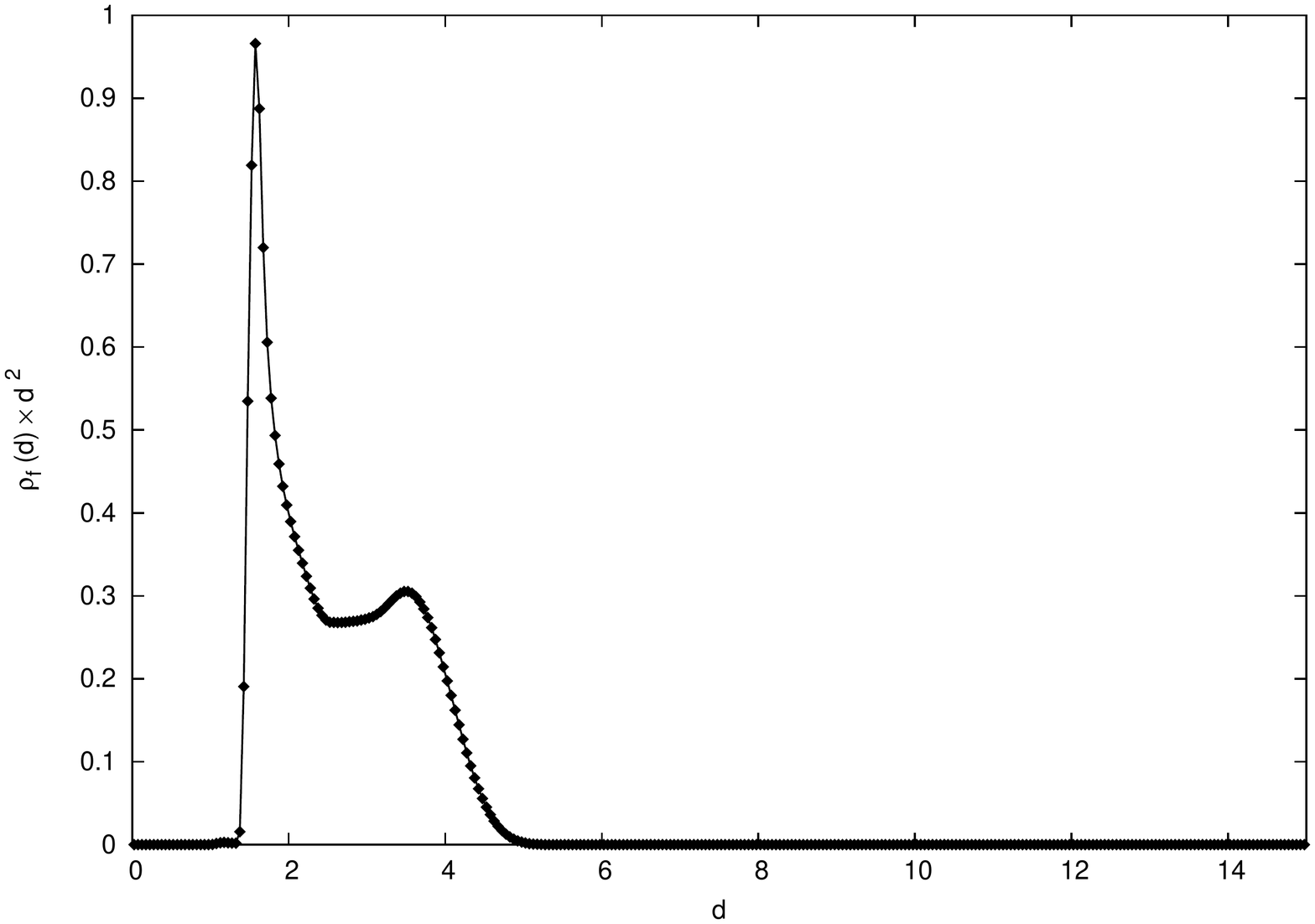}}
\subfigure{\includegraphics[scale=0.2, angle=0]{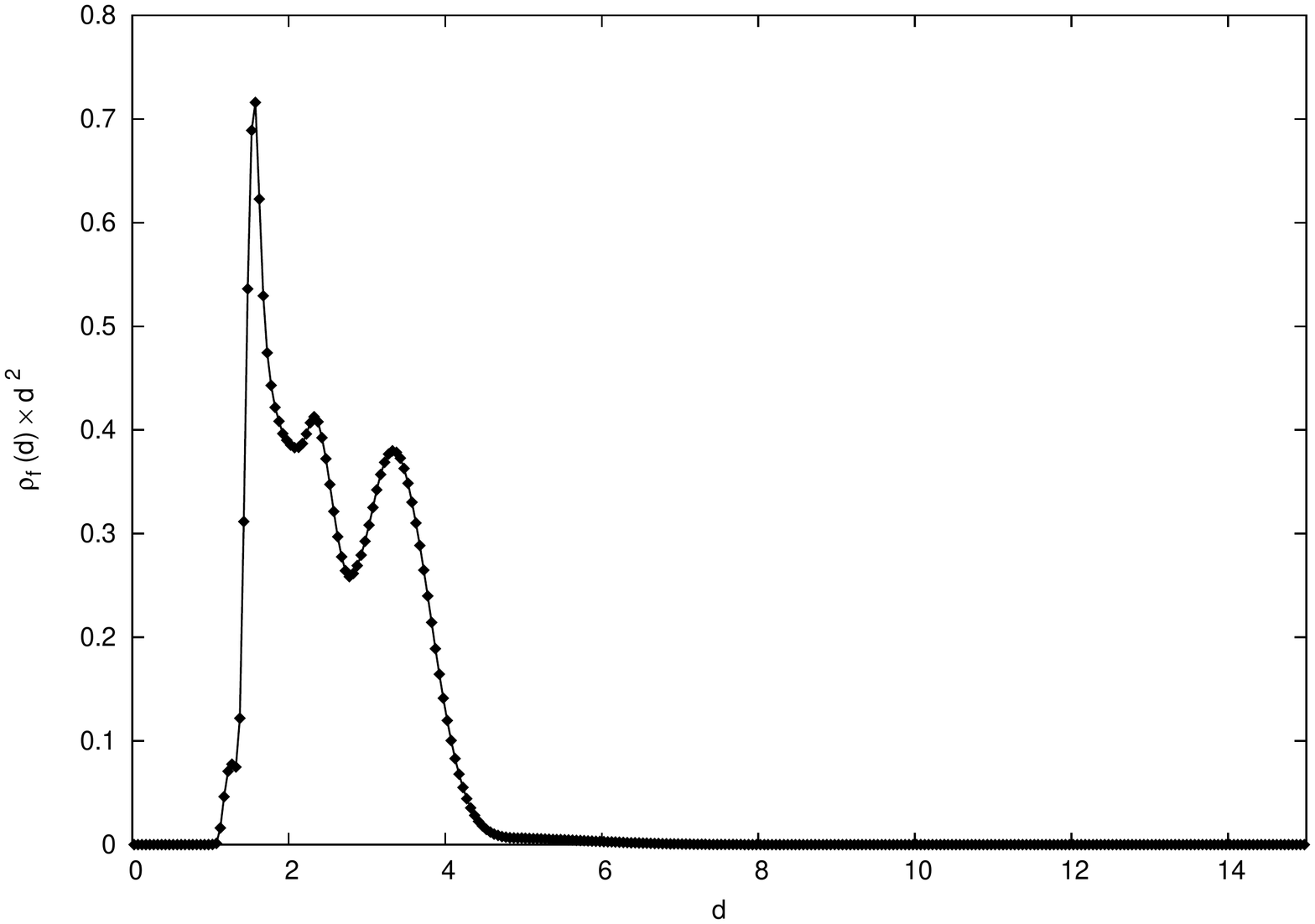}}
\subfigure{\includegraphics[scale=0.2, angle=0]{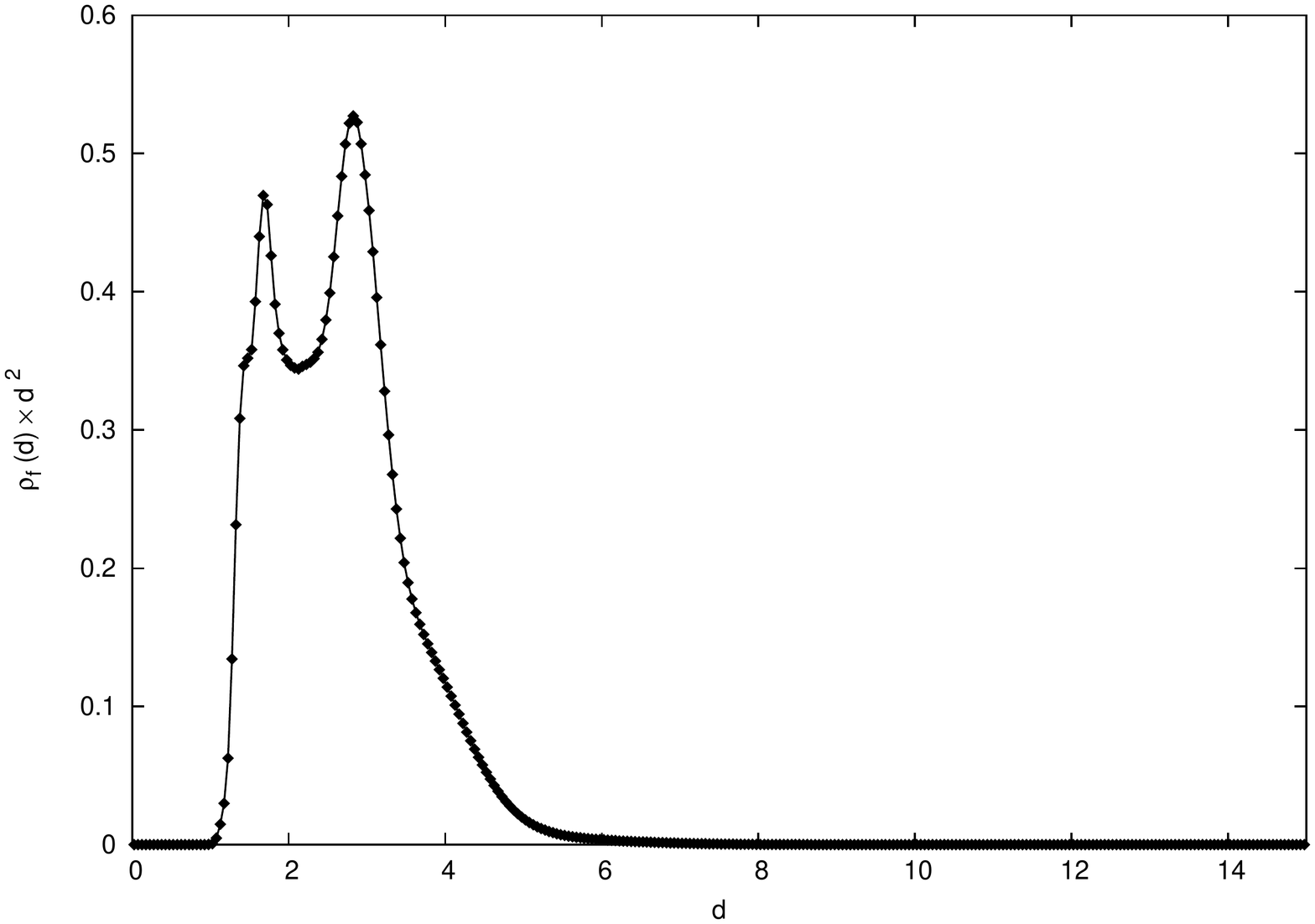}}
\subfigure{\includegraphics[scale=0.2, angle=0]{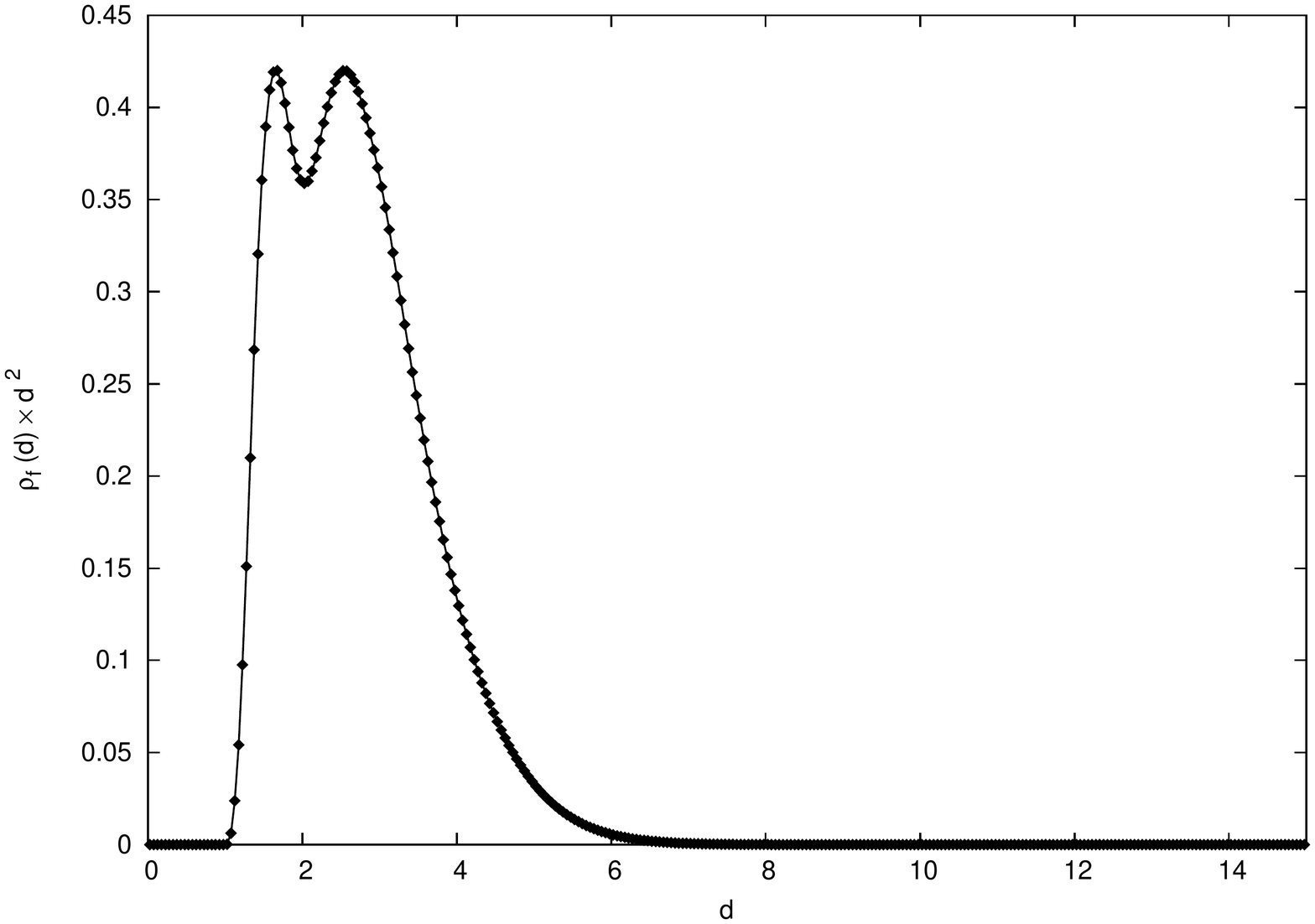}}
\caption{
  Time evolution of the integrated particle density $\rho_f(d)$ multiplied for the 
  $d^2$ factor for the isothermal model with angular momentum
  $\mathcal{L}=3.5$. 
  The different panels correspond to times $\mathcal{T}=0, 9.33, 18.84,
  30.79, 45.71, 200.39$, same that those in
  Figure~\ref{pics:PS_iso}.}
\label{pics:D_iso}
\end{figure}

\begin{figure}[H]
\centering
\subfigure{\includegraphics[scale=0.2, angle=0]{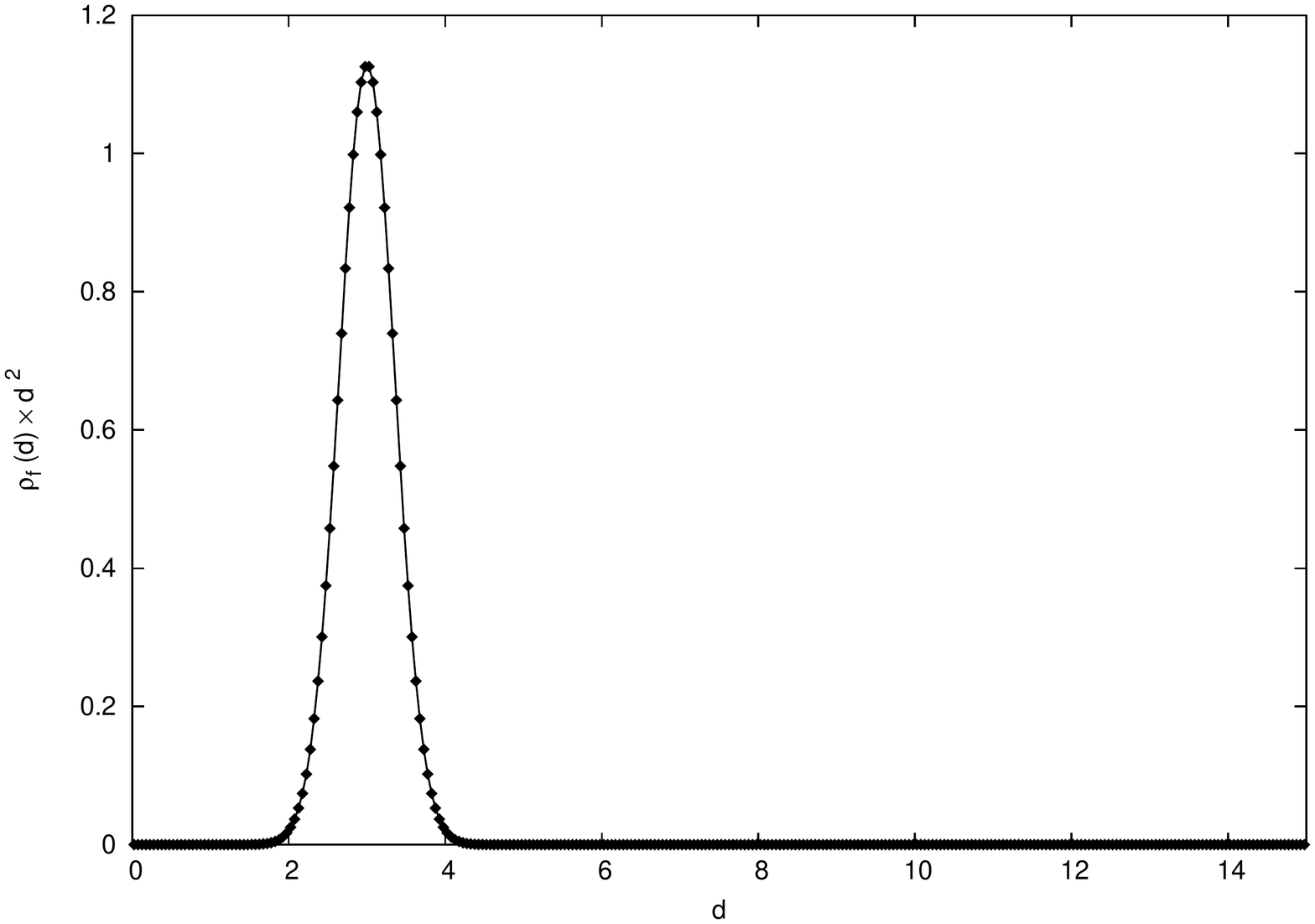}}
\subfigure{\includegraphics[scale=0.2, angle=0]{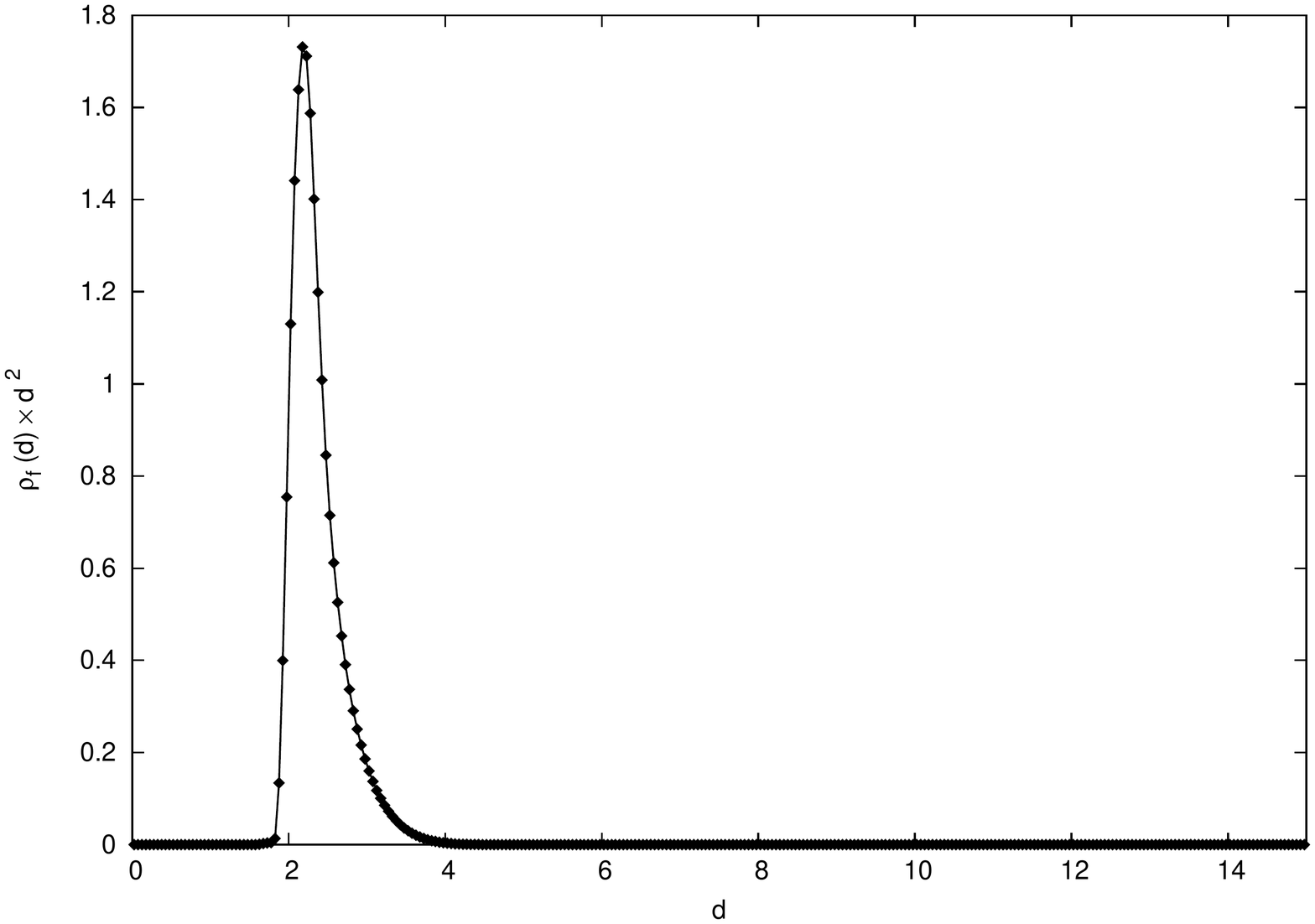}}
\subfigure{\includegraphics[scale=0.2, angle=0]{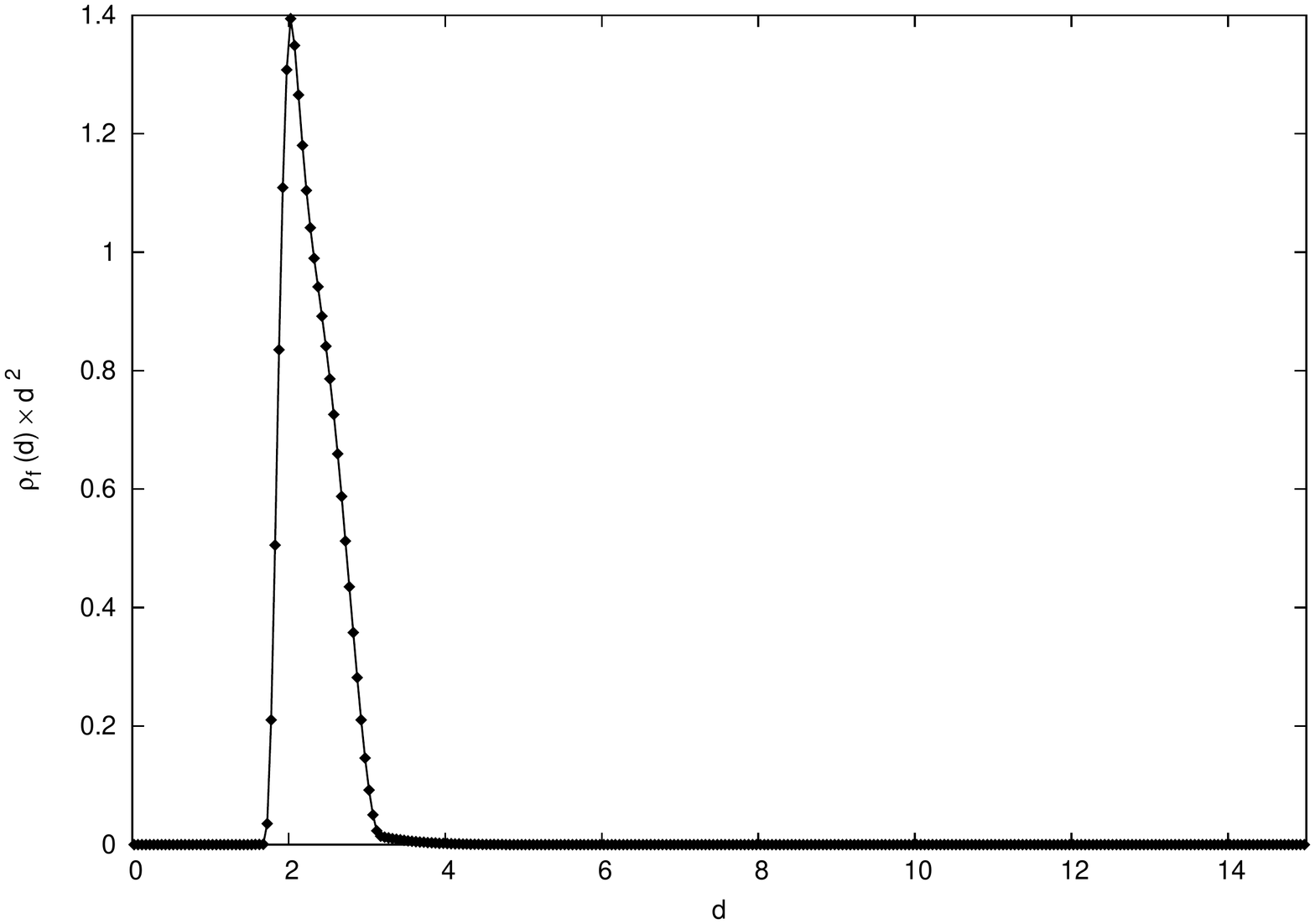}}
\subfigure{\includegraphics[scale=0.2, angle=0]{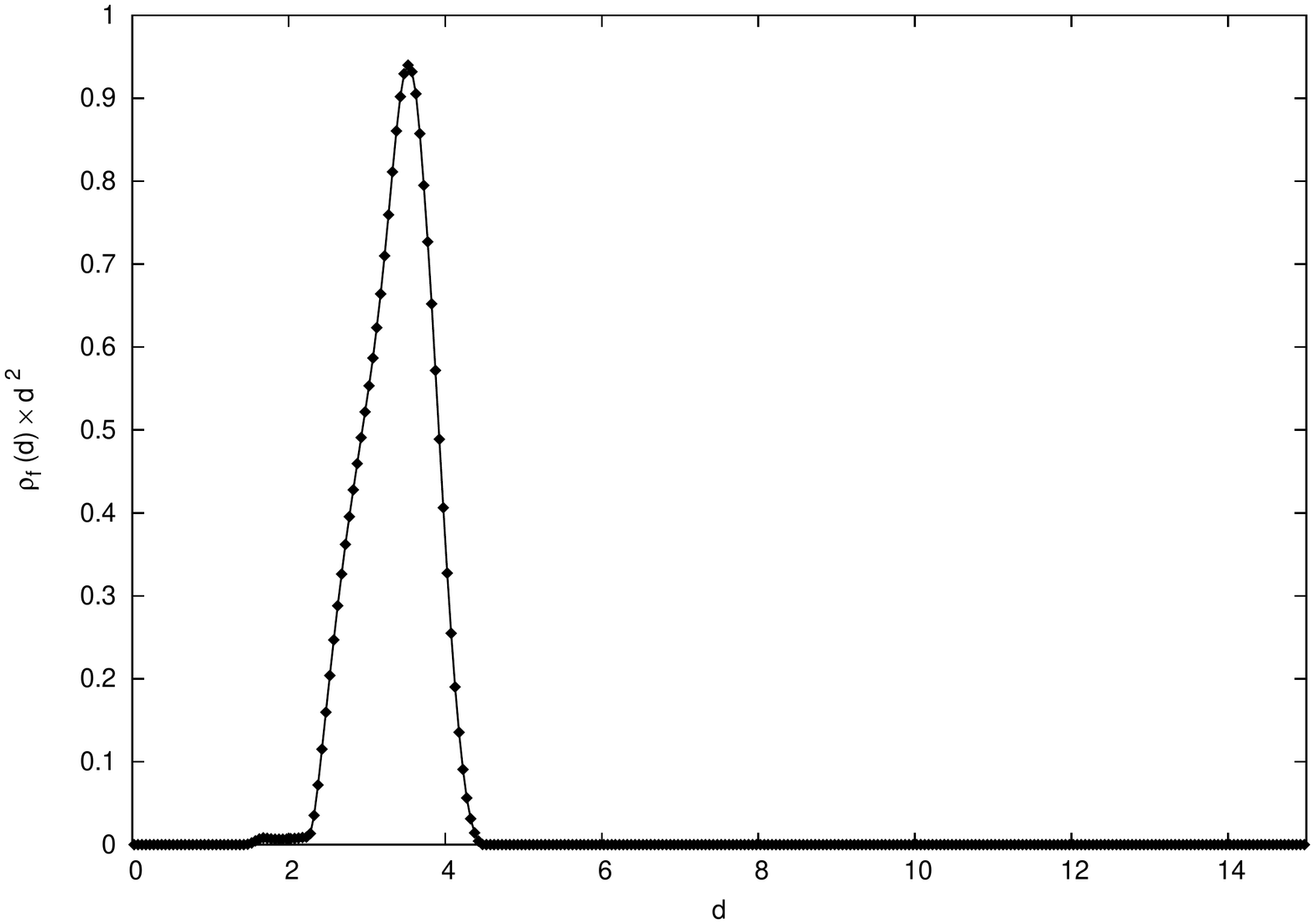}}
\subfigure{\includegraphics[scale=0.2, angle=0]{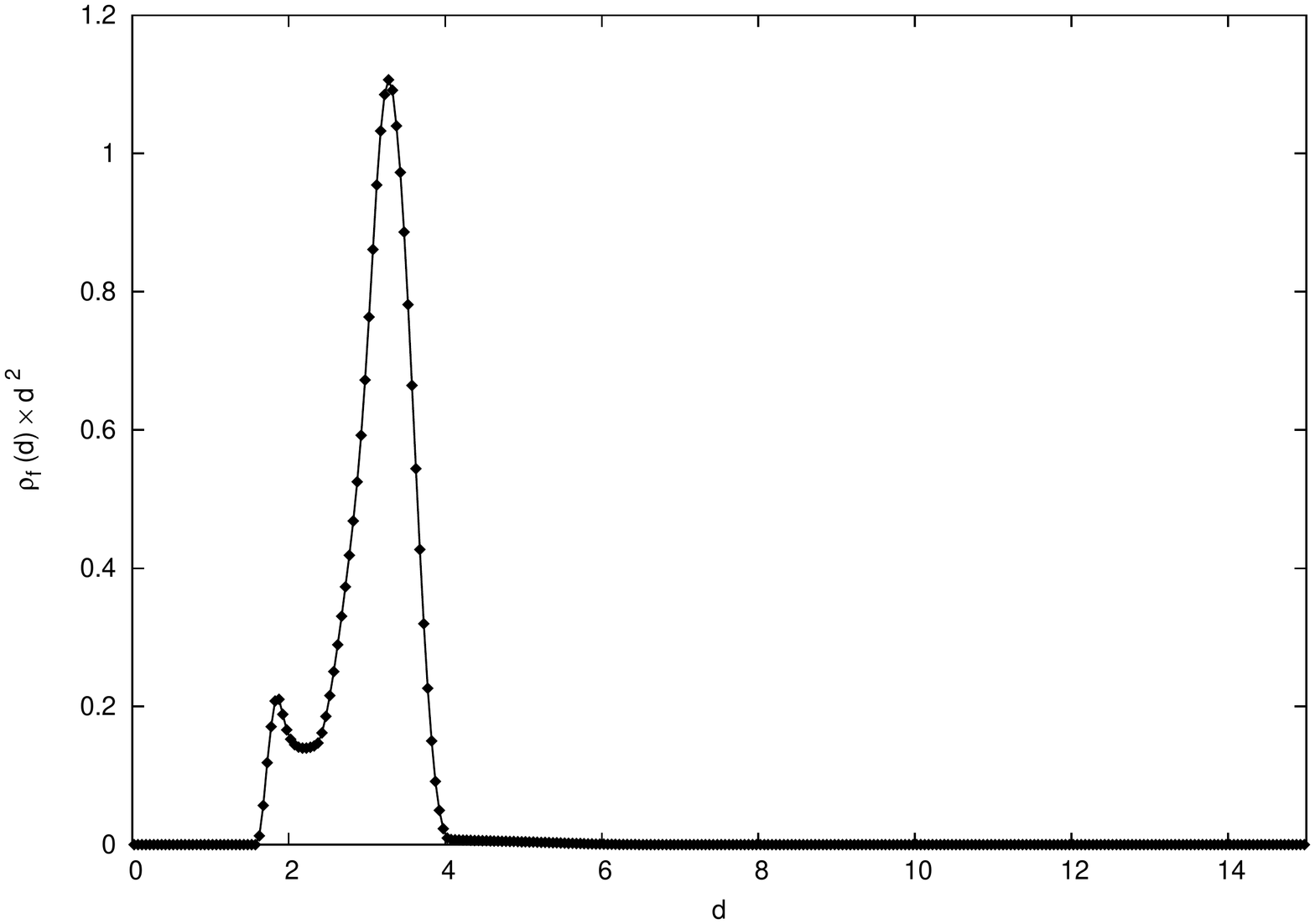}}
\subfigure{\includegraphics[scale=0.2, angle=0]{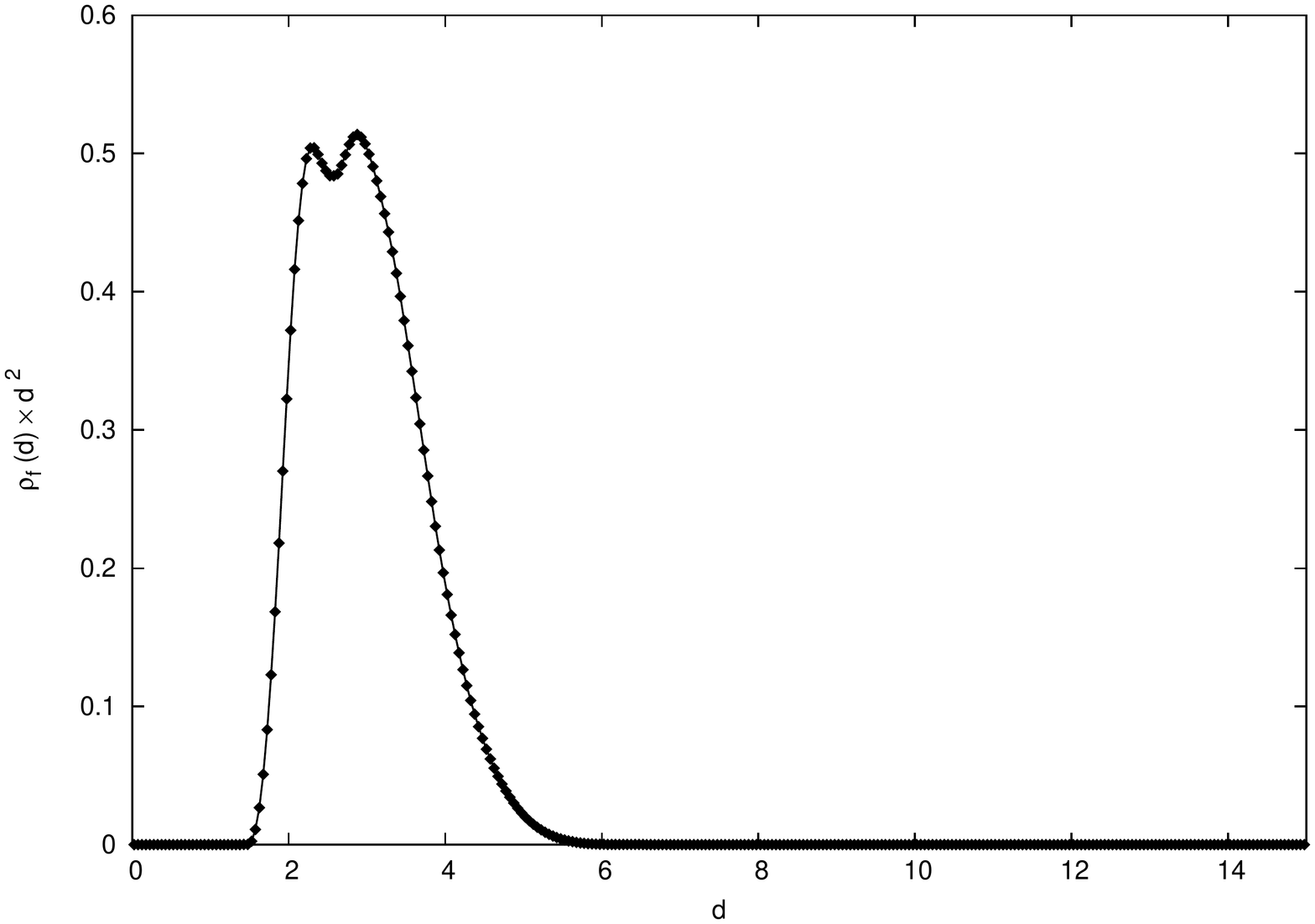}}
\caption{Time evolution of the integrated particle density $\rho_f(d)$
multiplied for the $d^2$ factor
  for the isothermal truncated model with angular momentum
  $\mathcal{L}=3.5$. 
  The different panels correspond to times
  $\mathcal{T}=0, 9.33, 18.84, 30.79, 45.71, 259.38$, same that those in
  Figure~\ref{pics:PS_trun}.}
\label{pics:D_trun}
\end{figure}

\begin{figure}[H]
\centering
\subfigure{\includegraphics[scale=0.2, angle=0]{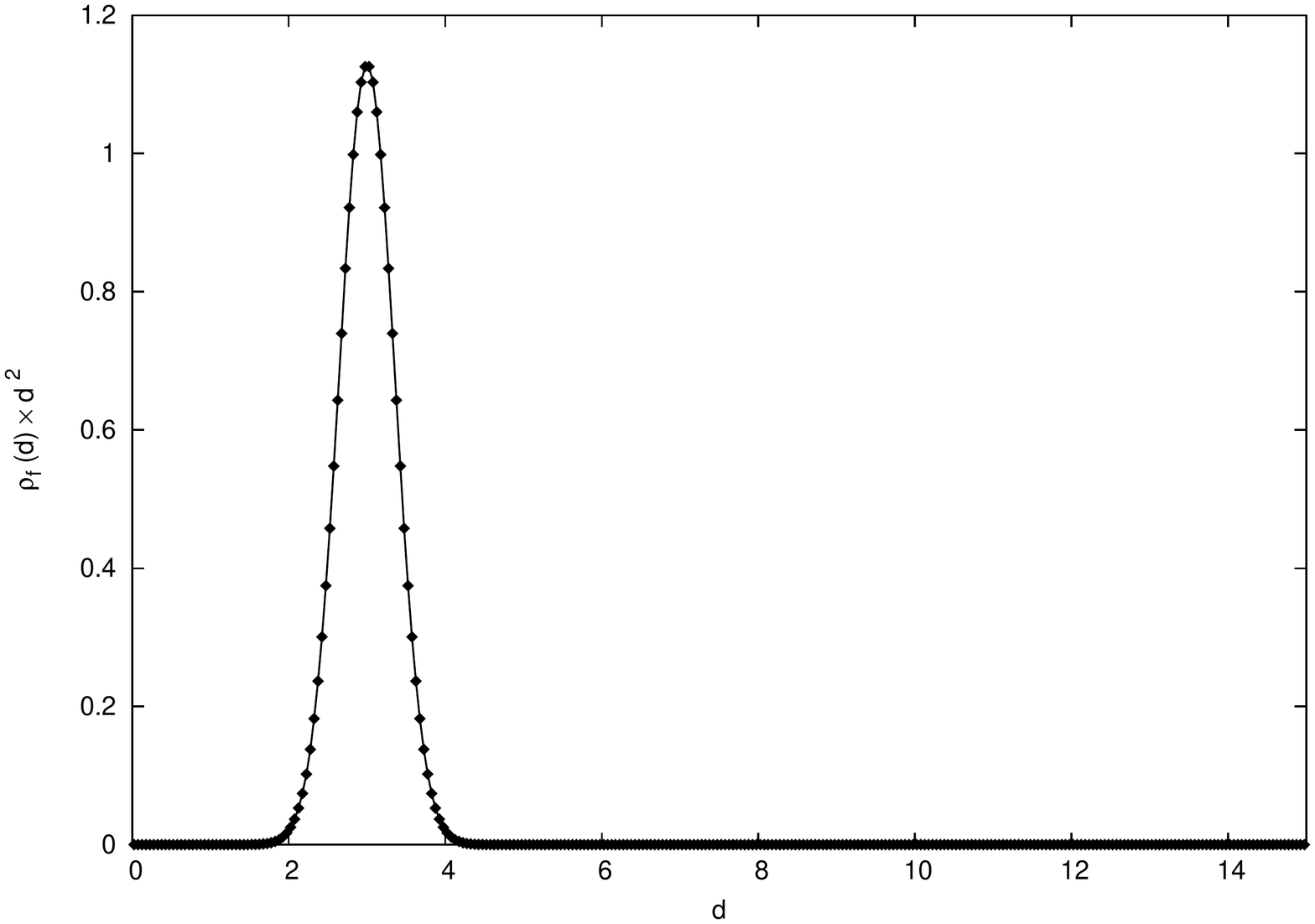}}
\subfigure{\includegraphics[scale=0.2, angle=0]{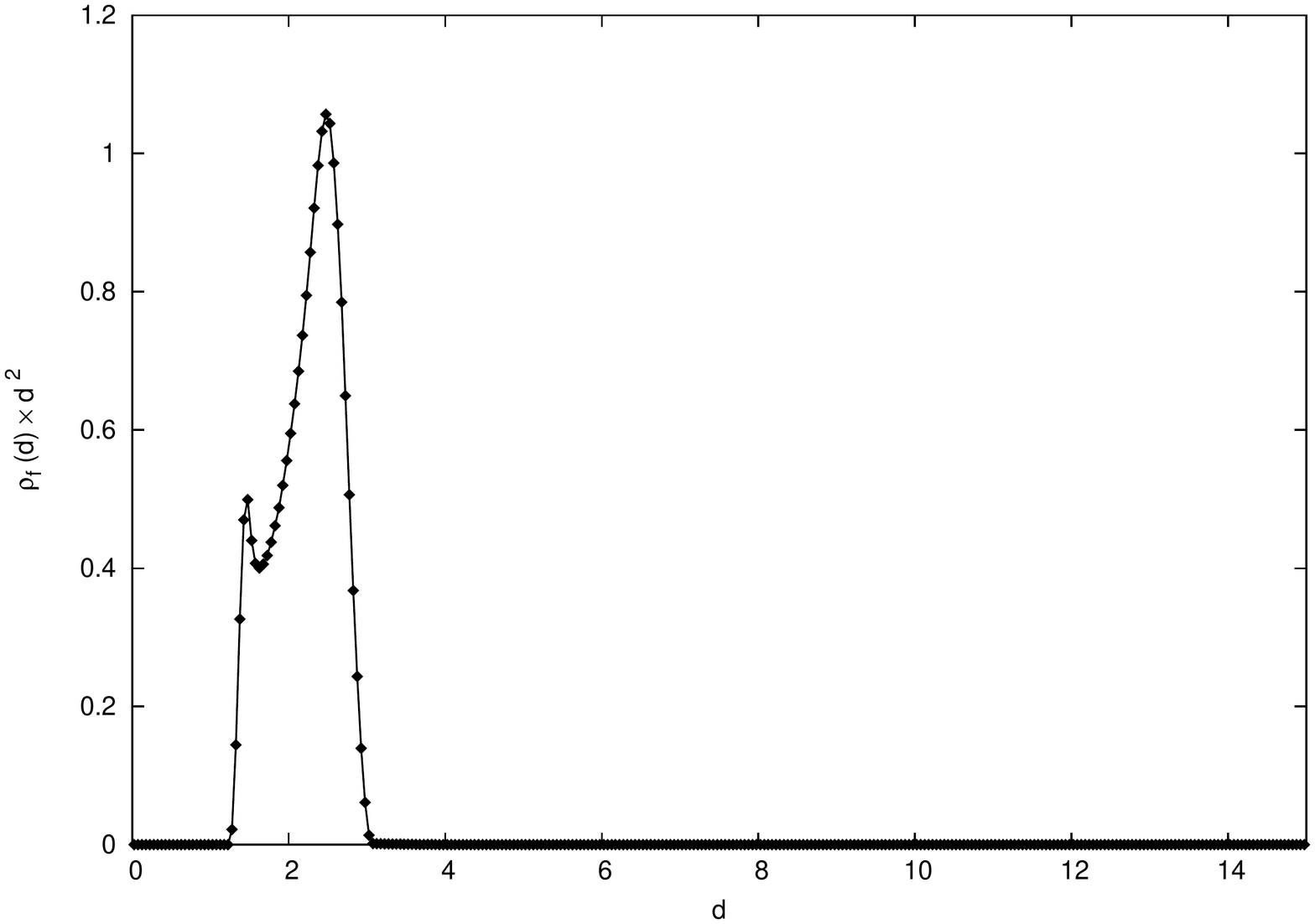}}
\subfigure{\includegraphics[scale=0.2, angle=0]{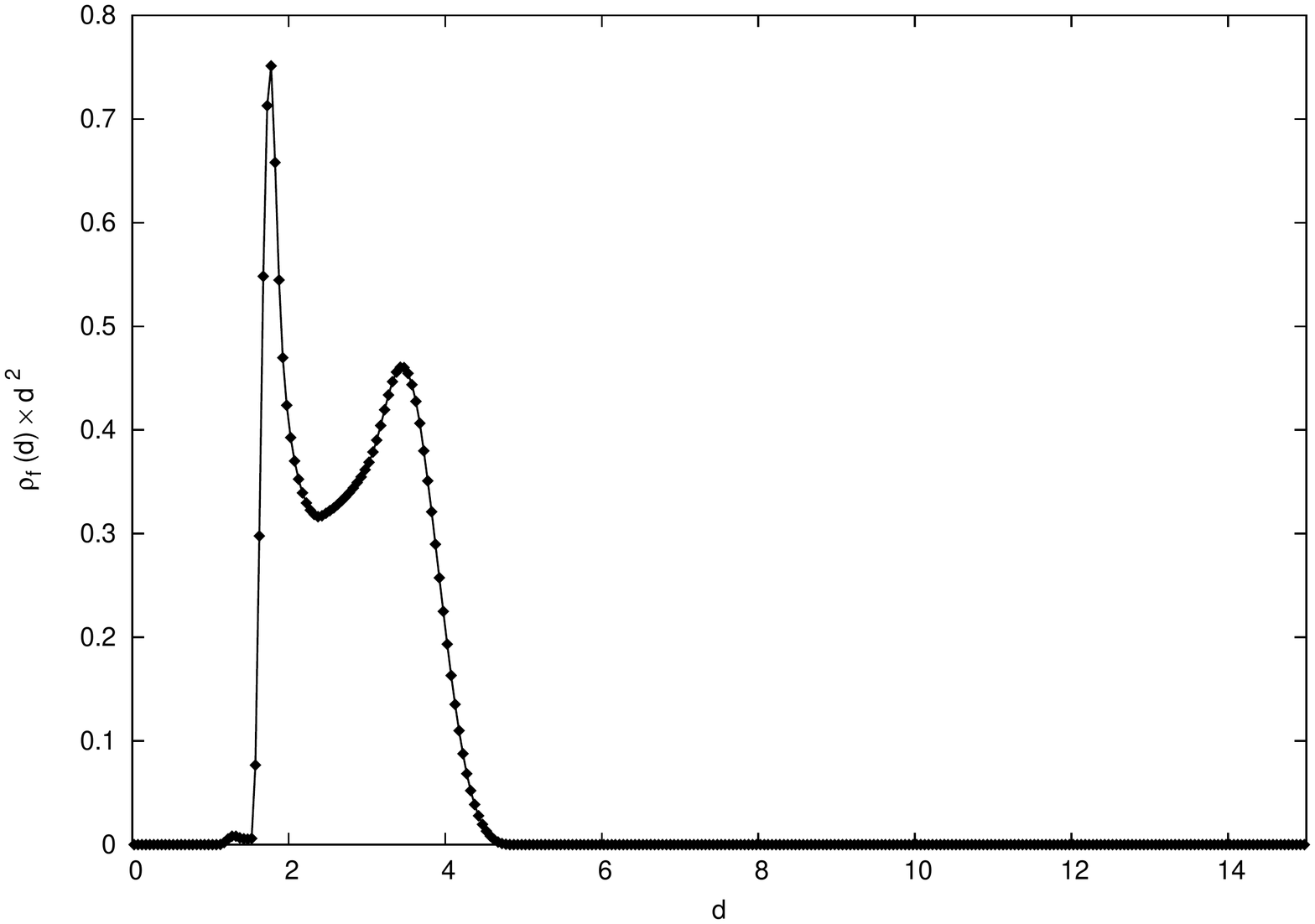}}
\subfigure{\includegraphics[scale=0.2, angle=0]{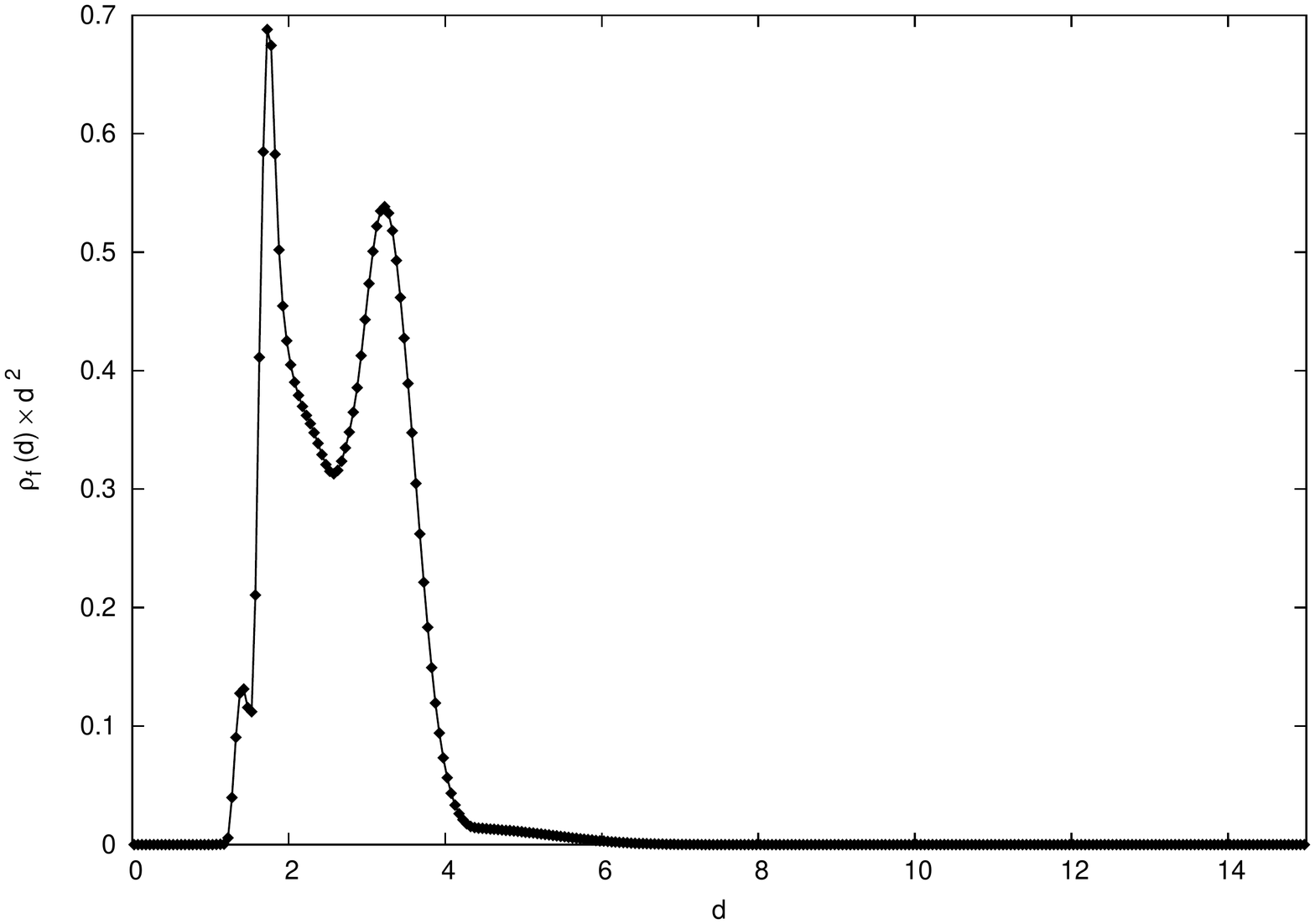}}
\subfigure{\includegraphics[scale=0.2, angle=0]{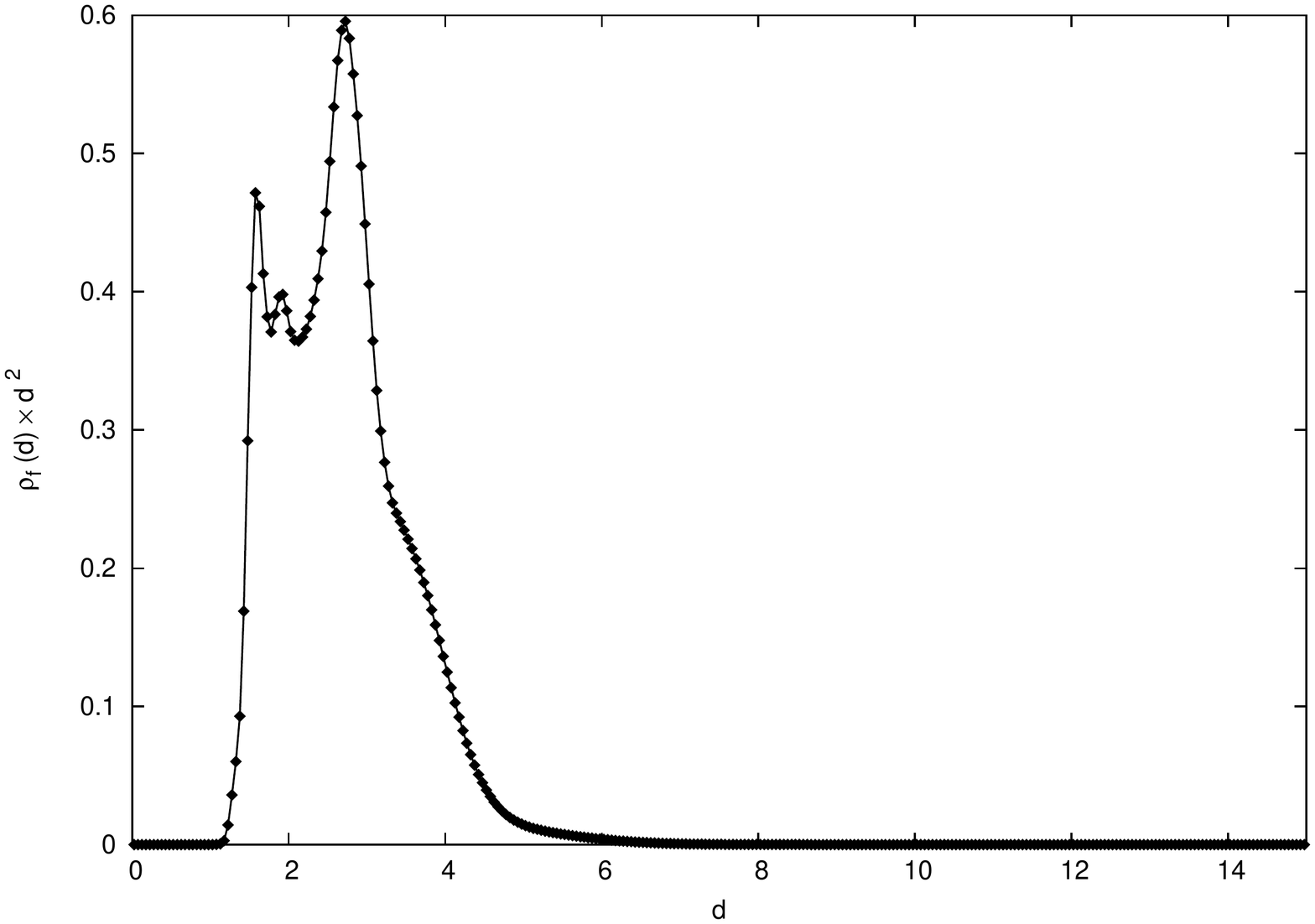}}
\subfigure{\includegraphics[scale=0.2, angle=0]{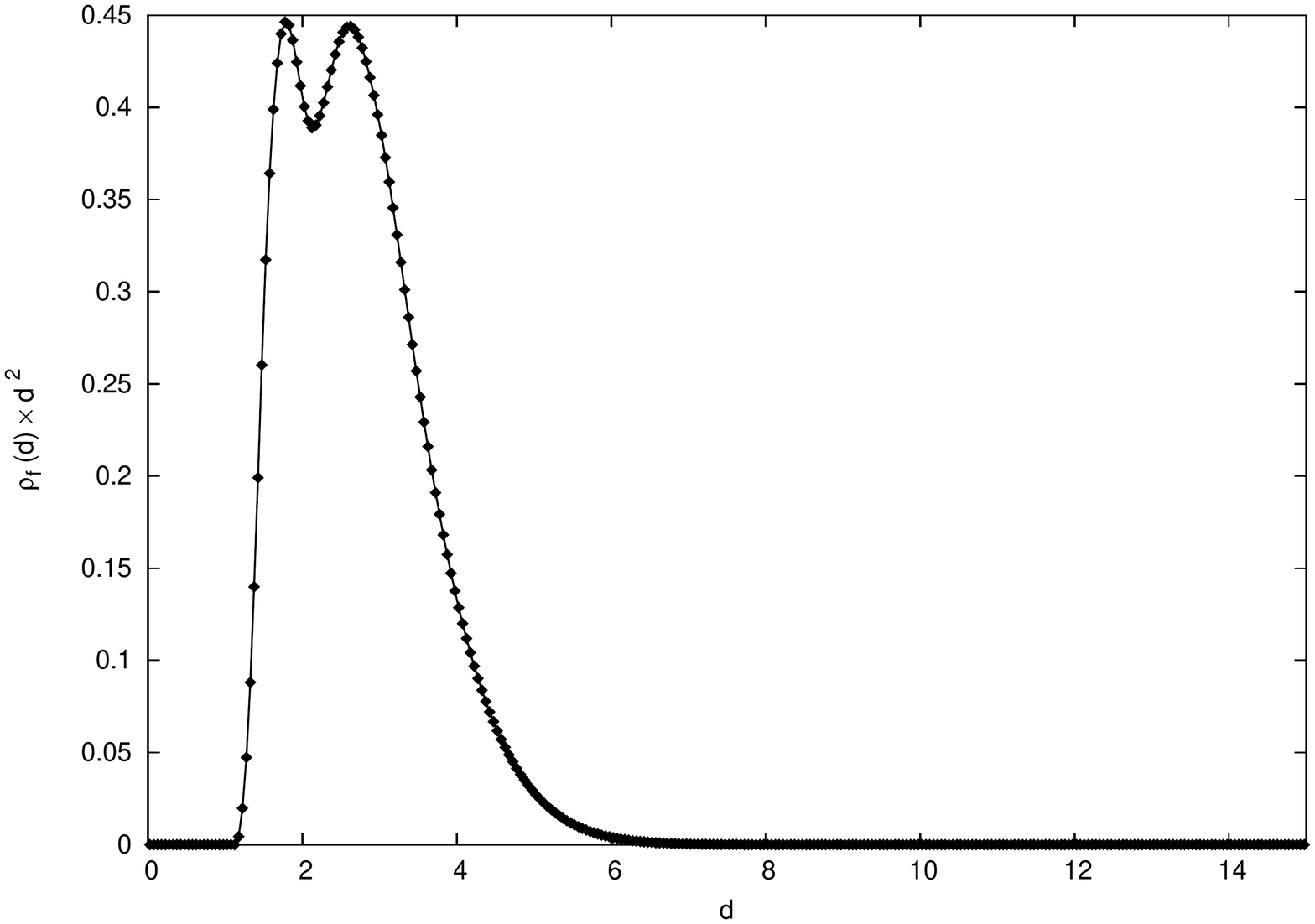}}
\caption{Time evolution of the integrated particle density $\rho_f(d)$
multiplied for the $d^2$ factor
  for the Burkert model with angular momentum $\mathcal{L}=3.5$. 
  The different panels correspond to times $\mathcal{T}=0, 9.33, 18.84,
  30.79, 45.71, 196.85$, same that 
 those in Figure~\ref{pics:PS_bur}.}
\label{pics:D_bur}
\end{figure}

\begin{figure}[H]
\centering
\subfigure{\includegraphics[scale=0.2, angle=0]{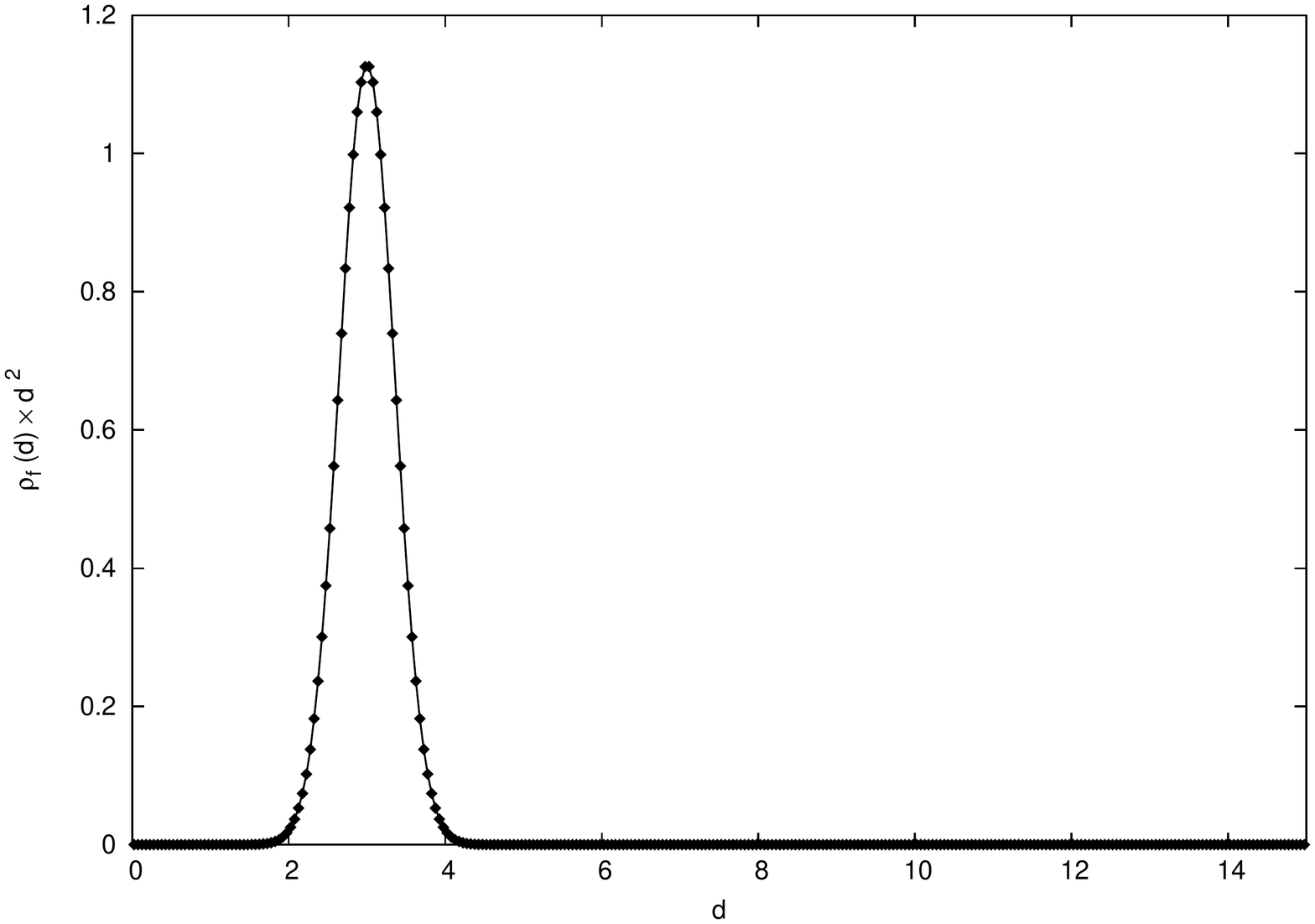}}
\subfigure{\includegraphics[scale=0.2, angle=0]{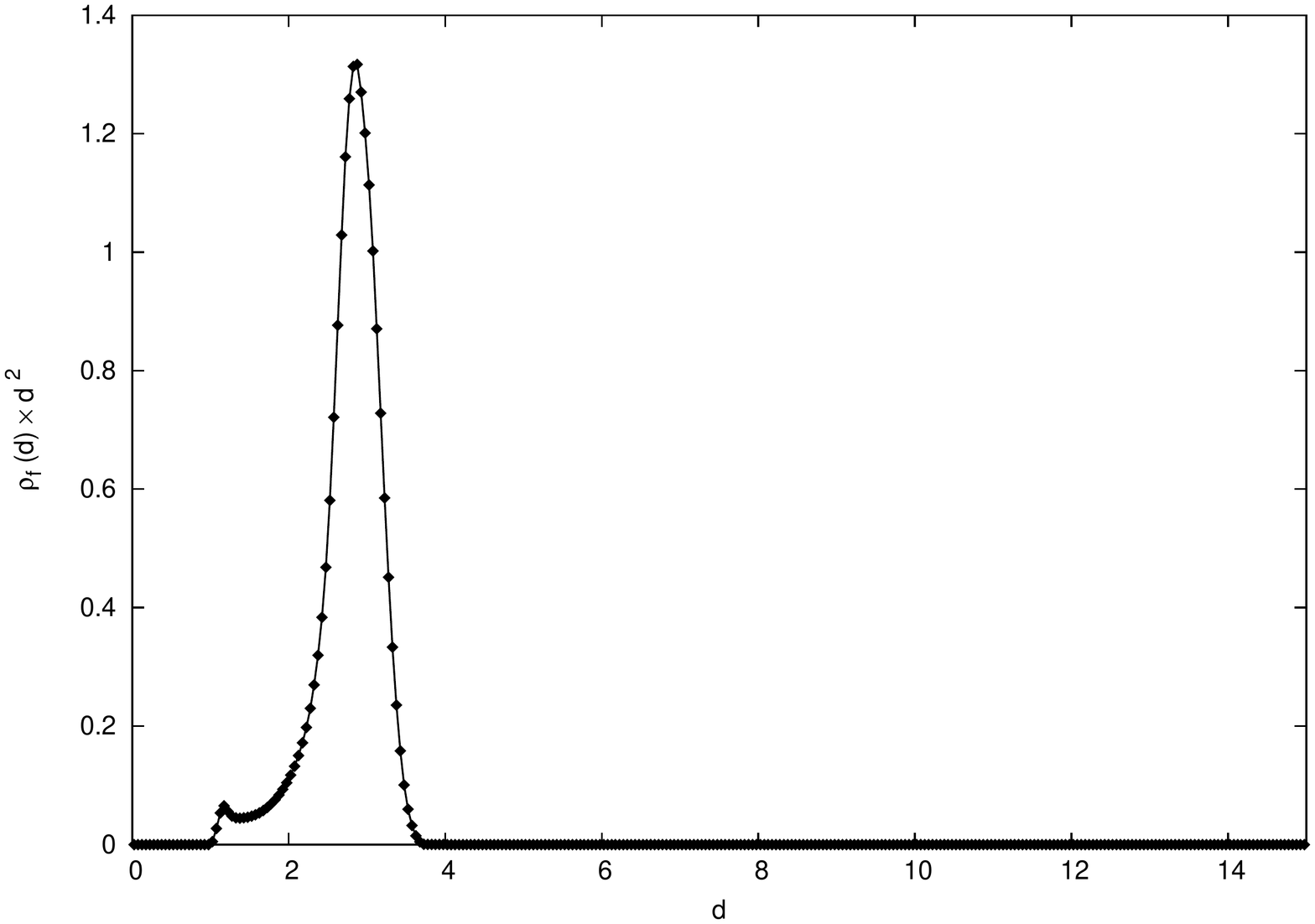}}
\subfigure{\includegraphics[scale=0.2, angle=0]{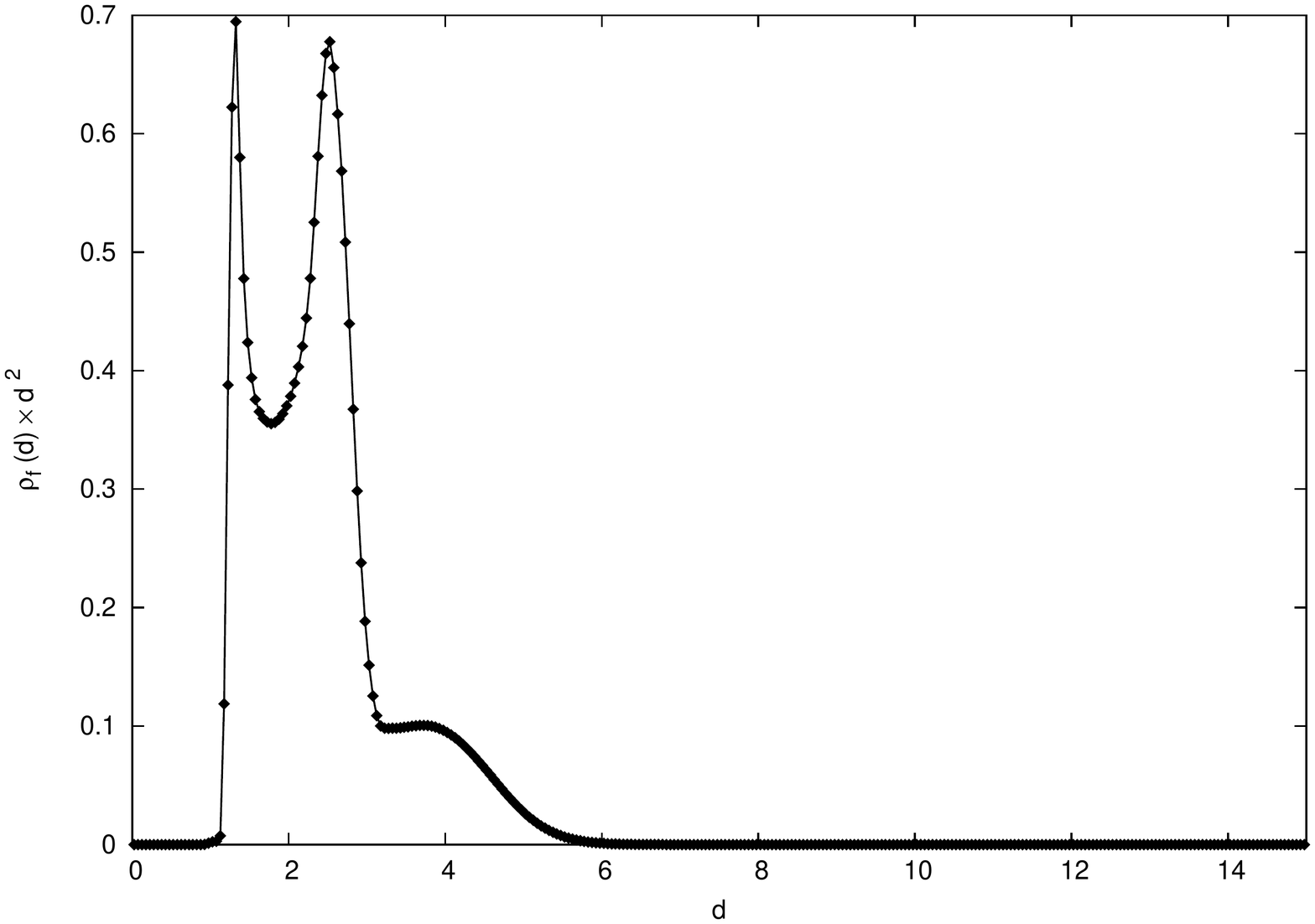}}
\subfigure{\includegraphics[scale=0.2, angle=0]{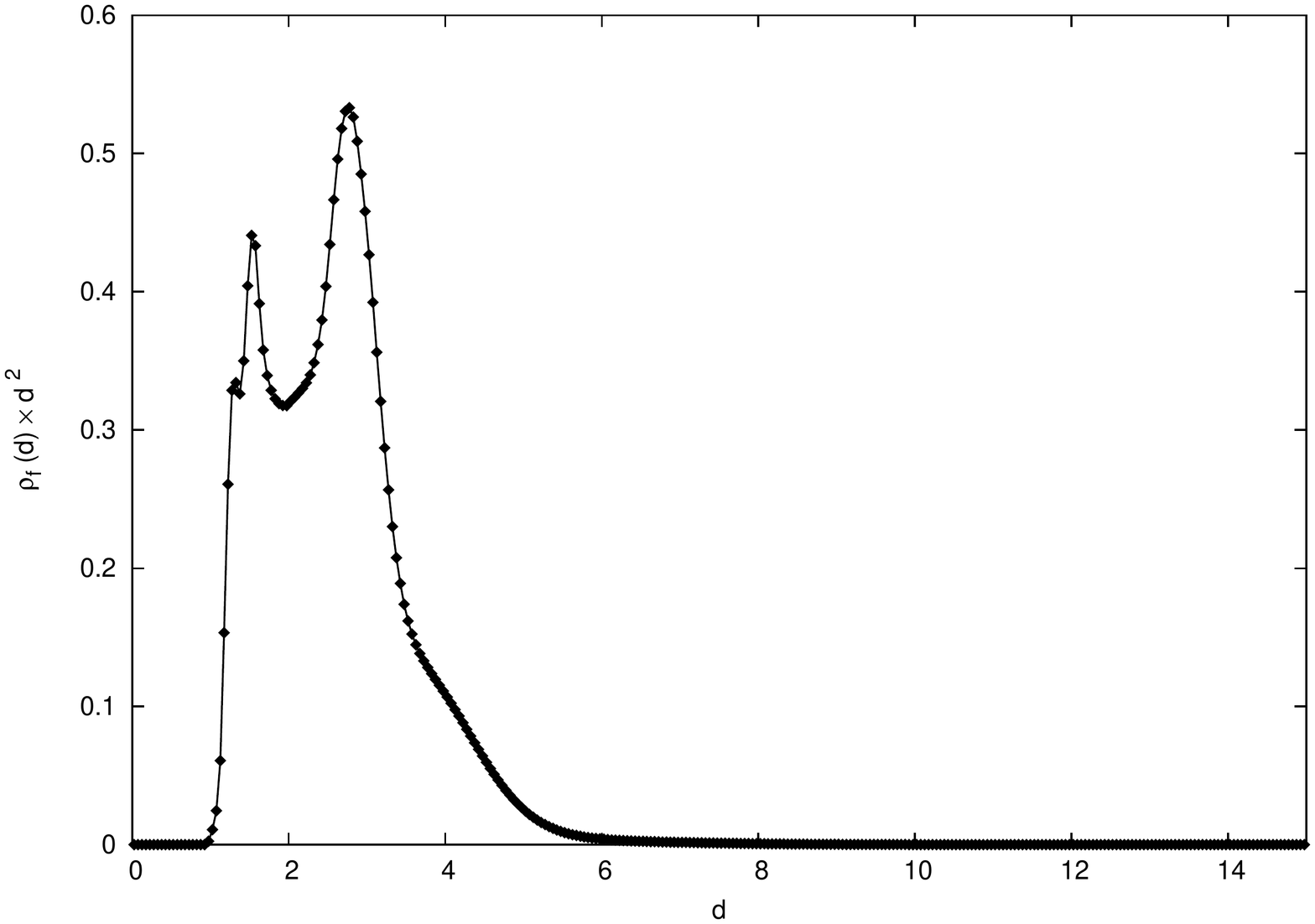}}
\subfigure{\includegraphics[scale=0.2, angle=0]{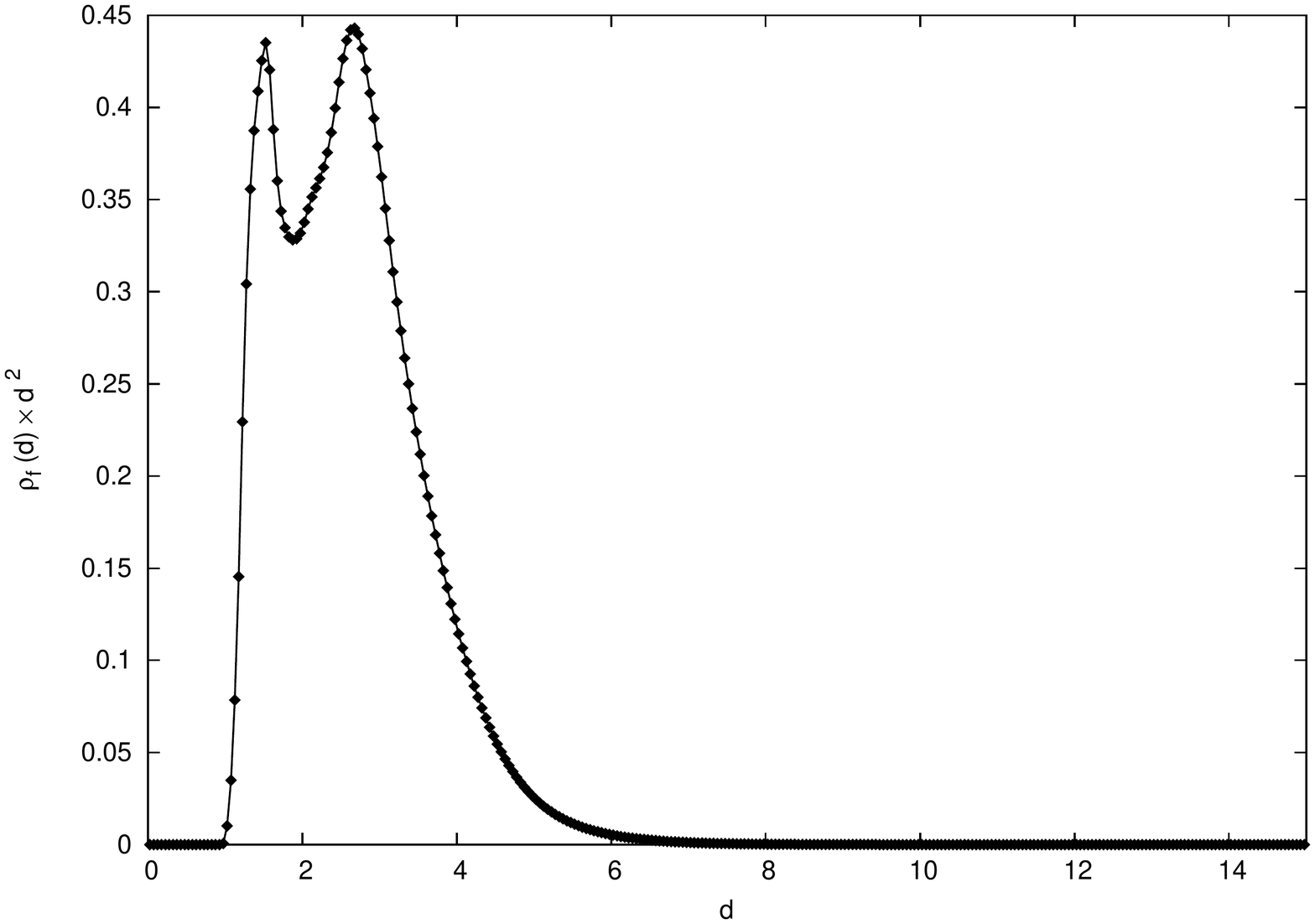}}
\subfigure{\includegraphics[scale=0.2, angle=0]{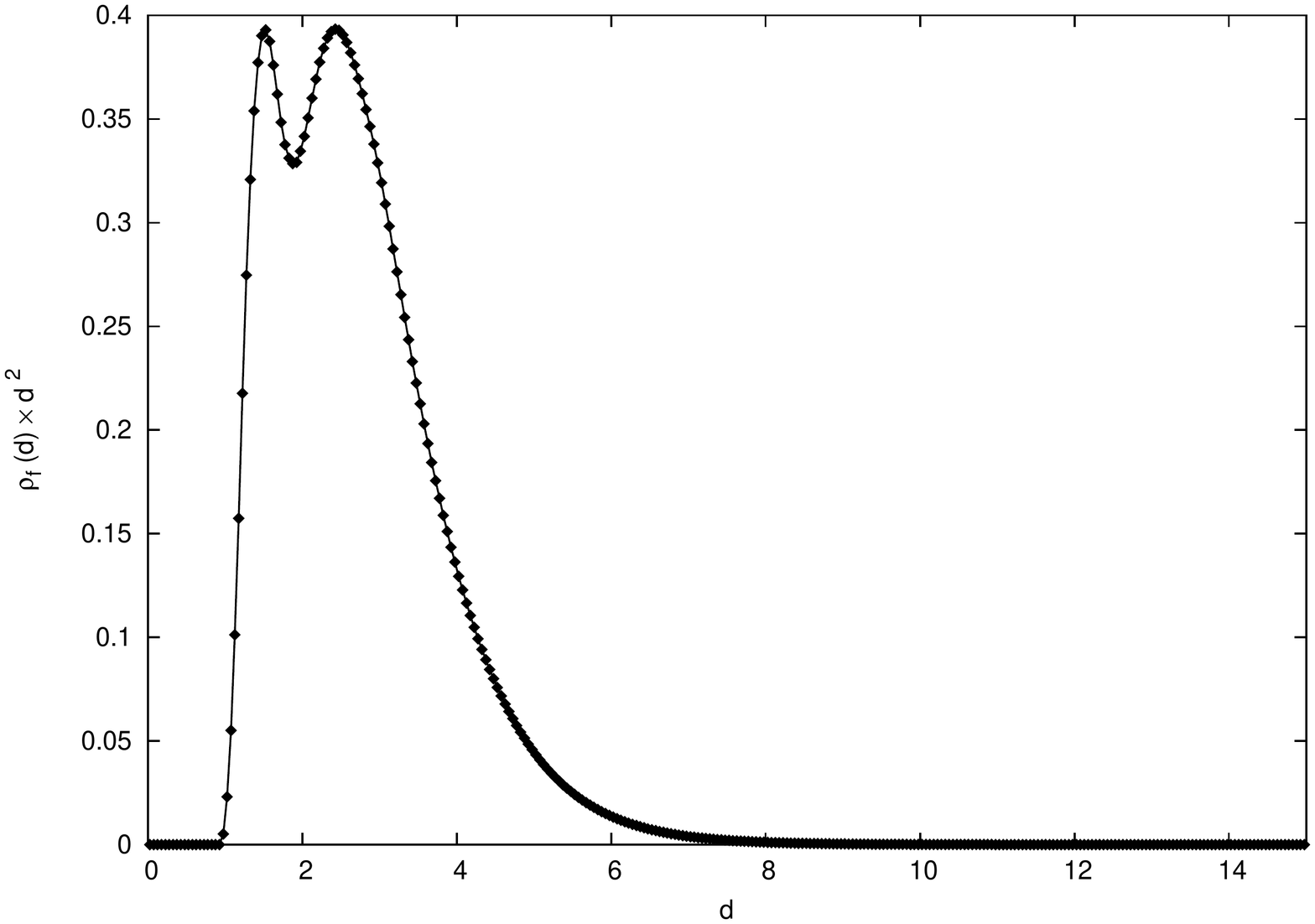}}
\caption{Time evolution of the integrated particle density $\rho_f(d)$
multiplied for the $d^2$ factor
  for the NFW model with angular momentum $\mathcal{L}=3.5$. 
  The different
  panels correspond to times $\mathcal{T}=0, 9.33, 18.84, 30.79, 45.71, 189.20$,
  same that those in Figure~\ref{pics:PS_NFW}.}
\label{pics:D_NFW}
\end{figure}

Figures~\ref{pics:PS_iso}-\ref{pics:PS_NFW} show the evolution of the
distribution function for our four different halo models. In each
figure we show six snapshots at different times during the evolution.
The first five snapshots correspond to the same times in all cases,
namely $\mathcal{T}=0,9.33, 18.84, 30.79,45.71$.  The time
corresponding to the last snapshot differs for each halo model because
in each case we show an epoch when the system has finally reached a
stationary state.  Figure~\ref{Fig:FD_all_ff} shows again the final
stationary distribution function in phase space for the four different
models.  We can see that all four halo models the distribution
function reaches a stationary state that corresponds to orbits in
phase space around a central point.  This is not surprising, since
individual particles with a given angular momentum are expected to
have orbits around the center of gravity with some minimum and maximum
radii (the individual orbits will not be precisely elliptical since
the potentials for the different halos are not simply $1/d$).  The
stationary states look similar for each halo model, even if they
are not reached at the same time; for the isothermal, Burkert and
NFW models the stationary state is reached at time $\mathcal{T} \sim 120$ 
and for the truncated isothermal this state is reached at time $\mathcal{T} \sim 260$.
Next, in Figures~\ref{pics:D_iso}-\ref{pics:D_NFW} we show the time
evolution of the integrated particle density $\rho_f(d)$ 
for the four
halo models and the same times as before.  Again we see how even
though the initial stages of the evolution are different for each halo
model, the final stationary particle density is quite similar, with a
characteristic two-hump shape corresponding to particles that describe
an orbit in phase space and accumulate mostly at the extrema of those
orbits as seen in physical space.

Even though we have only shown the case with angular momentum
$\mathcal{L}=3.5$, simulations with different values of the angular
momentum behave in a very similar way.

\subsection{One particle motion}

In order to understand the results from the previous section, it is
perhaps instructive to consider the motion of a single particle in a
gravitational potential determined by the different halo models
considered.

For a particle starting at an initial radius $d_0$, with initial
radial momentum ${{\cal P}_d}_0$, and with a given angular momentum $L$,
conservation of energy implies that
\begin{equation}
  \frac{{{\cal P}_d}^2}{2 m} + \frac{{\cal L}^2}{2 m d^2} + \Phi(d) =
  \frac{{{{\cal P}_d}_0}^2}{2 m} + \frac{{\cal L}^2}{2 m {d_0}^2} + \Phi(d_0) \; .
\end{equation}
From this one can determine the turning points of the orbit, or the
radius of a circular orbit for a given value of the angular
momentum.

\begin{table}[]
\centering
\begin{tabular}{|c|c|c|c|c|}
\hline
\multirow{2}{*}{$\cal L$} & \multirow{2}{*}{$V$} & \multicolumn{2}{c|}{Turning points} & $\rho_{max}$ \\ \cline{3-5} 
                   &                            & $d_{min}$        & $d_{max}$        & $d$            \\ \hline
2                  & Isothermal                 & 1.03             & 3.18             & 0.8          \\ \hline
2                  & Iso. Trun.                 & 1.59             & 3.33             & 1.25         \\ \hline
2                  & Burkert                    & 1.16             & 3.19             & 0.92        \\ \hline
2                  & NFW                        & 0.95             & 3.15             & 0.75         \\ \hline
2.5                & Isothermal                 & 1.44             & 3.23             & 1.02         \\ \hline
2.5                & Iso. Trun.                 & 2.06             & 3.49             & 1.67         \\ \hline
2.5                & Burkert                    & 1.56             & 3.25             & 1.17         \\ \hline
2.5                & NFW                        & 1.31             & 3.19             & 0.97        \\ \hline
3                  & Isothermal                 & 1.88             & 3.33             & 1.32         \\ \hline
3                  & Iso. Trun.                 & 2.46             & 3.85             & 2.02         \\ \hline
3                  & Burkert                    & 2.00             & 3.37             & 1.47         \\ \hline
3                  & NFW                        & 1.72             & 3.26             & 1.23         \\ \hline
3.5                & Isothermal                 & 2.35             & 3.58             & 1.65         \\ \hline
3.5                & Iso. Trun.                 & 2.69             & 4.54             & 2.32         \\ \hline
3.5                & Burkert                    & 2.41             & 3.67             & 1.77        \\ \hline
3.5                & NFW                        & 2.17             & 3.41             & 1.52         \\ \hline
\end{tabular}
\caption{Positions of the turning points for each configuration of angular moment and halo potential and position of the maximum 
density value at the final virialized distribution function.}
\label{tbl:turning points}
\end{table}

It is interesting to note 
that the turning points for a single
particle with initial position and momentum corresponding to the
maximum of the initial phase space distribution does not coincide with
the position of the maxima of the integrated energy density in the
equilibrium state (Table \ref{tbl:turning points}).  The reason for this, may be the so-called
statistical pressure: even though the particles are non interacting,
it seems that the collective motion is different from the individual
motion of particles initially at the center of the distribution.


\subsection{Virialization}

As mentioned above, in all our simulations the evolution of the
distribution function eventually reaches a stationary state.  Such a
state should satisfy the virial theorem, equation~\eqref{eq:virial2}.
In order to verify this we compute the average of the radial kinetic
energy $\left <{{\cal P}_d}^2/2m \right>$, and the average of the virial
$\left< F_{\rm eff}(d)\,d \right>$, and compare them during the
evolution.

\begin{figure}[H]
\centering
\includegraphics[scale=0.65]{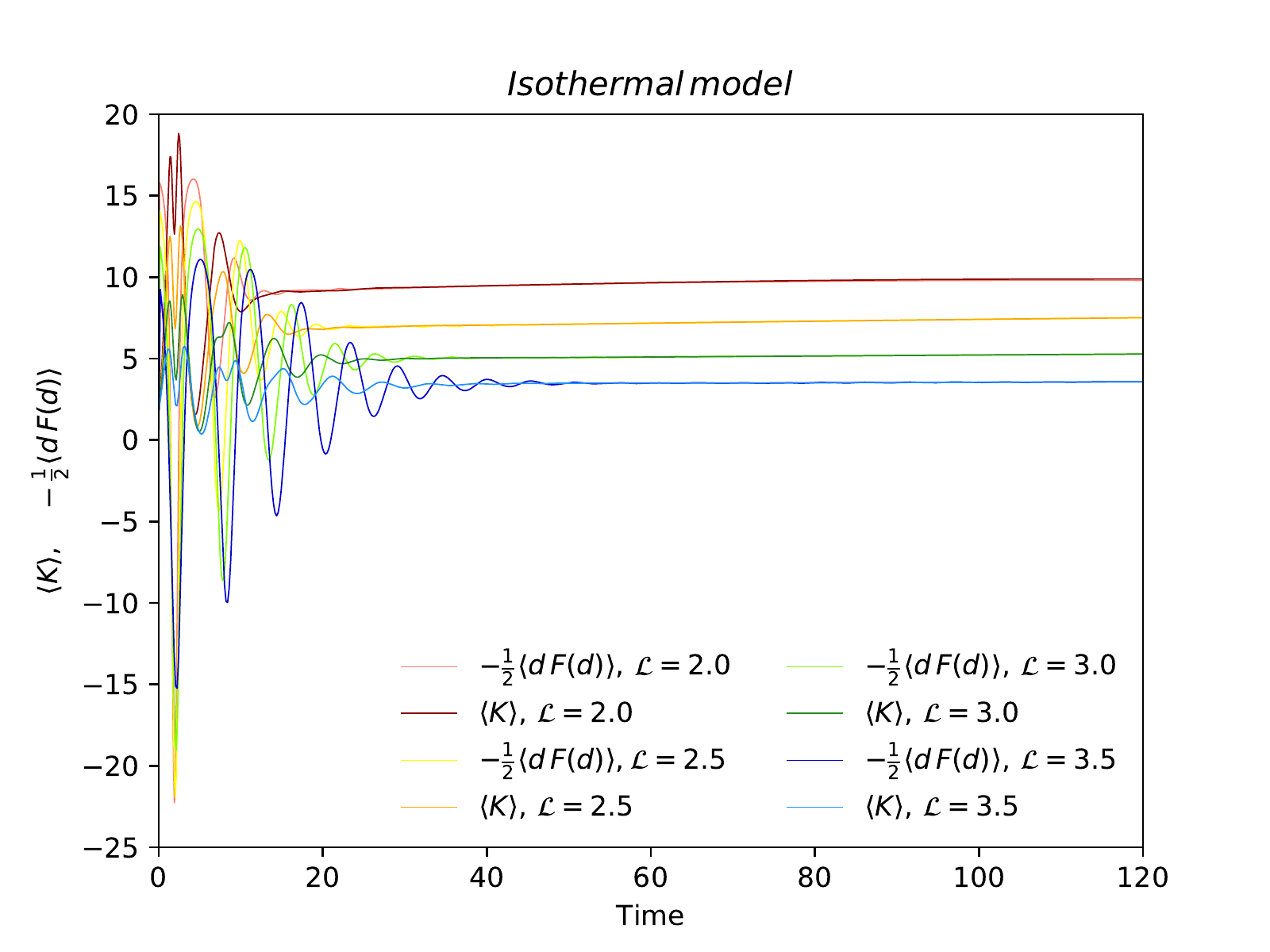}
\caption{
Virialization for the case of the isothermal model for
  different values of the angular momentum \mbox{$\mathcal{L}=2.0,2.5,3.0,3.5$}.}
\label{Fig:virialIso}
\end{figure}

\begin{figure}[H]
\centering
\includegraphics[scale=0.65]{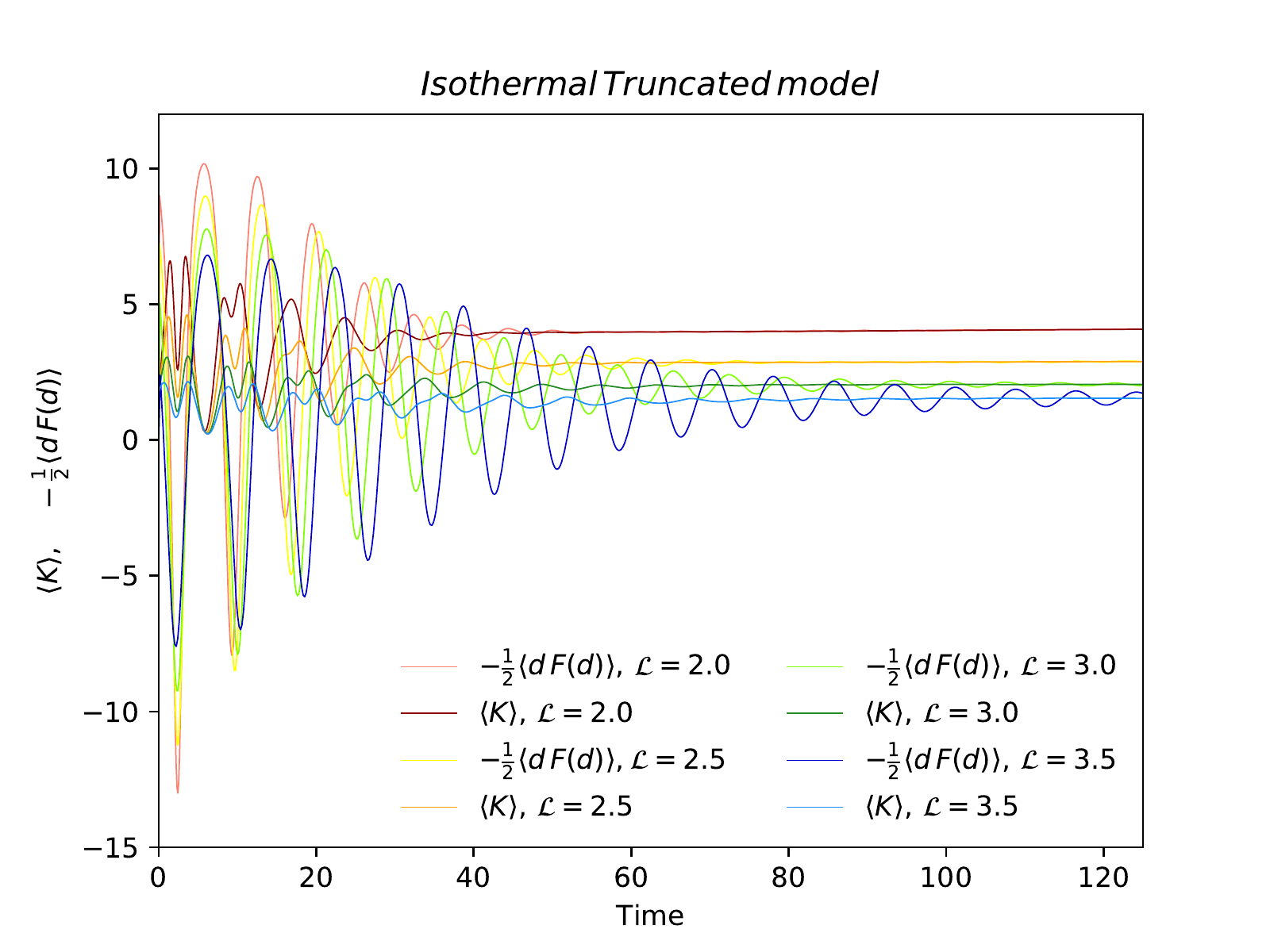}
\caption{
Virialization for the case of the truncated isothermal model
  for different values of the angular momentum \mbox{$\mathcal{L}=2.0,2.5,3.0,3.5$}.}
\label{Fig:virialTrun}
\end{figure}

\begin{figure}[H]
\centering
\includegraphics[scale=0.65]{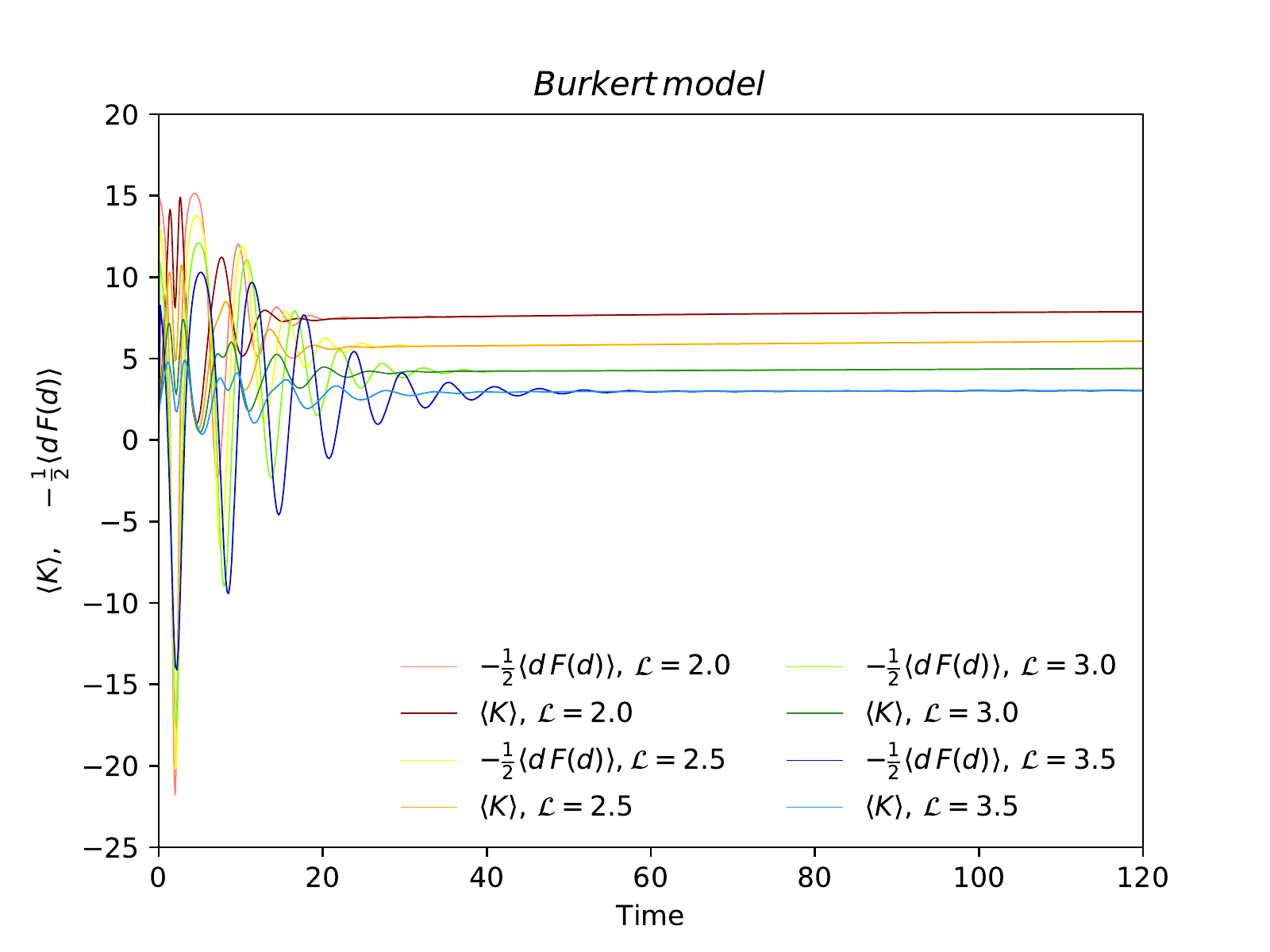}
\caption{
Virialization for the case of the Burkert model for different
  values of the angular momentum \mbox{$\mathcal{L}=2.0,2.5,3.0,3.5$}.}
\label{Fig:virialBurk}
\end{figure}

\begin{figure}[H]
\centering
\includegraphics[scale=0.65]{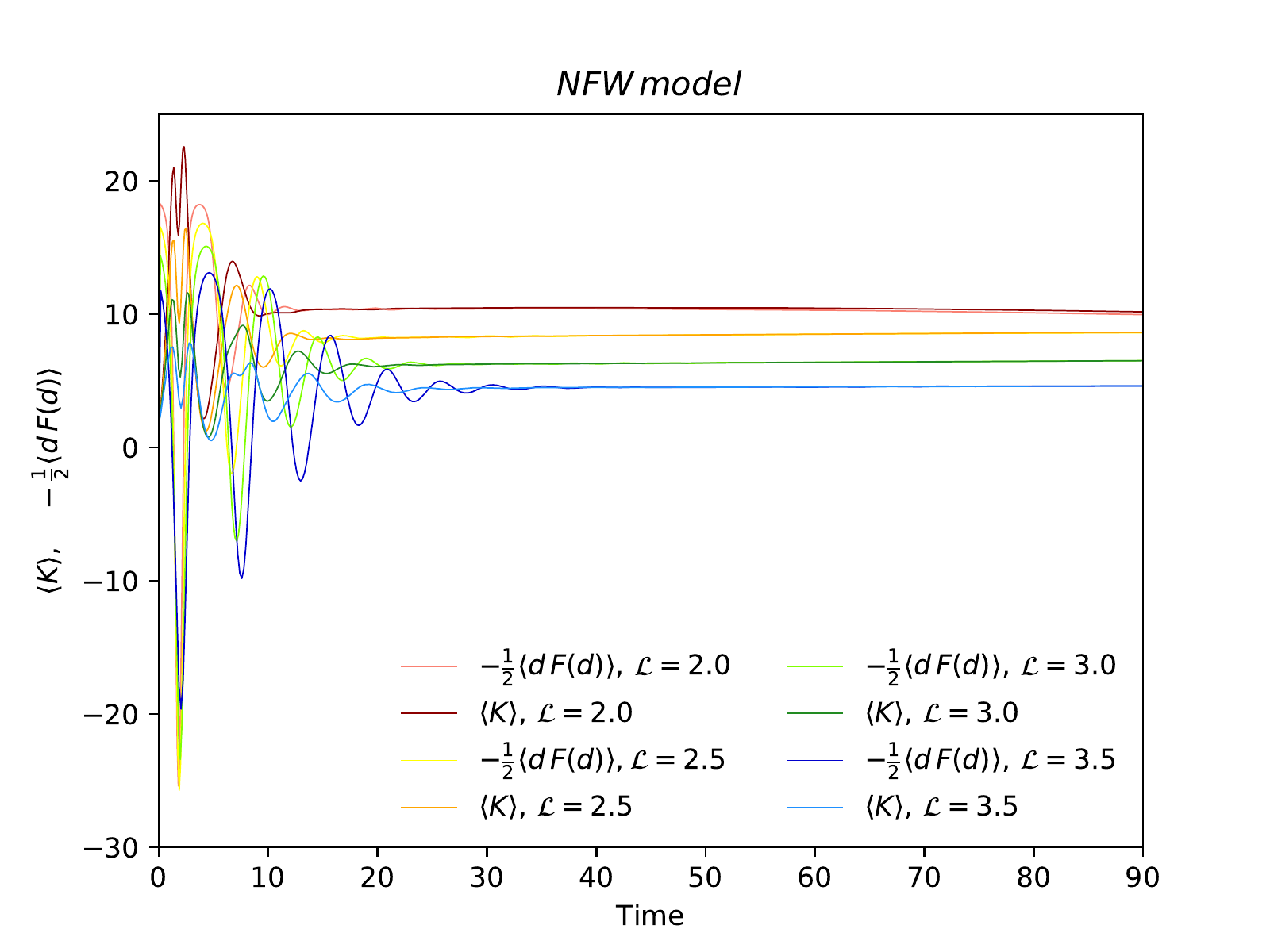}
\caption{
Virialization for the case of the NFW model for different
  values of the angular momentum \mbox{$\mathcal{L}=2.0,2.5,3.0,3.5$}.}
\label{Fig:virialNFW}
\end{figure}

We find that in each case, as the initial distribution evolves toward
a stationary state, the average of the radial kinetic energy and the
average of the virial both reach constant values that satisfy the
virial theorem~\eqref{eq:virial2}.
Figures~\ref{Fig:virialIso}-\ref{Fig:virialNFW} show plots of $\left<
{{\cal P}_d}^2/2m \right>$ and $\left< - F_{\rm eff}(d)\,d/2 \right>$ for each
of the different halo models, and for four values of the angular
momentum in each case.  We find that in each case, after strong
initial oscillations, the average kinetic energy and virial reach a
constant value satisfying the virial theorem.  We also find that this
virialization process takes longer for larger values of the angular
momentum, while that the final constant value of the average kinetic
energy becomes smaller.  Notice also that while $\left< {{\cal P}_r}^2/2m
\right>$ is always positive, $\left< - F_{\rm eff}(d)\,d/2 \right>$ can
become negative during the initial portions of the evolution.

\begin{figure}[H]
\centering
\includegraphics[scale=0.65]{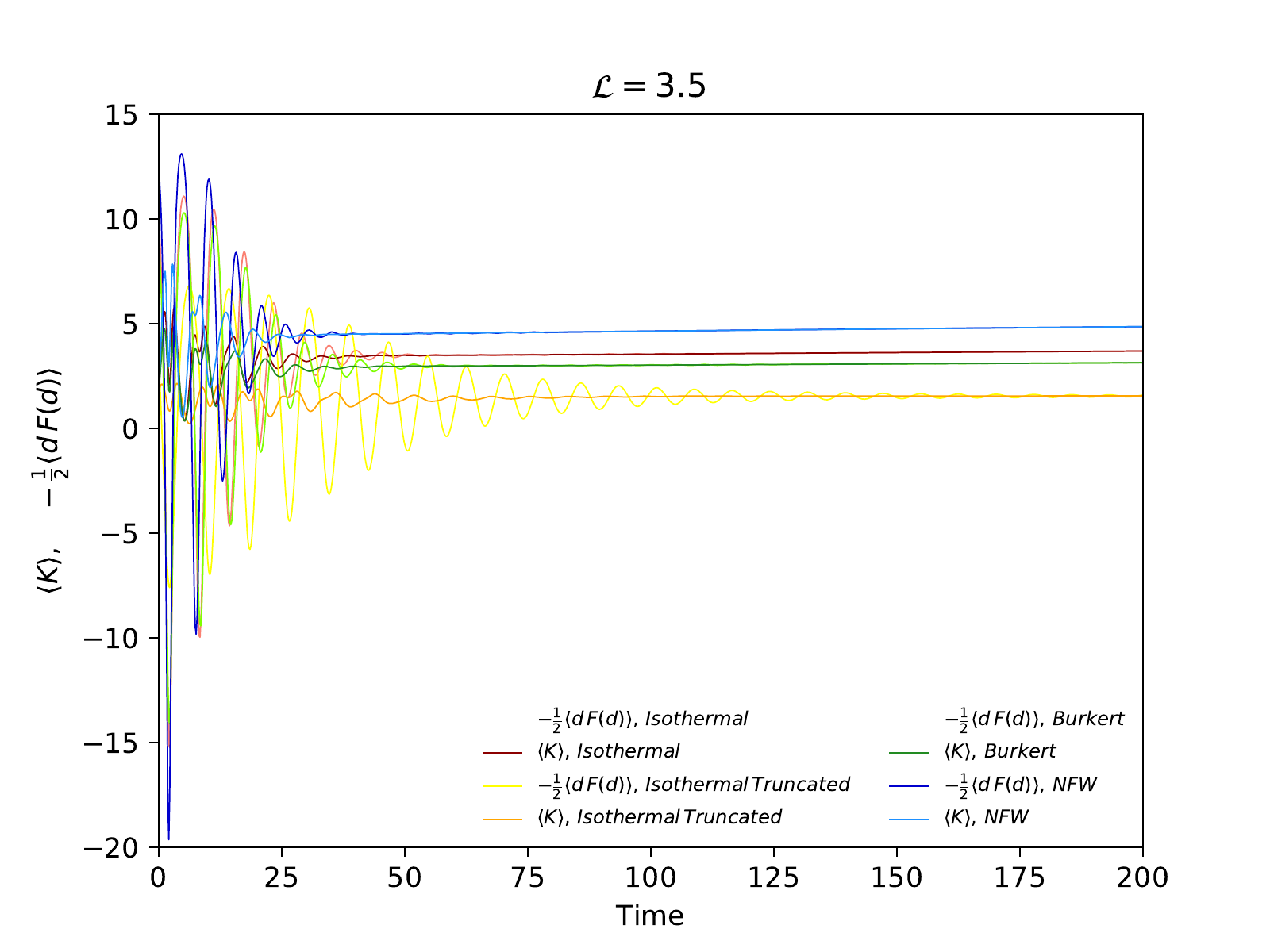}
\caption{Comparison of the virialization process for the different
  halo models in the case with $\mathcal{L}=3.5$.}
\label{Fig:virialcompare}
\end{figure}

Figure~\ref{Fig:virialcompare} shows a comparison of the virialization
process for the four different halo models in the specific case with
angular momentum ${\cal L}=3.5$.  We can see that the NFW model
virializes very rapidly, while the truncated isothermal model takes
much longer than all other models to virialize.  Again we see that
models that take longer to virialize reach smaller values of the final
average kinetic energy.

\subsection{Conservation of particles}

As mentioned above, since we are using a conservative numerical method,
the total number of particles defined in~\eqref{eq:totalN} should be
conserved up to machine round off error in our simulations.  Changes
in the total number of particles should reflect only the particles
that leave the computational domain through the external boundaries.
We find, however, that numerical diffusion can cause particles to
reach the boundaries that would not do so physically.

\begin{figure}[H]
\centering
\subfigure{\includegraphics[scale=0.4]{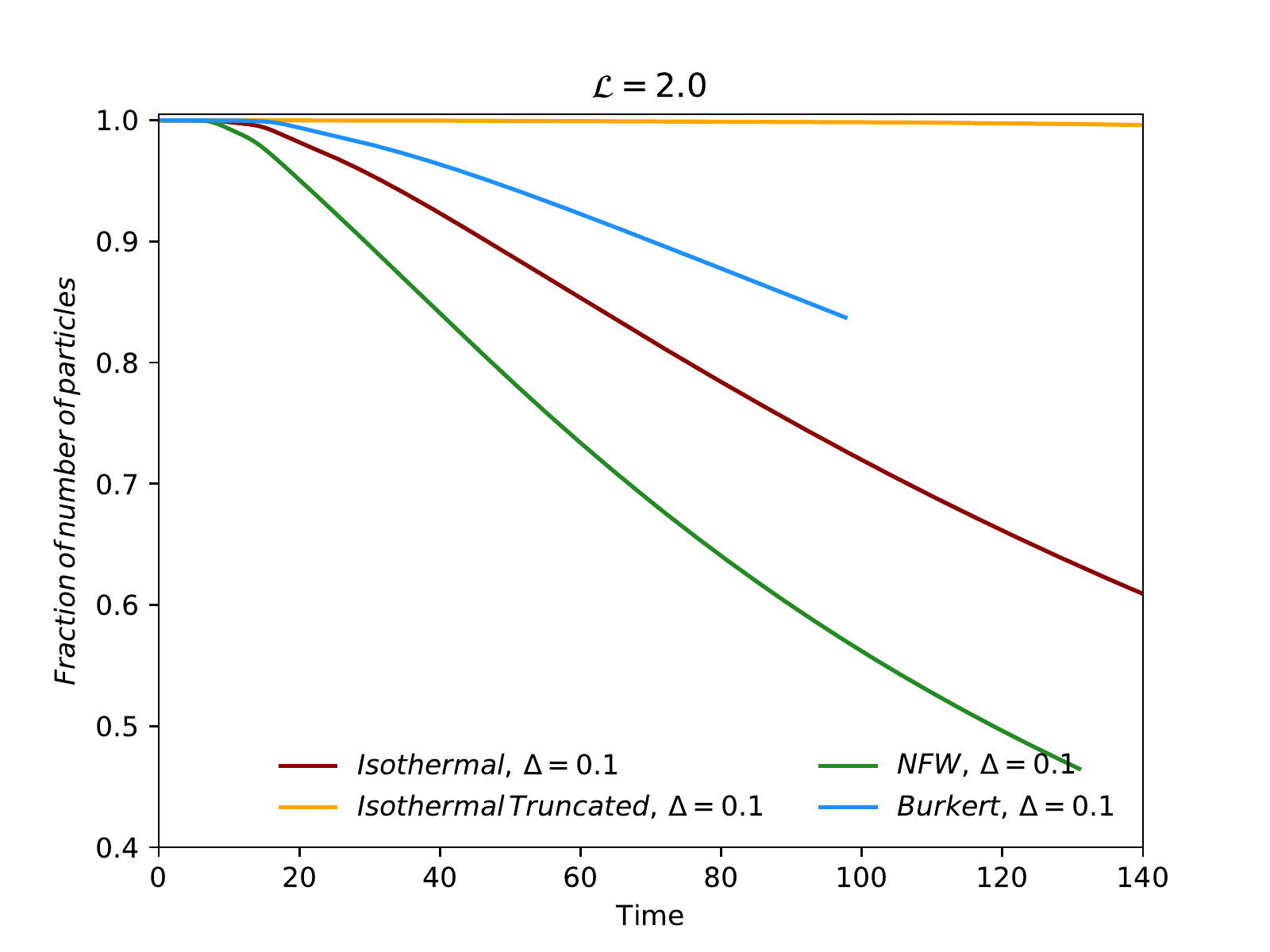}}
\subfigure{\includegraphics[scale=0.4]{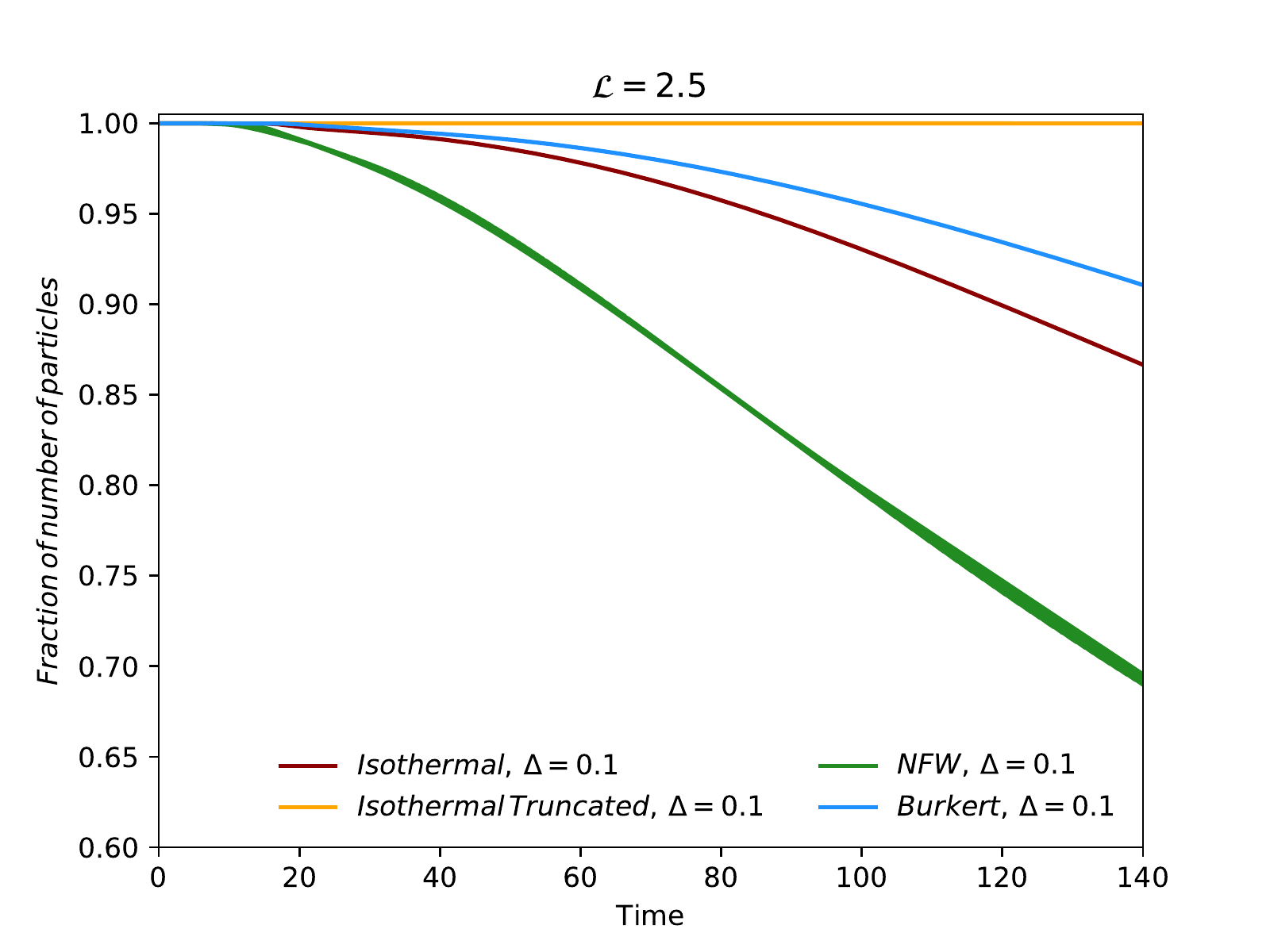}}
\subfigure{\includegraphics[scale=0.4]{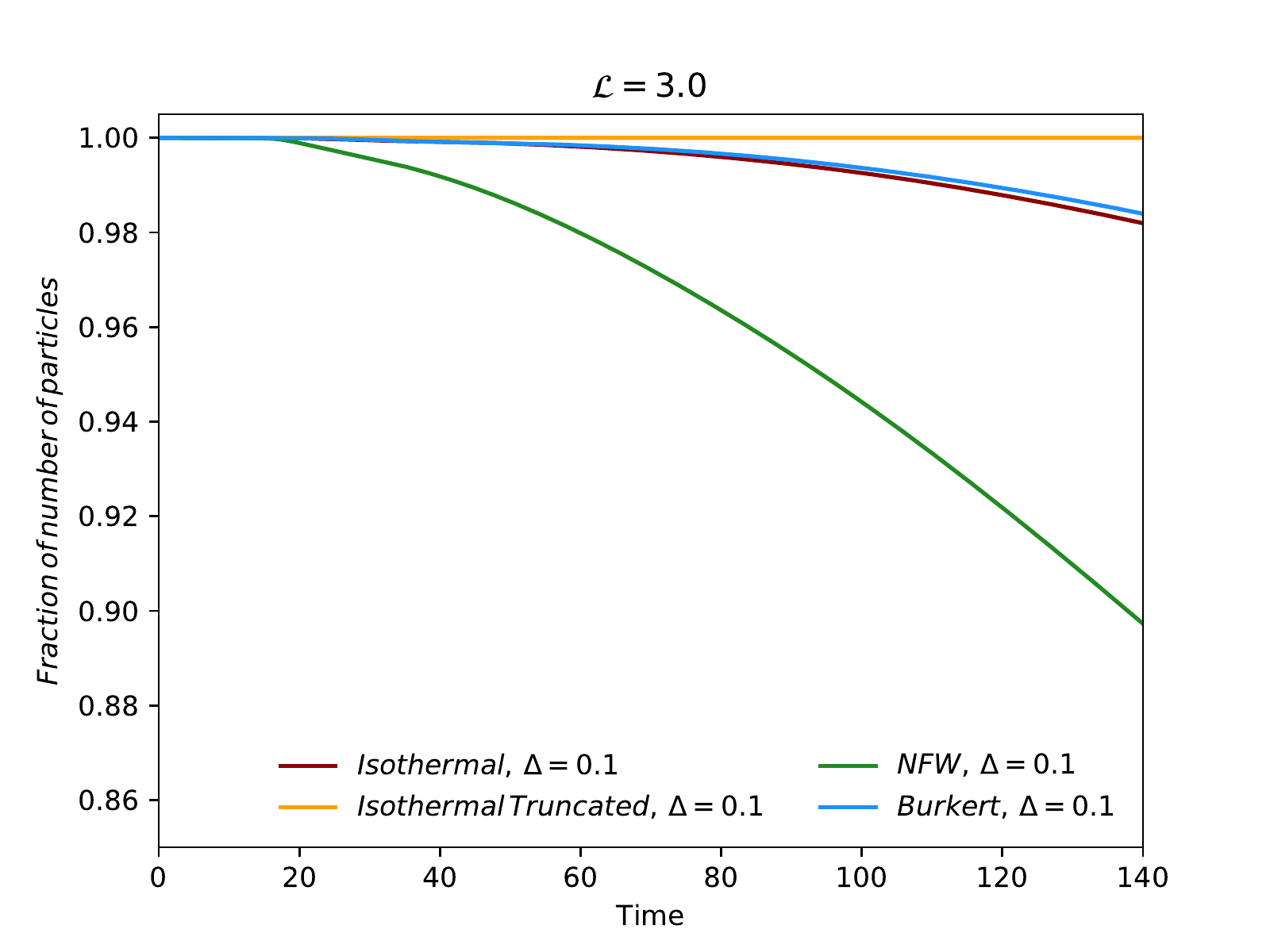}}
\subfigure{\includegraphics[scale=0.4]{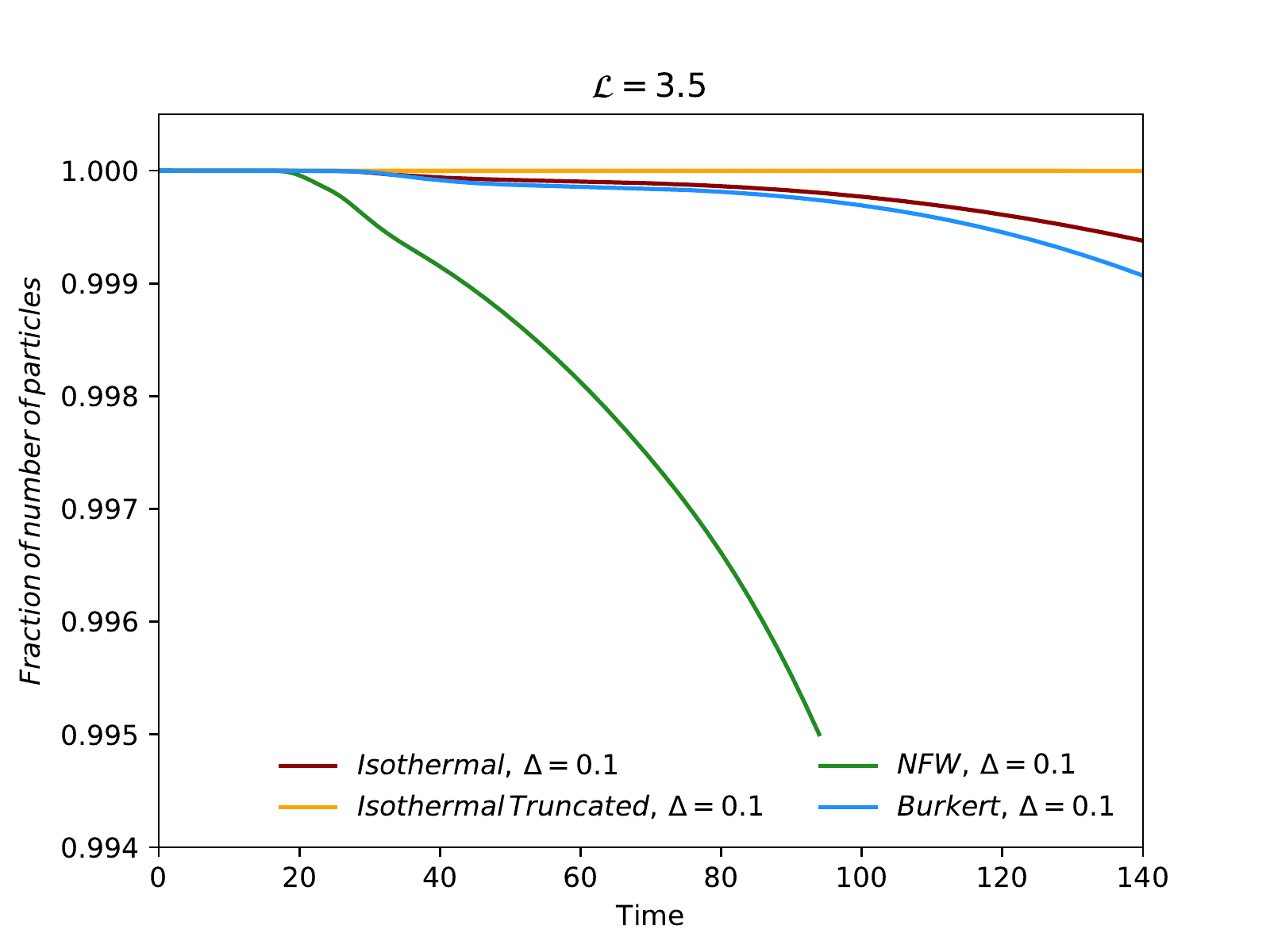}}
\caption{Total number of particles integrated over the complete
  computational domain for the different halo models, and different
  values of the angular momentum ${\cal L}$, for a resolution of $\Delta
  d = \Delta {\cal P} = 0.1$.}
\label{pics:Npart_low}
\end{figure}

\begin{figure}[H]
\centering
\subfigure{\includegraphics[scale=0.4]{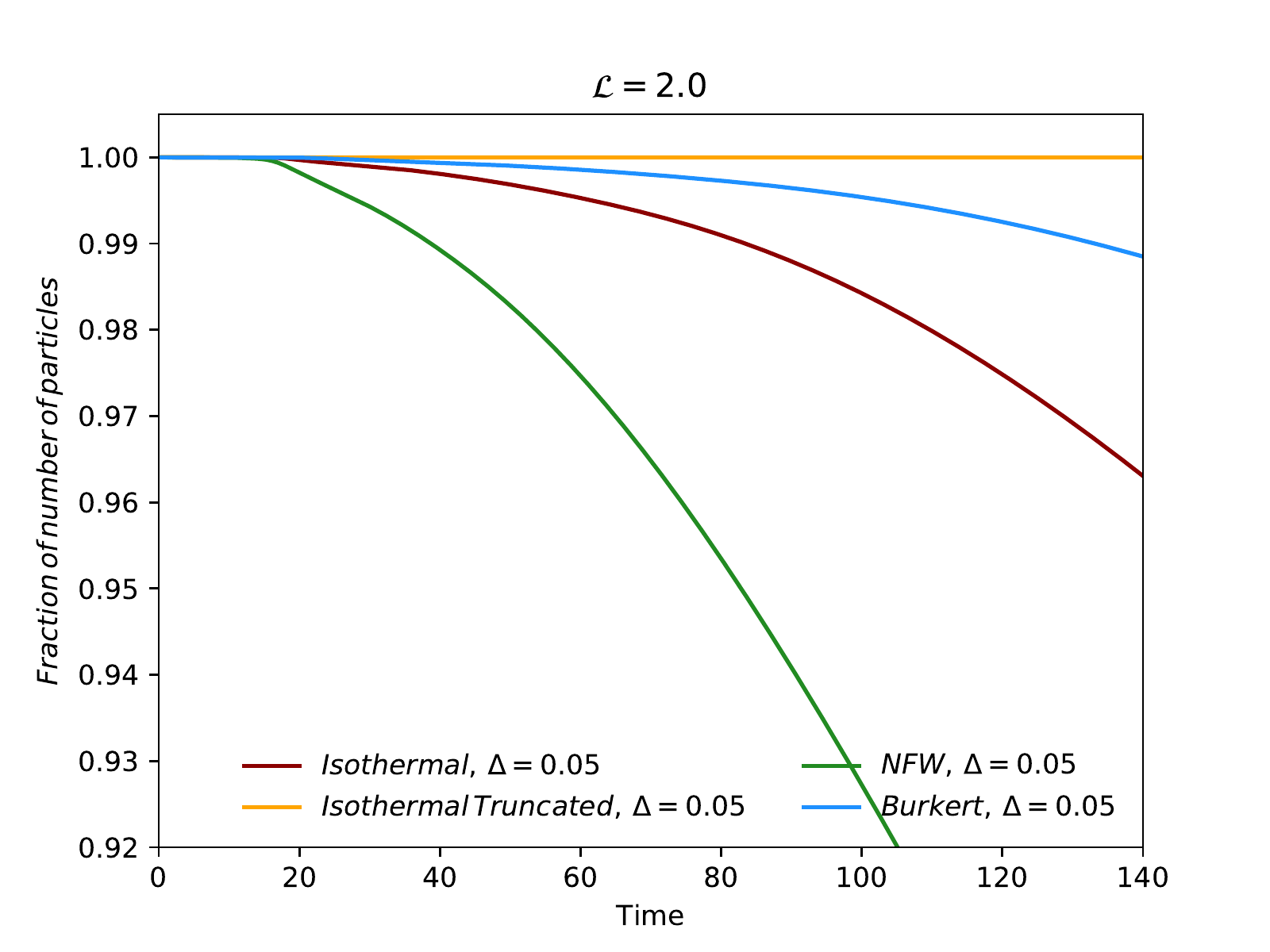}}
\subfigure{\includegraphics[scale=0.4]{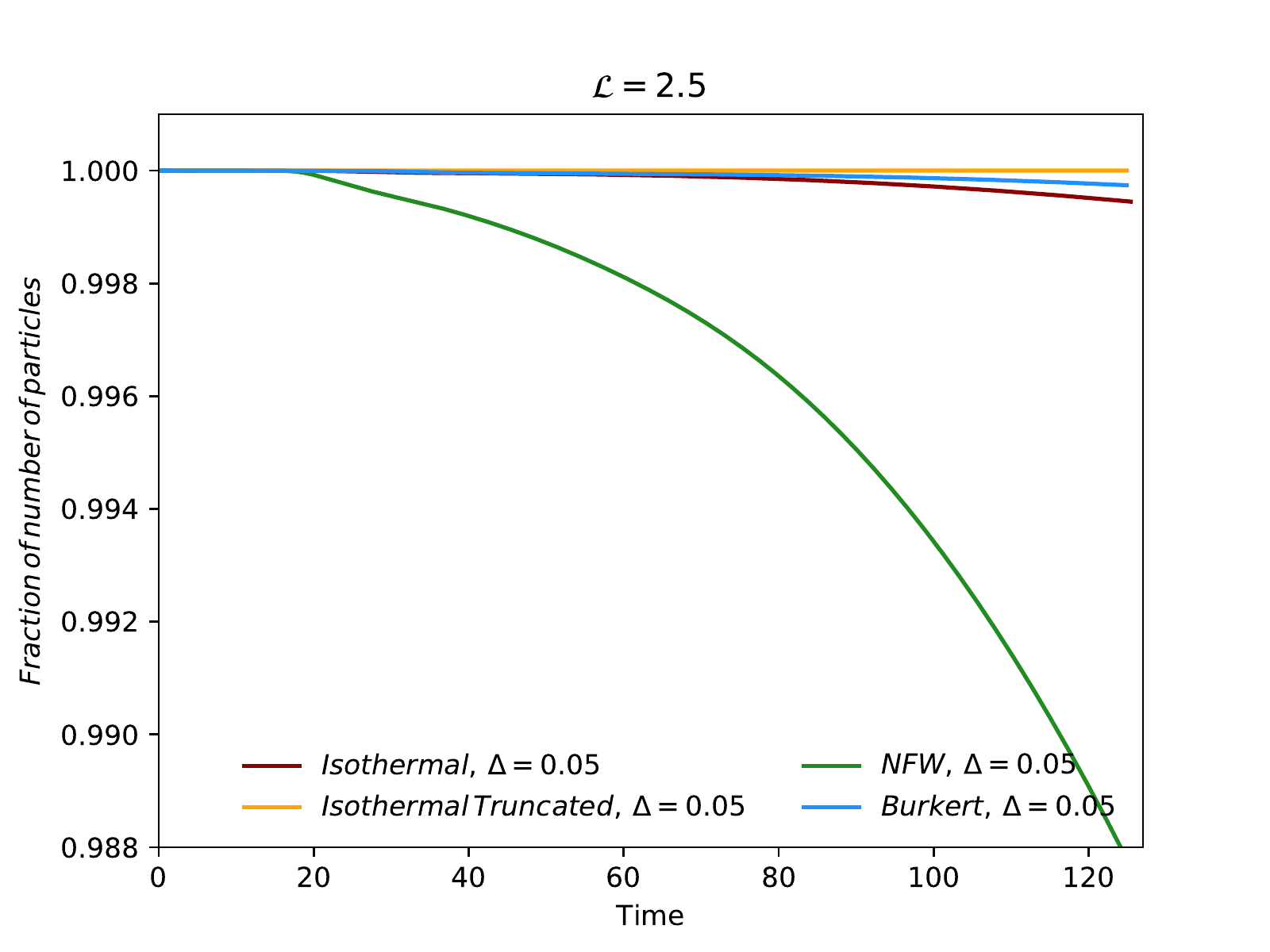}}
\subfigure{\includegraphics[scale=0.4]{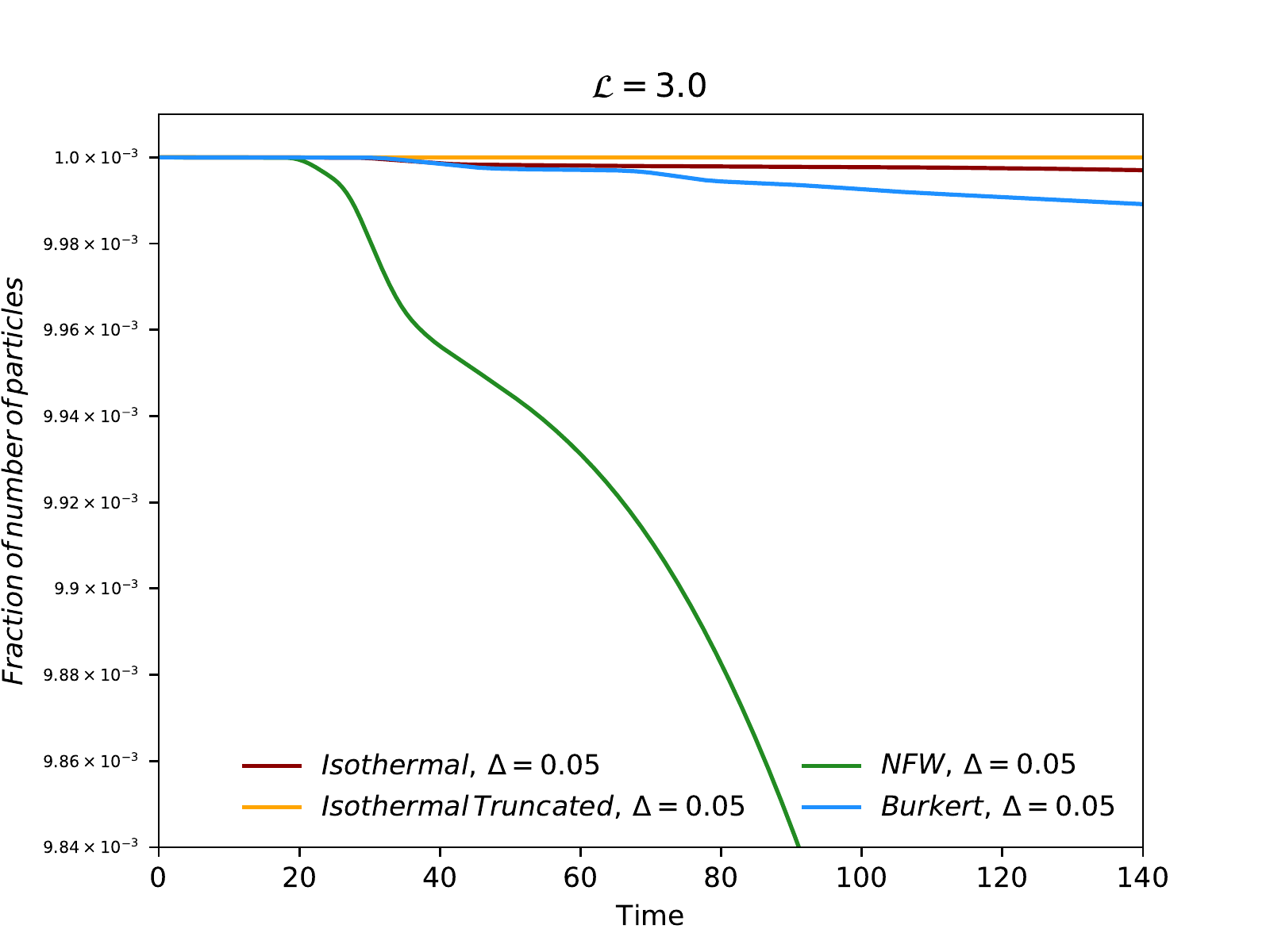}}
\subfigure{\includegraphics[scale=0.4]{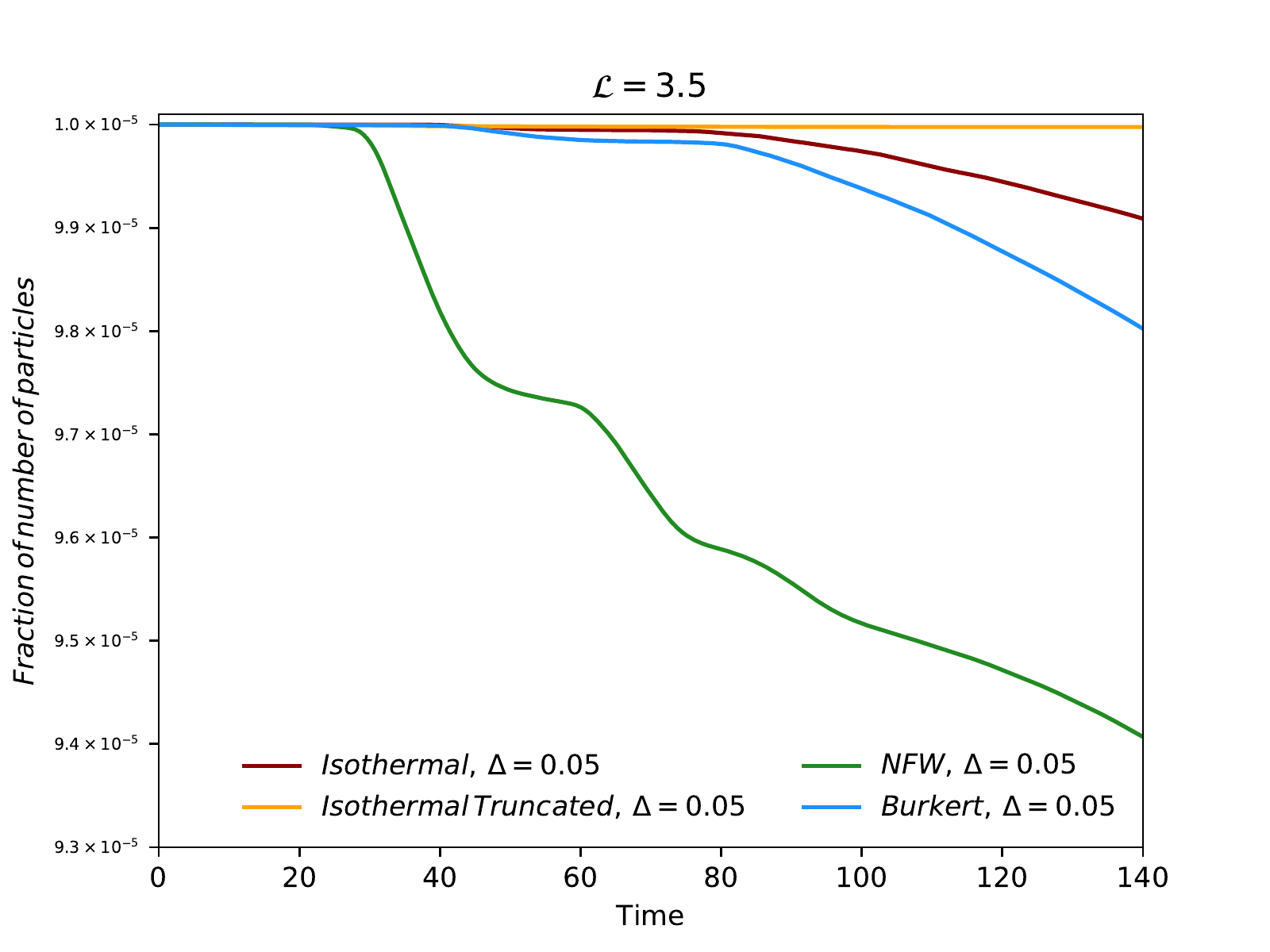}}
\caption{Total number of particles integrated over the complete
  computational domain for the different halo models, and different
  values of the angular momentum ${\cal L}$, for a resolution of $\Delta
  d = \Delta {\cal P} = 0.05$.}
\label{pics:Npart_high}
\end{figure}

In order to illustrate this behavior, in Figures~\ref{pics:Npart_low}
and~\ref{pics:Npart_high} 
we show the time evolution of the total
number of particles integrated over the whole computational domain,
for the different halo models and different values of the angular
momentum (we remind the reader that this number has been normalized to
$1$ in the initial data). Figure~\ref{pics:Npart_low} shows the results
for low resolution runs with $\Delta d = \Delta {\cal P} = 0.1$,
while Figure~\ref{pics:Npart_high} shows the results for the higher
resolution runs with $\Delta d = \Delta {\cal P} = 0.05$.

There are several things to notice from the figures.  First, at large
times we are indeed loosing particles through the boundaries. A more
detailed analysis of the fluxes through the different boundaries shows
that we are loosing them mostly through the boundary at
negative momenta.  Second, even though at the
lower resolution the loss of particles is very significant by the end
of the simulations (in one case we loose as much as $70\%$ of the
particles), at the higher resolution this loss is much lower (in the
worst case only about $8\%$), which indicates that this loss of
particles is mostly due to numerical diffusion which becomes smaller
at higher resolutions.

Considering only the results from the higher resolution,
Figure~\ref{pics:Npart_high}, we notice that more particles are lost
for low values of the angular momentum (for ${\cal L}=2.0$ the NWF
halo looses $8\%$ of the particles), while for larger angular momentum
the loss of particles becomes extremely small (for ${\cal L}=3.5$ the
NFW halo looses only about $0.0007\%$).  We also notice that the NFW
halo is the one which has more trouble keeping the particles inside
the computational domain whereas the truncated isothermal is the one
which best preserves the total number of particles.

\section{Discussion}
\label{Sec5}

We have studied the (Newtonian) Vlasov equation in spherical symmetry
for the case of particles moving in a background gravitational
potential.  We have shown that in this case the problem effectively
reduces to that of a three-dimensional phase-space corresponding to
the radial coordinate $r$, the radial momentum $p_r$, and the total
angular momentum $L$. Moreover, since the angular momentum is
conserved, it can effectively 
be decoupled from the other phase space
coordinates, so that one can consider the motion of particles all of
which have the same value of the angular momentum $L_0$, 
reducing the problem to a two-dimensional case. We have
constructed a numerical code to solve directly the Vlasov
equation given several values of angular momenta $L_0$
using a conservative TVD (flux limiter) scheme. 

We have used our code to study the evolution of an initial localized
distribution of particles in phase space in the background
gravitational potential of four different models for a dark matter
halos, namely the isothermal, truncated isothermal, Burkert and NFW
models.  We interpret this initial distribution as a inhomogeneity in
the dark matter halo. Knowing that the
resulting Vlasov equation implies the continuity equation and also
the standard virial theorem for stationary solutions, we have used
these results for testing the corresponding temporal evolution. 
We find that, for all the cases considered,
during evolution the initial distribution leads to non-trivial
stationary final distributions that satisfy the virial theorem. The
detailed properties of such final distributions change from one type of
halo to another, but the overall features are very similar. We believe
the detailed study of the evolution of such inhomogeneities could be
relevant to characterize the observed halos.

Even though the virial theorem has to be satisfied for equilibrium
states, by itself it does not tell us under what conditions an arbitrary
initial configuration will evolve toward a stationary configuration,
nor how such an evolution will develop.  It is well known that an
arbitrary distribution that depends only on conserved quantities like
the total energy $E$ and angular momentum $L$, $f=f(E,L)$, is
automatically a stationary solution of the Vlasov equation.  However,
for arbitrary initial configurations such as those studied here, we
can't predict a priori what stationary solution will be reached, or
even if such a stationary situation will be reached. As we have seen,
the increase of entropy is not a useful concept in this case, since
the Vlasov equation preserves entropy.  We therefore believe that the
use of ergodic theory might be useful in this case.

It is important to mention that the dynamical description of dark
matter when seen from the point of view of kinetic theory, as has been
done here, differs from the standard description in terms of N-body
simulations \cite{Harker:2005um}, though of course both descriptions should coincide in the
limit of a very large number of bodies~\cite{Colombi:2015eia}

Still, a statistical description in terms of kinetic theory allows us to
define a continuous distribution of matter, as opposed to a discrete
number of point particles, which could have important advantages and
in particular would be easier to generalize to the case of general
relativity (where point particles are conceptually problematic).
Also, our description here is very different from other approaches
that consider dark matter as a type of fluid as
in~\cite{Barranco:2013wy}, or as a ultra-light scalar field as
in~\cite{Burt:2011pv,Barranco:2012qs,Barranco:2013rua}.

The results obtained and presented in this manuscript are encouraging,
showing that dark matter inhomogeneities can be well described by
kinetic theory, leading to non-trivial equilibrium configurations.
With these results, there are two clear avenues for further research.
The first is to consider the case of particles with a distribution of
different values of the angular momentum, which would require a
three-dimensional code for the phase space coordinates ($r,p_r,L$).
On the other hand, one can consider the case of a system of
self-gravitating particles, which would require us to solve the
Poisson equation to obtain the gravitational potential.  And of
course, one can move away from the simple Newtonian description to a
general relativistic one (see
e.g.~\cite{MartinGarcia:2001nh,Akbarian:2014gna}).  All these
problems are currently being investigated, and we will report their
progress in the near future.


\bibliographystyle{bibtex/prsty}
\bibliography{referencias}
\bibliographystyle{unsrt}


\end{document}